\documentclass[12pt]{article}
\usepackage{amsmath}
\usepackage{amssymb}
\usepackage[dvips]{graphics}
\usepackage{epsfig}
\usepackage{calc}
\usepackage{amsfonts}
\usepackage{graphicx}
\usepackage[ansinew]{inputenc}
\usepackage{color}

\usepackage{amsfonts}
\usepackage{amsthm}
\usepackage[svgnames,hyperref]{xcolor}
\newcommand{\mycolor}{Navy}
\usepackage[pdfauthor={Ada}, colorlinks, linktocpage, citecolor = \mycolor, linkcolor = \mycolor, urlcolor = \mycolor]{hyperref}

\def\be{\begin{equation}}
\def\ee{\end{equation}}
\def\ba{\begin{eqnarray}}
\def\ea{\end{eqnarray}}
\def\bdm{\begin{displaymath}}
\def\edm{\end{displaymath}}

\def\bra#1{\left\langle #1\right|}

\def\ket#1{\left| #1\right\rangle}

\def\del{\partial}
\def\d{{\rm d}}
\def\tr{\,{\rm tr}\,}
\def\str{\,{\rm str}\,}
\def\Tr{\,{\rm Tr}\,}
\def\Det{\,{\rm Det}\,}

\def\a{\alpha}
\def\b{\beta}
\def\g{\gamma}
\def\G{\Gamma}
\def\dd{\delta}
\def\D{\Delta}
\def\e{\epsilon}
\def\f{\phi}

\def\p{\psi}
\def\pb{\bar\p}
\def\P{\Psi}

\def\r{\rho}
\def\th{\theta}

\def\t{\tau}
\def\m{\mu}
\def\n{\nu}
\def\o{\omega}
\def\O{\Omega}
\def\l{\lambda}
\def\L{\Lambda}
\def\s{\sigma}

\def\vf{\varphi}

\def\Ub{\bar U}

\def\trR{{\rm tr}_\cR\,}
\def\strR{{\rm str}_\cR\,}
\def\trD{{\rm tr}_D\,}

\def\cN{{\cal N}}
\def\cL{{\cal L}}
\def\cR{{\cal R}}

\def\cC{{\cal C}}
\def\cF{{\cal F}}

\def\cO{{\cal O}}
\def\cD{{\cal D}}
\def\cA{{\cal A}}
\def\cU{{\cal U}}
\def\cUb{\bar{\cal U}}
\def \cH {{\cal H}}

\def\cE{{\cal E}}
\def\cM{{\cal M}}

\def\wt{\widetilde}
\def\wh{\widehat}

\def\no{\noindent}

\def\qsl{{q\hskip-1.8mm /}}
\def\ksl{{k\hskip-1.9mm /}}
\def\psl{{p\hskip-1.6mm /}}

\def\Asl{{A\hskip-2.2mm /}}
\def\dsl{{\del\hskip-2.0mm /}}

\def\Dsl{{D\hskip-2.5mm /}}

\def\ov{\overline}

\def\footksl{{k\hskip-1.6mm /}}
\def\footpsl{{p\hskip-1.5mm /}}

\begin{document}
\topmargin0.4cm

%
%
%
%
\begin{titlepage}
\begin{flushright}
August 2026\\
\end{flushright}
\vskip 0.5cm

\begin{center}
{\Large\bf A Pedagogical Introduction to}\\
\vskip2.mm
{\Large\bf  Anomalies in Quantum Field Theory }\\

\vskip12.mm

{\large\bf Ada \c Cubuk\c cu and Adel Bilal}
\\

\medskip
\it {Laboratoire de Physique de l'\'Ecole Normale Sup\'erieure\\
PSL University, CNRS, Sorbonne Universit\'e, Universit\'e Paris Cit\'e\\
24 rue Lhomond, F-75231 Paris Cedex 05, France}

\end{center}
\vskip .3cm

\begin{center}
{\bf \large Abstract}
\end{center}
\begin{quote}
These notes are aimed at providing a relatively concise and self-contained introduction to anomalies in quantum field theory. A first part reviews the necessary notions of quantum field theory, in particular its functional integral formulation which we use throughout. Then we discuss in some detail how and why anomalies arise as non-invariance of the functional integral measure or through the absence of a regulator preserving all symmetries, and how they manifest themselves as current non-conservation in anomalous loop diagrams and anomalous Ward identities, or equivalently as a non-invariance of the effective action. Specifically, we deal with the so-called abelian anomaly under chiral transformations, and the so-called chiral anomaly which is a violation of non-abelian gauge invariance in the presence of chiral fermions. We perform a detailed triangle Feynman diagram computation for both cases. Then we move on to provide a brief overview of some important properties of anomalies (locality, finiteness, one-loop exactness), and  how they are characterized in general by the Wess-Zumino consistency condition. We show how this condition can be reformulated in terms of BRST cohomology and how it can be solved, for arbitrary (even) space-time dimensions,  by the descent equations. 
Finally, we briefly discuss the general relation between anomalies in arbitrary even dimensions $2n$ and index theorems in $2n+2$ dimensions, generalizing also to gravitational anomalies. We conclude with the exhibition of the highly non-trivial anomaly cancellation in ten-dimensional IIB supergravity.
\end{quote}

\end{titlepage}


\newpage
\thispagestyle{empty} 
\begin{small}
\setlength{\baselineskip}{.38cm}
\tableofcontents{}
\end{small}

\newpage
\setcounter{footnote}{0} 
\setcounter{page}{1}
\setcounter{section}{0}

\setlength{\baselineskip}{.54cm}

\section{Introduction\label{sec-introduction}}
\setcounter{equation}{0}

An anomaly is the breaking of a classical symmetry by quantum effects. This was first observed in 1969 as a non-conservation of the axial-vector current by Adler \cite{Adler} and Jackiw and Johnson \cite{Jackiw-Johnson}. Since (global) symmetries in quantum mechanics impose selection rules, the breaking of such a symmetry implies that forbidden transitions actually become allowed. It was quickly realized that this current non-conservation provided an explanation for the observed disintegration rate of the neutral pion into two photons \cite{Bell-Jackiw-pi0}. 
While tree-level Feynman diagrams essentially only reflect classical physics, typically quantum effects show up in loop diagrams. In particular, the current non-conservation is seen by explicitly computing a certain one-loop diagram with three vertices (triangle diagram). It has been shown by Adler and Bardeen that there are no higher-loop corrections to this anomaly \cite{Adler-Bardeen, Adler-all}. This is quite amazing, as it means that one can get an exact expression for this anomaly by solely computing the one-loop diagram - a rare situation in quantum field theory ! This anomaly is now referred to as the abelian anomaly or the Adler-Bardeen-Jackiw (ABJ) anomaly.

But anomalies also occur for local symmetries such as gauge symmetries in (abelian or non-abelian) gauge theories, as well as in quantum field theories on curved space-times, where they manifest themselves as a lack of general covariance or of local Lorentz symmetry (gravitational anomalies). A very nice overview is given in the first 5 chapters of a classical paper by Alvarez-Gaum\'e and Witten \cite{Alvarez-Gaume-Witten}. These anomalies again show up in one-loop diagrams in the presence of chiral fermions, and are often referred to as chiral anomalies (although ``gauge anomalies" would be a more fitting name). A rather detailed discussion of these anomalies can be found in many places, including \cite{Weinberg-book}-\cite{BvN}, as well as \cite{ABanomaly}.\footnote{
Although reference \cite{ABanomaly} by one of the  authors has inspired several parts of these notes, the present introduction to anomalies is written in a different spirit~: it is meant to be at the same time more concise and more self-contained, omitting certain aspects and emphasizing others.
} 
Again, only one-loop diagrams contribute to the anomaly, although there are different such diagrams that are anomalous (triangle, box).\footnote{
The same is true for the abelian anomaly.
} 
However, the form of the anomaly is constrained by the so-called Wess-Zumino consistency condition \cite{Wess-Zumino} which relates the contributions of the different diagrams, so that only an explicit computation of the triangle is necessary. 

The Wess-Zumino consistency condition is very conveniently reformulated as the anomaly being the BRST variation of the effective action and, hence, it is BRST closed. The freedom 
to add local counterterms to the classical action appears as the possibility to add BRST-exact terms to the anomaly. Relevant anomalies then appear as BRST cohomology classes at ghost number one. These cohomology classes are related by the descent equations \cite{Zumino}  to certain  characteristic classes in two more dimensions. The latter  uniquely characterize the anomaly. This provides an intriguing link with index theorems for the Dirac operator \cite{indextheoremAS}, as spelled out in \cite{AGG}, which is particularly useful when discussing gravitational anomalies in higher dimensions.

Contrary to anomalies in global symmetries, an anomaly of a local (gauge) symmetry is a real problem, as it would eventually ruin the renormalizability \cite{GrossJackiw}, unitarity and/or Lorentz invariance of the theory. (Similar remarks apply to gravitational anomalies.) Luckily, not all gauge groups lead to anomalies \cite{GeorgiGlashow}, and when they do, different chiral fermions contribute additively to the anomaly. Thus with the right combination of chiral fermions  it may happen that the individual anomalies cancel and the total anomaly vanishes. This is precisely the case in the standard model of particle physics as was first shown by Bouchiat, Iliopoulos and Meyer \cite{BIM}, as well as in several prominent higher-dimensional theories that appear as low-energy limits of certain superstring theories \cite{GreenSchwarz}, such as the ten-dimensional type I supergravity coupled to super Yang-Mills theory for specific gauge groups, or type IIB supergravity in 10 dimensions.

These notes are organized as follows : Section 2 is dedicated to a quick introduction to the functional integral formulation of quantum field theory which we will use throughout, and how this is related to standard Feynman diagrams.
The reader who is well acquainted with this material may safely skip this section.
The next Section 3 discusses how the various manifestations of an anomaly (current non-conservation, anomalous variation of the effective action, non-invariance of the functional integral measure) are related via  Ward type identities. We present a detailed computation of the abelian anomaly from the transformation of the functional integral measure, following Fujikawa and Suzuki \cite{Fujikawa,Fujikawa-Suzuki}. A first relation to the Atiyah-Singer index theorem \cite{indextheoremAS,EGH, indextheorem} is outlined. In Section~4, we present the explicit Feynman diagram computation of the triangle diagrams, both for the abelian and the chiral (gauge) anomalies, following closely what is done in \cite{ABanomaly, Weinberg-book}. In Section 5, we briefly discuss some general properties of the chiral anomaly (locality, finiteness, one-loop exactness), and their characterization by the Wess-Zumino consistency condition, as well as the issue of anomaly cancellation. Section 6 contains some more formal developments, including the reformulation of the consistency condition in terms of BRST cohomology and how to solve this using the descent equations which relate the anomaly in arbitrary even dimensions to certain characteristic classes in two more dimensions. Following Alvarez-Gaum\'e and Ginsparg \cite{AGG}, the final Section 7 briefly discusses the general relation between anomalies in $2n$ dimensions and index theorems in $2n+2$ dimensions, providing also a brief introduction to gravitational anomalies. We conclude by exhibiting the highly non-trivial anomaly cancellation in ten-dimensional IIB supergravity, as first found in \cite{Alvarez-Gaume-Witten}.



\section{Some elements of quantum field theory and the functional integral}
\setcounter{equation}{0}

This section aims to give an overview of the functional integral formulation of quantum field theory we will be using throughout these notes, and how it is related to the (perturbative) Feynman diagram expansion.  While this is standard material and can be found in many textbooks, the presentation here will follow mostly \cite{Weinberg-book,ABM3}. Many readers may safely skip this section.

\subsection{Functional integrals}

Functional integrals are the generalization to QFT of path integrals in ordinary QM. Although not necessarily mathematically rigorous, functional integrals are a useful tool for many computations whose accuracy can be confirmed by independent ``proper" Feynman diagram computations. Our signature is $(-+++)$ and, of course, $\hbar=c=1$.

\subsubsection{Path integrals in QM }

Let us first briefly introduce path integrals. We will be working in the Heisenberg picture and $\ket{q,t}$ denotes the eigenstate of the position operator $Q_H(t)$ at time $t$, corresponding to the eigenvalue $q$, and similarly $\ket{p,t}$ for the momentum operator $P_H(t)$ and eigenvalue $p$.
For a small time interval $\D t$, we can approximate the scalar product by the following integral \footnote{
Note that $H=H(Q_H(t),P_H(t))=H(Q,P)$ does not depend on $t$.
}: 
\begin{eqnarray}
\langle q',t+\D t \ket{q,t} &=& \bra{q',t}e^{-i\D t H(Q,P)}\ket{q,t}\simeq  \bra{q',t}\big(1-i\D t H(Q,P)\big)\ket{q,t}\nonumber\\
&&\hskip-2.5cm =\int\d p \bra{q',t}\big( 1-i\D t H(Q,P)\big)\ket{p,t} \langle p,t\ket{q,t}
=\int\d p \bra{q',t}\big( 1-i\D t H(q',p)\big)\ket{p,t} 
\langle p,t\ket{q,t}\nonumber\\
&&\hskip-2.5cm\simeq
\int\frac{\d p}{2\pi} e^{-i \D t H(q',p)\D t+ip(q'-q)} \ ,
\end{eqnarray}
where we assumed a certain operator ordering in $H$ such that $ \bra{q',t} H(Q,P)\ket{p,t} = H(q',p) \frac{e^{ipq'}}{\sqrt{2\pi}}$.
For a finite time interval $t_f-t_i$, we proceed by dividing it into small chunks $t_f-t_i=N \D t$ and let $t_k=t_i+k \D t$ ($t_0=t_i,\ t_N=t_f$). Repeating the previous process at each step  yields the following formula, where $\prod \d q_k \simeq  {\cal D} q$ and $\prod \frac{\d p_k}{2\pi} \simeq  {\cal D} p$: 
\begin{eqnarray}
&&\langle q_f,t_f\ket{q_i,t_i}
=\int \prod \d q_1\ldots \d q_{N-1} 
\langle q_f,t_f\ket{q_{N-1},t_{N-1}}\ldots  \langle q_1,t_1\ket{q_i,t_i}
\nonumber\\
&&\simeq \int \prod \d q_1\ldots \d q_{N-1}
\prod {\d p_1\over 2\pi} \ldots {\d p_N\over 2\pi}
\exp\left\{ -i \sum_{k=1}^{N} H(q_k,p_k)\D t + i \sum_{k=1}^{N} p_k(q_k-q_{k-1})\right\} \nonumber\\
&&\simeq \int_{q(t_i)=q_i,\ q(t_f)=q_f} 
 \cD q\, \cD p
\exp\left\{-i \int_{t_i}^{t_f} \d \tau\, H(q(\tau),p(\tau)) 
+i \int_{t_i}^{t_f} \d \tau\, p(\tau) \dot q(\tau) \right\}  \ ,
\end{eqnarray}
where in the last step we have taken the limit $N\to\infty$, so that one integrates over all paths $q(t)$ and $p(t)$ in phase space with the prescribed initial and final conditions.

This can easily be generalized to obtain the matrix elements of the time-ordered product of operators $O_a(Q(t_a),P(t_a))$ constructed from the Heisenberg operators $Q$ and $P$ at times $\{t_a\}_a$: 
\be
\bra{q_f,t_f} T \prod_a O_a(Q(t_a),P(t_a)) \ket{q_i,t_i} 
= \int_{q(t_i)=q_i}^{q(t_f)=q_f} \hskip-2.mm
\cD q \, \cD p\
\prod_a O_a(q(t_a),p(t_a))\  e^{i  \int_{t_i}^{t_f} \d \tau\, \big( p(\tau) \dot q(\tau)  - H(q(\tau),p(\tau)) \big)}\, .
\ee
This formulation is called the Hamiltonian formulation. If the operators $\cO_a$ do not depend on momenta and the Hamiltonian is at most quadratic in the latter, the integration over $\cD p$ is Gaussian and can be easily performed. This is the only case we will consider from here on out. Thus, one gets: 
\be
\bra{q',t'} T \prod_a O_a(Q(t_a)) \ket{q,t} 
= \cN \int_{q(t_i)=q_i}^{q(t_f)=q_f} \hskip-1.mm
\cD q \,
\prod_a O_a(q(t_a))\, \big(\Det \hat A\big)^{-1/2}  e^{i \int_{t_i}^{t_f}  \d \tau\, \big( p_*(\tau) \dot q(\tau) - H(q(\tau),p_*(\tau)) \big)} \ ,
\ee
where $\hat A$ denotes the integral kernel of the Hamiltonian involving the momentum,\footnote{
More precisely, if the eigenvalues of the integral kernel are $a_k$, one gets $\prod_k \frac{1}{2\pi} \sqrt{\frac{2\pi}{i a_k}} = \big(\Det \big( 2\pi i \hat A\big)\big)^{-1/2}$. But we include all the numerical factors into the normalization constant ${\cal N}$ and simply write 
$\big(\Det \hat A\big)^{-1/2}$.
} 
$p_*(t)$ the extremal argument of the exponent and $\cN$ a potentially infinite normalization constant. 
Eventually, we will be interested in normalized quantities, i.e.~quotients, and then this constant will drop out. 
It should be noted that performing this Gaussian integration corresponds to doing the Legendre transformation from the Hamiltonian to the Lagrangian, and yields the Lagrangian formulation of path integrals: 
\begin{eqnarray}
    \bra{q',t'} T \prod_a O_a(Q(t_a)) \ket{q,t} 
= \int_{q(t_i)=q_i,\ q(t_f)=q_f}\hskip-7.mm
\cD q \hskip2.mm
\prod_a \cO_a(q(t_a))\ \big( \Det \hat A\big)^{-1/2}  e^{i\int_{t_i}^{t_f}  \d \tau\, L(q(\tau),\dot q(\tau)) } \ .
\end{eqnarray}
If $\hat A$ is independent of $Q$, $\Det \hat A$ will be absorbed into the normalization constant. If not, we will rewrite this quantity as $\big( \Det \hat A\big)^{-1/2}  =\exp(-\frac{1}{2} \log \Det \hat A)=\exp(-\frac{1}{2} \Tr \log\hat A)$ and consider it an additional contribution to the Lagrangian.

\subsubsection{Functional integrals in QFT}

In QFT, the quantities of interest are the vacuum expectation values of time-ordered products of field operators (or functions of the field operators) and not so much transition amplitudes between eigenstates of the field operators. This induces two major modifications in the integral formalism: prescribed initial and final conditions become obsolete, and one needs to take the ``vacuum wave function" into account, which results in a purely imaginary, infinitesimally small contribution to the Lagrangian. These are the $i\e$-terms which will tell us that the propagators are the Feynman propagators.

When generalizing path integrals to functional integrals, we will start with the Lagrangian formulation and not preoccupy ourselves with its derivation from the Hamiltonian formulation. In particular, when dealing with (non-abelian) gauge theories, the necessary gauge-fixing and ghost terms will be added in when necessary. The Lagrangian (more precisely Lagrangian density) $\cL$ at a given space-time point $x$ depends only on the fields $\p_l(x)$ and their derivatives $\del_\m\p_l(x)$ at the same point. For brevity, we will simply denote it by $\cL(\p(x))$. For bosonic fields, this yields the following generic formula for  expectation values between the in and out vacua
\footnote{States with particle labels are hard to interpret in an interacting theory. This is why we define ``in" and ``out" states: these are eigenstates of the Hamiltonian of the interacting theory which asymptotically (as $t \to - \infty$ for in-states and $t \to + \infty$ for out-states) behave like a collection of non-interacting particles with the same labels. This ``asymptotic behavior" is, of course, to be understood with regards to time-dependent operators as the states themselves are time-independent in the Heisenberg picture. We will define in and out states more precisely below in subsection \ref{Dysonsec}.
} 
of
time-ordered products of Heisenberg-picture operators $O_a(\P)$ (that only depend on the Heisenberg fields $\P_l(x)$ and not on
their momenta $\Pi_l(x)$): 
\ba\label{bosint}
&&\hskip-2.cm \bra{{\rm vac,\ out}}   T\Big\{ 
\prod_a O_a\big(\P(x_a)\big) \Big\}
\ket{{\rm vac,\ in}} \nonumber\\
&&=\cN\, \int\prod_l \cD \p_l  
\ \big( {\rm Det}\hat{\cA}(\p)\big)^{-1/2}\ 
\prod_a O_a\big(\p(x_a)\big) \ 
\exp\left\{i \int\d^4 x \,\cL(\p(x)) 
 \right\} ,
\ea
where the notation $\p$  (and $\P$) is used to indicate a dependence in all fields $\p_l$ (and $\P_l)$, and $\hat\cA$ is again the integral kernel of the term in the Hamiltonian that is quadratic in the momentum fields. In most theories, $\hat\cA$ does not depend on the fields $\p$, and its determinant is an irrelevant constant.

For anti-commuting Dirac fields $\p$ and $\pb$, the analogue of the bosonic formula \eqref{bosint} becomes the following: 
\ba\label{matrixFIferm}
&&\hskip-2.cm\bra{{\rm vac,\ out}} \,
T\Big\{ \prod_a O_a\big(\P(x_a),\ov\P(x_a)\big)\ \Big\}
\ket{{\rm vac,\ in}} \nonumber\\
&& =\cN\,\int\prod_l \cD \p_l  \cD \ov\p_l\ 
\prod_a O_a\big(\p(x_a),\ov\p(x_a)\big) \exp\left\{i \left[\int\d^4 x\, \cL(\ov\p,\p) \right]\right\} \ .
\ea
Note that the normalization constant $\cN$ depends on how exactly one defines the functional integral measure, but does not depend on which operator insertions are present.

\subsection{Propagators}
The free propagators are defined as: 
\be\label{prop}
-i\D_{lk}(x,y)=\bra{\rm vac,\, out} T(\P_l(x) \P_k(y))\ket{\rm vac,\ in} \vert_{\rm no\ interactions} \ .
\ee
In a free theory, Heisenberg and interaction pictures coincide. Free propagators are determined by the free part of the action, i.e.~the part of the Lagrangian density that is quadratic in the fields. Hence, the computation of the propagators reduces, yet again, to computing a Gaussian integral. 
In the bosonic case, the free action is of the form \footnote{
One should not confuse the kernel $\cD(x,y)$ acting on $\p$ with the functional integral measure $\cD\p$.
}:
\be\label{freelag}
\int\d^4x\, \cL_0 = - {1\over 2} \int\d^4x\, \d^4y\, \sum_{l,l'} \p_l(x) \cD_{l,l'}(x,y) \p_{l'}(y) \ .
\ee
For a hermitian scalar field,  $\cL_0=-\frac{1}{2}\del_\m\f\del^\m\f-\frac{m^2}{2}\f^2$, so that
$\cD(x,y)=(-\del^\m\del_\m+m^2) \dd^{(4)}(x-y)$.
In the free theory the generic formula (\ref{bosint}) becomes: 
\be\label{Deltalk}
-i\D_{lk}(x,y)=\cN\ \int\prod_{l'} \cD \p_{l'}  
\ \p_l(x)\, \p_k(y)
\exp\left\{i\int\d^4x\, \cL_0 \right\} .
\ee
In a free theory, the ``in" and ``out"  states coincide  and one has:
\be
1=\langle{\rm vac,\, out} \ket{\rm vac,\ in} \vert_{\rm no\ interactions} \  =\ \cN\ \int\prod_{l'} \cD \p_{l'}  
\ \exp\left\{i\int\d^4x\, \cL_0 \right\}\ ,
\ee
which allows us to eliminate the normalization constant by rewriting \eqref{Deltalk} as 
\be
-i\D_{lk}(x,y)= {\int\prod_{l'} \cD \p_{l'}  \ \p_l(x)\, \p_k(y)
\exp\left\{i\int\d^4x\, \cL_0 \right\}
\over
\int\prod_{l'} \cD \p_{l'}  
\ \exp\left\{i\int\d^4x\, \cL_0 \right\} }\ .
\ee

Let us now introduce the generating functional for the free theory: 
\be\label{Z0J}
Z_0[J]= \int\prod_{l'} \cD \p_{l'} \ \exp\left\{i\int\d^4x\, \big[ \cL_0(x) + J_l(x) \p^l(x)\big] \right\} \ .
\ee
One can then easily see that, in a free theory, the $n$-point function can be rewritten using the functional integral and derivative formalism, as well as the generating functional, as: 
\begin{eqnarray}\label{Z0def}
&&\hskip-0.7cm\bra{\rm vac,\, out} T(\P_{l_1}(x_1) \ldots \P_{l_n}(x_n))\ket{\rm vac,\, in} \Big\vert_{\rm no\ interaction} \hskip-2.mm={
 \int\prod_{l'} \cD \p_{l'}  \ \p_{l_1}(x_1) \ldots \p_{l_n}(x_n)
\exp\left\{i\int\d^4x\, \cL_0 \right\} 
\over
\int\prod_{l'} \cD \p_{l'}  
\ \exp\left\{i\int\d^4x\, \cL_0 \right\} } \nonumber
\\
&&\hskip3.cm= \big(Z_0[0]\big)^{-1}\ (-i)^n\ {\dd\over \dd J_{l_1}(x_1)}\ldots {\dd\over \dd J_{l_n}(x_n)} Z_0[J] \Bigg\vert_{J=0}\ ,
\end{eqnarray}
One can then compute $Z_0[J]$ as a Gaussian integral since $\cL_0$ is quadratic. This is done by ``completing the square" in the exponent, and yields: 
\be\label{ZoJres}
Z_0[J]= Z_0[0]\
\exp \left( {i\over 2} \int\d^4 x\, \d^4 y\, J_l(x) \cD^{-1}_{lk}(x,y) J_k(y)\right) \ ,
\ee
where $Z_0[0]=\Big( \Det \big[ {i\wh \cD\over 2\pi} \big] \Big)^{-1/2}$.
This formula yields a simple expression of the free propagator: 
\be
-i\D_{lk}(x,y)=(-i)^2\ \big(Z_0[0]\big)^{-1}\  {\dd\over \dd J_{l}(x)} {\dd\over \dd J_{k}(y)} Z_0[J] \Bigg\vert_{J=0}
= - i \cD^{-1}_{lk}(x,y) \ .
\ee

We can now use the Fourier transform to determine this expression explicitly. Translation invariance reduces the number of variables and allows us to write:
\be 
\cD_{lk}(x,y)\equiv \cD_{lk}(x-y) = \int{\d^4p\over (2\pi)^4} e^{ip(x-y)} \cD_{lk}(p)
\ee 
The inverse operator  $(\cD^{-1})_{lk}(x,y)$ is given by the Fourier transform of  $(\cD^{-1})_{lk}(p)$, which is the inverse matrix of $\cD_{lk}(p)$:
\be\label{prop2}
\D_{lk}(x,y)\equiv  \D_{lk}(x-y)= \int{\d^4p\over (2\pi)^4} e^{ip(x-y)} (\cD^{-1})_{lk}(p) \ .
\ee
For the example of the hermitian scalar field, with the replacement $m^2 \to m^2-i\e$, we get
\begin{equation*}
    \cD(x,y)=(-\del^\m\del_\m+m^2-i\e)\, \dd^{(4)}(x-y)=\int{\d^4 p\over (2\pi)^4} e^{ip(x-y)} (p^2+m^2-i\e) \ ,
\end{equation*} 
with inverse
\be\label{scalarpropa}
   \D_{\rm scalar}(x-y)=\int\frac{\d^4 p}{(2\pi)^4} \, e^{ip(x-y)}  \, {1\over p^2+m^2-i\e} \ ,
\ee
so that the momentum space propagator is 
\be\label{scalarprop}
\frac{-i}{(2\pi)^4}\, \D_{\rm scalar}(p)=\frac{-i}{(2\pi)^4}  {1\over p^2+m^2-i\e} \ .
\ee

Let us note that for fermions, i.e.~for anti-commuting fields, Gaussian functional integrals can be defined similarly as
\be\label{fermionicGaussian}
\int \cD \p \cD \ov\p\ \exp \Big( \int \big(\ov\p K \p + \ov\eta \p + \ov\p \eta\big)\Big) = \cN\ \Det K \ \exp\Big( -\int \ov\eta K^{-1}\eta  \Big) \ ,
\ee
which shows  very analogously that the fermionic propagator is also given by the inverse of the kernel of the quadratic part of the Lagrangian, i.e.~for Dirac fields by $(\dsl+m-i\e)^{-1}$, or after Fourier transform by 
\be\label{Diracpropa}
 \D_{\rm Dirac}(x-y)=\int{\d^4 p\over (2\pi)^4} e^{ip(x-y)}\ (i\psl+m-i\e)^{-1}
 =\int{\d^4 p\over (2\pi)^4} e^{ip(x-y)}\ \frac{-i\psl+m}{p^2+m^2-i\e} \ ,
 \ee
and the momentum space propagator is
\be\label{Diracprop}
\frac{-i}{(2\pi)^4}\,  \D_{\rm Dirac}(p)= \frac{-i}{(2\pi)^4} \, (i\psl+m-i\e)^{-1}
 =\frac{-i}{(2\pi)^4} \, \frac{-i\psl+m}{p^2+m^2-i\e} \ .
 \ee
Note that in the fermionic Gaussian integral \eqref{fermionicGaussian} the determinant of the kernel $K$ appears with a positive power.

\subsection{Interactions and Feynman diagrams}

We now go back to the full interacting theory and explain how to compute the vacuum expectation values \eqref{bosint} of time-ordered Heisenberg-picture operators. Without much loss of generality, we will assume that the products of the $O_a(\P(x_a))$ are simply products of  the $\P(x_b)$ (and $\Det \hat\cA={\rm const}$). Then, much as in \eqref{Z0def}, these vacuum expectation values $\bra{\rm vac,\, out} T(\P_{l_1}(x_1) \ldots \P_{l_p}(x_p))\ket{\rm vac,\, in}$ can be obtained by multiple functional derivatives with respect to a source function $J(x)$ of a generating functional
\be\label{fullZdef}
Z[J]= \int \cD \p \ \exp\left\{i\int\d^4x\, \big[ \cL(x) + J_l(x) \p^l(x)\big] \right\} \ ,
\ee
where now the full Lagrangian  is present in the exponent. 
Moreover, to eliminate the unknown normalization constant $\cN$ we take again a ratio with $\langle\rm vac,\, out \ket{\rm vac,\, in}$. Note that, contrary to the free theory, the latter no longer equals one, since the in- and out-vacua now differ by a non-trivial phase. Then the ``normalized $p$-point function" is given by
\be\label{ZJformula}
\frac{\bra{\rm vac,\, out} T(\P_{l_1}(x_1) \ldots \P_{l_p}(x_p))\ket{\rm vac,\ in}} {\langle\rm vac,\, out \ket{\rm vac,\, in} }
= \big(Z[0]\big)^{-1}\ (-i)^p\ {\dd\over \dd J_{l_1}(x_1)}\ldots {\dd\over \dd J_{l_p}(x_p)} Z[J] \Bigg\vert_{J=0}\ .\
\ee

\subsubsection{Feynman diagrams from functional integrals}

In order to relate this quantity to the sum of Feynman diagrams, we would like to do a perturbative expansion in the coupling constant. Let us begin by separating the Lagrangian into its free - the quadratic - part, and the interaction Lagrangian: 
$\cL=\cL_0 +\cL_{\rm int}$
and proceed to develop $e^{i\int\cL_{\rm int}}$ in a Taylor series: 
\ba\label{ZJdef}
Z[J]&=&\int \cD \p \ e^{i\int\d^4x\, \big( \cL_0(x) + \cL_{\rm int}(x) +  J_l(x) \p^l(x)\big)} \nonumber\\
&=& \int\cD\p \sum_{n=0}^\infty {i^n\over n!} \Big[\int\d^4 x\, \cL_{\rm int}\big(\p(x))\Big]^n\ e^{ i \int\d^4 x\ \big( \cL_0(x) + J_l(x) \p^l(x)\big)}
\nonumber\\
&=&
\sum_{n=0}^\infty {i^n\over n!}   
\Big[\int\d^4 x\, \cL_{\rm int}\big(-i{\dd\over \dd J(x)}\big)\Big]^n\ \int \cD\p\, e^{ i \int\d^4 x\ \big( \cL_0(x) + J_l(x) \p^l(x)\big)} 
\nonumber\\
&=&\ \sum_{n=0}^\infty {i^n\over n!}   
\Big[\int\d^4 x\, \cL_{\rm int}\big(-i{\dd\over \dd J(x)}\big)\Big]^n\ Z_0[J] \ ,
\ea
where $Z_0[J]$ is the generating functional of the free theory as computed above in \eqref{ZoJres} with $\cD_{lk}^{-1} \equiv \D_{lk}$~:
\be
Z_0[J]= Z_0[0]\ \exp \left( {1\over 2} \int\d^4 x\, \d^4 y\, \big(i J_l(x)\big) \big( -i\D_{lk}(x,y)\big) \big( iJ_k(y)\big)\right)
\ .
\ee
This expression allows us to give a diagrammatic interpretation of the generating functional $Z[J]$. Each functional derivative $-i\dd/\dd J_l(x)$ acting on $Z_0[J]$ yields a propagator $-i\D_{lk}(x,y)$ attached to a vertex at $x$ and multiplied with a $J_k(y)$, integrated over $y$ (which reads $\int \d^4 y (-i\D_{lk}(x,y) (iJ_k(y)))$). The number of propagators attached to each vertex is equal to the number of fields in the interaction Lagrangian, which translates into the number of functional derivatives $-i\dd/\dd J(x)$ applied to $Z_0[J]$. The number of vertices of a diagram coincides with the order of expansion $n$. Every propagator is either attached to a vertex or a  $iJ_k(z_i)$. Thus, $Z[J]$ can be interpreted as the sum of all Feynman diagrams, with any number of external lines attached to a factor $iJ(z_i)$ (integrated $\d^4 z_i$), times $Z_0[J]$. In particular, the contribution from each vertex can be read from the ``numerical factors" contained in the corresponding interaction Lagrangian. These factors typically include some coupling constant $g$, and the expansion then is an expansion in powers of $g$.

Furthermore, the additional functional derivatives in \eqref{ZJformula} act similarly, but replace a $J_{l_i}(z_i)$ by an external point $x_i$. Since in the end one is instructed to set $J=0$, there cannot be left any propagators with $J$'s attached. This means that all propagators have their ends attached either to a vertex or to an external point. Very similarly, the denominator $Z[0]$ in \eqref{ZJformula}  has all propagators connected to vertices with no external points. This equals one plus the contribution from all so-called vacuum bubble diagrams. Thus, the ratio 
\eqref{ZJformula}  appears as the ratio of all Feynman diagrams with $p$ external lines and of all Feynman diagrams without external lines (one plus all vacuum bubbles). Since the sum of all diagrams in the numerator  can be organized as the product of (i) the sum of the diagrams without vacuum bubbles and (ii) one plus all vacuum bubbles, we see that these vacuum bubble diagrams cancel between the numerator and the denominator, and the normalized $p$-point function \eqref{ZJformula} is given exactly by the sum of all Feynman diagrams with $p$ external propagators and containing {\it no} vacuum bubbles.

We should note that in \eqref{ZJdef} we have interchanged the order of the (convergent) sum for the exponential series and of the functional integration. In general this cannot be justified and  in most cases results in a divergent series. This means that $Z[J]\equiv Z[J](g)$ considered as a function of the coupling constant $g$ is not Taylor expandable, and the corresponding series is a diverging one. However, in many theories, if $g$ is small enough, the sum truncated at low order still yields very good approximations. Such series are known as asymptotic series.\footnote{Probably the best-known asymptotic series is Stirling's approximation for the $\G$-function for large arguments.}
This is the case in particular in QED where perturbation theory at order $e^9$ (four loops) yields  results for the magnetic moment of the muon that are in impressive agreement with the experimental values.

\subsubsection{Relation with Dyson's formula and Wick's theorem}\label{Dysonsec}

We will now re-derive the previous result from a somewhat more conventional perspective, using Dyson's formula and Wick's theorem. We begin with the relation between the Heisenberg-picture fields $\P_l(x)$ and the interaction-picture fields \footnote{
We use the same symbol $\p_l$ for the interaction picture fields as we used for the functional integration variables, which should hopefully not cause any confusion.
} 
$\p_l(x)$ which is
\be\label{Heis-int-rel}
\P_l(t, \vec{x})=e^{iHt}e^{-iH_0t}\ \p_l(t, \vec{x}) \ e^{iH_0t}e^{-iHt} =\O(t) \ \p_l(t, \vec{x}) \ \O^{-1}(t) \quad , \quad \O(t) \equiv  e^{iHt}e^{-iH_0t}\ .
\ee
The unitary operators $\O(t)$ provide the desired relation between the two pictures. They also allow us to define the in- and out-states. Here we only need the in- and out-vacuum states~:
\be
\ket{\rm vac,in}= \lim_{\t \to - \infty} \O(\t)\ket{0}
\quad \ , \quad
\ket{\rm vac,out} = \lim_{\t \to + \infty} \O(\t)\ket{0} \ .
\ee
Then, assuming $x_1^0>x_2^0> \ldots x_p^0$ without loss of generality, we have
\begin{eqnarray} \label{Dyson}
&&\bra{{\rm vac,\ out}}   T\Big\{ \prod_a\P_{l_a}(x_a) \Big\} \ket{{\rm vac,\ in}}
=\bra{{\rm vac,\ out}}   \P_{l_1}(x_1) \P_{l_2}(x_2) \ldots \P_{l_p}(x_p) \ket{{\rm vac,\ in}}\nonumber\\
&&=  \bra{0}\ \O ^{-1} (+ \infty) \O(t_1) \p_{l_1}(x_1) \O^{-1}(t_1) ... \p_{l_p}(t_p)\O^{-1}(t_p)\O(- \infty) 
\ket{0} \\ \nonumber
&&= \bra{0} \ \cU(+ \infty, t_1) \p_{l_1}(x_1) \cU(t_1, t_2)  \p_{l_2}(x_2) ... \p_{l_p}(x_p) \cU(t_p, - \infty) \ket{0} \ , 
\end{eqnarray}
\vskip2.mm
\noindent
where $\cU$ is the interaction picture evolution operator~:
\be\label{evolutionop}
\cU(t, t') = \O^{-1}(t)\O(t') = e^{iH_0 t}e^{-iH(t-t')}e^{-iH_0 t'} \ .
\ee
It satisfies a simple differential equation and initial condition
\be\label{evoldiffeq}
i \frac{\d}{\d t}\cU(t, t')=e^{i H_0 t} (-H_0 + H)e^{iH(t-t') } e^{-i H_0 t'} = e^{i H_0 t} V e^{-iH_0 t} \ \cU(t, t') = V_I(t) \ \cU (t, t') \ , \quad
U(t,t)={\bf 1} \ ,
\ee
$V_I(t)$ being the interaction Hamiltonian $H-H_0$ in the interaction-picture. In quantum field theory, $V_I$ is given by a space-integral of a local interaction Hamiltonian density $V_I(t)=\int \d^3 x \, \cH_{\rm int}(t,\vec{x})$. The latter is given by a sum of products of interaction-picture fields (and possibly their derivatives).
The solution of \eqref{evoldiffeq} is then given by the time-ordered exponential
\be\label{DysonU}
\cU(t, t')=T \exp\Big( -i\int_{t'}^t \d\tau V_I(\tau) \Big)
=T \exp\Big( -i\int_{t'}^t \d\tau  \int\d^3 x\, \cH_{\rm int}(\tau,\vec{x}) \Big) \ .
\ee
This evolution operator obeys the composition law
$\cU(t_1,t_2) \cU(t_2,t_3)=\cU(t_1,t_3)$.
It is then easy to see that \eqref{Dyson} can be very simply rewritten as
\be\label{Dyson2}
\bra{{\rm vac,\ out}}   T\Big\{ \prod_a\P_{l_a}(x_a) \Big\} \ket{{\rm vac,\ in}}
= \bra{0} T \Big\{ \p_{l_1}(x_1) ... \p_{l_p}(x_p) \exp{(-i \int d^4 z\, \cH_{\rm int} (z))}\Big\}\ket{0}\ ,
\ee
where the integral in the exponent is over all space and all times (from $-\infty$ to $+\infty$).

This can be evaluated in perturbation theory by expanding the exponential into a Taylor series,
\ba\label{Dyson3}
&&\hskip-2.cm\bra{{\rm vac,\ out}}   T\Big\{ \prod_a\P_{l_a}(x_a) \Big\} \ket{{\rm vac,\ in}}\nonumber\\
&&= \sum_{n=0}^\infty \frac{(-i)^n}{n!} \ T \ \Big( \int d^4z_1 ... d^4z_n \bra{0} \p_{l_1}(x_1) ... \p_{l_p}(x_p) \cH_{\rm int} (z_1) ... \cH_{\rm int} (z_n) \ket{0} \Big) \ ,
\ea
and using Wick's theorem. Wick's theorem allows us to rewrite any product of interaction-picture fields in terms of a sum of terms with an arbitrary number of pairs of fields ``contracted" and the remaining fields being in normal order. A normal-ordered expression of interaction-picture fields (i.e.~free fields) has all its annihilation operators $a$ to the right and all its creation operators $a^\dag$ to the left, so that its vacuum expectation value vanishes (since $a\ket{0}=0$ and $\bra{0}a^\dag=0$). As a result, the only terms that have a non-vanishing contribution are the ones with {\it all} fields contracted.
Each contraction between a field at $y_A$ and a field at $y_B$ (the $y$'s being either a $x_l$ or a $z_l$) is given by the  time-ordered two-point function of interaction-picture fields. Since these interaction-picture fields evolve with the free Hamiltonian $H_0$ they are free fields and the two-point function
\be\label{prop3}
\bra{0} T \p_l(y_A) \p_k(y_B)\ket{0} = - i \D_{lk}(y_A-y_B) 
\ee
coincides with the propagator as determined above. These propagators are all connected either to an external point $x_i$ or to an internal point $z_j$ which are again the vertices of the corresponding Feynman diagram. In the simplest quantum field theories, the interaction Hamiltonian and interaction Lagrangian are simply related as
$\cH_{\rm int} = - \cL_{\rm int}$,
and then it is not too difficult to convince oneself that Wick's theorem applied to Dyson's formula \eqref{Dyson3} exactly generates the same Feynman diagrams as the expansion from the functional integral above \eqref{ZJdef}.\footnote{
If the interaction Hamiltonian contains derivatives of the fields, there are some subtleties, but we will not discuss this any further.
}

\subsection{Generating functional for connected Green functions} \label{sec-conngreen}

The $p$-point  Green function $\wh G_{(p)}(x_1,\ldots ,x_p)$ without vacuum-bubbles is a  sum of connected and disconnected pieces. The {\it connected} $p$-point Green functions $G_{(p)}^C(x_1,\ldots ,x_p)$  can be recursively defined as~: $G_{(1)}^C(x)\equiv\wh G_{(1)}(x)$ and  $G_{(2)}^C(x_1,x_2) \equiv \wh G_{(2)}(x_1,x_2)-G_{(1)}^C(x_1)G_{(1)}^C(x_2)$, etc.\footnote{
For the general $p$-point function one needs to consider all non-trivial partitions ${\cal P}$ of $(x_1,\ldots x_p)$ and subtract $\sum_{\cal P} (\pm) G^{C} \ldots G^C$ from $G_{(p)}(x_1,,\ldots x_p)$, where the possible minus signs take into account the exchange of fermionic labels.}
$G_{(p)}^C(x_1,\ldots ,x_p)$  then is the sum of the corresponding connected Feynman diagrams. 

This recursive relation can also be rewritten as a relation between generating functionals. We denote $iW[J]$ the generating functional of connected Green functions: 
\be\label{WJdef}
iW[J]=iW[0]+ \sum_{p=1}^\infty {1\over p!} \int\d^4 x_1 \ldots \d^4 x_p\ G^{C,\ l_1 \ldots l_p}_{(p)}(x_1,\ldots ,x_p)\ iJ_{l_1}(x_1) \ldots iJ_{l_p}(x_p) \ .
\ee
The term of order 0, namely $iW[0]$, corresponds to connected 0 point Green function, i.e.~to connected vacuum-bubbles. It is worth noting that, for any $p\ge 1$, $G_p^C$ cannot contain vacuum-bubbles. 

One can show that this generating functional $W[J]$ of all connected Green's functions relates to the generating functional $Z[J]$ of all Green's functions  in the fully interacting theory as follows: 
\be\label{ZJWJrel}
Z[J]=\exp\big( iW[J]\big)  \ .
\ee

\subsection{Quantum effective action}\label{queffacsec}

We define the functional $\G[\vf]$ as the Legendre transform of $W[J]$. For simplicity, we suppress again all labels on the fields. First, let
\be\label{phiJdef}
\f_J(x)\ \equiv\  {\dd\over\dd J(x)} W[J]\  =\  -i {\dd\over\dd J(x)} \log Z[J]\ =\ {1\over Z[J]}\Big( -i {\dd\over\dd J(x)} Z[J]\Big) \ .
\ee
The expression on the r.h.s.~is similar to the one-point Green function without vacuum bubbles (which is the connected one-point function) $\wh G_{(1)}(x)\equiv G^C_{(1)}(x)$ but computed with $J\ne 0$, thus keeping the additional interaction terms $\f J$ in the Lagrangian. We see that $\f_J(x)$ is the connected one-point function in the presence of the additional interactions generated by the sources. 

The relation  $\f_J(x)={\dd\over\dd J(x)} W[J]$ can be inverted to get $J(x)$ as a function of $\f(x)$.  We denote $j_{\vf}(x)$ the function such that $\f_J(x)=\vf(x)$ if $J(x)=j_{\vf}(x)$. We can now use $\vf$ as variable and define the Legendre transform of $W$ as
\be\label{GammaLeg}
\G[\vf]=W[j_\vf]-\int\d^4 x\, \vf(x) j_{\vf}(x) \ .
\ee
It can be shown (cf the remark below) that this is the generating functional of one-particle-irreducible diagrams, also called 1PI, which are connected diagrams that do not become disconnected when a single line is cut.  More specifically, if one develops $\G[\vf]$ in powers of the $\vf(x_i)$, just as we developed $W[J]$ in powers of the $J(x_i)$, the coefficients $\G_{(p)}(x_1,\ldots, x_p)$ are the 1PI $p$-point functions which are given by the sum of all corresponding 1PI Feynman diagrams with $p$ external points at $x_i,\ i=1,\ldots p$.

$\G[\vf]$ also bears the name of {\it quantum effective action} because its tree level contribution coincides with the classical action.
As a matter of fact, let us introduce a loop-counting parameter $\l$ by replacing the action as $S\to {1\over \l}S$ and the sources as $J\to {1\over \l}J$. This means that all vertices are multiplied by ${1\over \l}$ and all propagators by $\l$. External lines get an overall factor $\l^0$ since they get a factor $\l$ from the propagator and a ${1\over \l}$ from the sources. The overall factor of a diagram is $\l^{I-V}$, where $I$ denotes the number of internal lines and $V$ the number of vertices. We also denote $L$ the number of loops and $C$ the number of connected components. Using the diagrammatic identity 
\be\label{IVidentity}
I-V=L-C \ ,
\ee 
the overall factor can also be written as $\l^{L-C}$. Since $C$ is fixed, $\l$ is a loop-counting parameter. We look at connected components, i.e. $C=1$. Thus, taking the limit $\l \to 0$ amounts to isolating the tree level contribution. One can then write the following loop expansions
\be
W[J,\l]=\sum_{L\ge 0} \l^{L-1} W_L[J]
\quad , \quad 
\G[\vf,\l]=\sum_{L\ge 0} \l^{L-1} \G_L[\vf] \ ,
\ee
where $W_0[J]$ and $\G_0[\vf]$ correspond to the classical limit \footnote{
If we did not use units such that $\hbar=1$, we would have $e^{\frac{i}{\hbar}S}$ and $\hbar$ would act as the loop-counting parameter $\l$. This is why $\l\to 0$ is the classical limit.
} 
(tree-level contributions), and the $W_L[J]$ and $\G_L[\vf]$ for $L\ge 1$ to $L$-loop contributions, which can be interpreted as quantum corrections.

We will now proceed by approximating the generating functional $W[J,\l]$ in the classical limit $\l \to 0$ using the saddle point method. We have, cf \eqref{ZJWJrel}, 
\be \label{WloopS}
\exp\Big\{i W[J,\l]\Big\} = Z[J,\l]=  \int \cD \p \exp\left\{\frac{i}{\l}\Big( S[\p] + \int\d^4x\,  J(x) \p(x)\Big) \right\} \ .
\ee
If we let $\l\to 0$, the integral is dominated \footnote{
For $\l\to 0$ we have the saddle-point approximation $e^{iW_0/\l}=(\Det \frac{i\cD}{2\pi})^{-1/2} e^{i \big(S(\p_J)+  \int J \p_J\big)/\l}$, or taking the logarithm, $\frac{W_0}{\l}=\frac{1}{\l}(\big(S(\p_J)+  \int J \p_J\big) +\frac{i}{2} \log \Det \frac{i\cD}{2\pi}$. As expected, the determinant only contributes at order $\l^0$ which corresponds to one loop.
} 
by the $\p_J$ that solve ${\dd S\over \dd \p}+J=0$. It follows that, at tree level,
\be\label{W0SLeg}
W_0[J]=S[\p_J]+\int\d^4 x J(x) \p_J(x) \ ,
\ee 
is the Legendre transform of the classical action.
On the other hand, since $W[J]$ is the inverse Legendre transform of $\G[\vf]$, 
we conclude that in the classical limit the quantum effective action equals the classical action~: 
\be
\G_0[\vf]=S[\vf] \ .
\ee
Since the $\G_L[\vf]$ for $L\ge 1$ correspond to loop corrections, $\G$ represents indeed a (quantum corrected or) quantum effective action.

One can show \cite{Weinberg-book, ABM3} by a similar saddle-point argument, that  doing the functional integral, with $S[\p]$ replaced by $\G[\vf]$, generates {\it at tree-level} the full generating functional $Z[J]=e^{i W[J]}$. This shows that all Feynman diagrams are generated as tree diagrams with (full) propagators that follow from the quadratic part of $\G$ and with effective vertices that follow from the non-quadratic parts of $\G$. Since a given diagram is obtained in only one way from 1PI-vertices and full propagators this means that these effective vertices are 1PI, and this is why $\G$ is the generating functional of 1PI diagrams. We will not go into further detail.

\subsection{Loop diagrams: regularization and renormalization}

We have shown above how to obtain the Feynman rules to compute a generic $n$-point Green function $\wh G_{(n)}(x_1,\ldots, x_n)$ without vacuum bubbles, or the corresponding connected $n$-point function $G_{(n)}^C(x_1,\ldots, x_n)$. If one further restricts to 1PI $n$-point diagrams, one obtains the 1PI $n$-point function $\G_{(n)}(x_1,\ldots, x_n)$. In practice one is more interested in the (four-dimensional) Fourier transforms, e.g.
\be\label{FTofGreen}
\G_{(n)}(p_1,\ldots, p_n)=\int \d^4 x_1\ldots \d^4 x_n \exp\Big(i\sum_{r=1}^n (p_r)_\m x_r^\m\Big)\, 
\G_{(n)}(x_1,\ldots, x_n) \ ,
\ee
where all momenta are counted as flowing into the diagram. It follows from invariance under translations $x_i^\m\to x_i^\m + a^\m$ that one of the integrations simply yields a $\dd^{(4)}(\sum p_i)$ of overall 4-momentum conservation.

Tree diagrams are always well-defined finite expressions of the external momenta $p_r$. In particular, all momenta carried by the external {\it and} internal propagators are perfectly determined by four-momentum conservation at each vertex.

On the other hand, for a diagram containing $L$ loops, $I$ internal lines and $V$ vertices, one has $I$ internal momenta, constrained by conservation at $V$ vertices. One of these constraints only reproduces the overall conservation of the external momenta, and hence there are $I-(V-1)=L$ unconstrained momenta~: there is exactly one  unconstrained momentum per loop, also referred to as the loop momenta. The integrations over these loop four-momenta $k_i^\m$ may be divergent at large values of $|k_i^\m|$. For example, if there is a single loop containing $N$ internal scalar lines, i.e.~$N$ scalar propagators, and if the vertices contain no momentum factors (interactions without derivatives) then the loop integrand will behave for large $k$ as $\d^4 k\, (k^2)^{-N}$. There is a standard procedure to replace $k^0$ by $i k^0_E$, called the Wick rotation, which makes the loop momentum a Euclidean 4-momentum and provides a factor of $i$ from the change of variables. For large Euclidean $k$ one can then use 4-dimensional spherical coordinates, so that $\d^4 k = k^3 \d k\, \d^3\O$ and the integrand behaves as $\d k\, k^{3-2N}$. One sees that this is logarithmically divergent for $N=2$ and convergent for $N\ge 3$.  On the other hand, if the $N$ internal lines are fermionic lines with Dirac propagators that behave as $\frac{1}{k}$, the large $k$ behavior now is $\d^4 k\, k^{-N}\sim \d k\, k^{3-N}$. This is then divergent for $N\le 4$ and convergent for $N\ge 5$.

These divergent integrals must be regularized. There are various ways to proceed. One may artificially ``improve" the large momentum behavior of the propagators in one way or the other. The Pauli-Villars regularization to be discussed in section 4 essentially does this. Another very convenient regularization is the dimensional regularization which formally continues the integrand to an arbitrary complex value $d$ of the dimension. The integral then appears as a meromorphic function of the space-time dimension $d=4-\e$ which has a pole at $d=4$ (if the original integral was divergent), i.e.~at $\e=0$. Unfortunately, dimensional regularization does not work in the presence of structures that are specific to 4 dimensions, like the $\e^{\m\n\r\s}$-tensor or the Dirac matrix $\g_5$.

Of course, once one removes the regulator, the loop expressions will diverge again. But in general, one has a clear separation between the diverging piece, i.e.~the polar part $\sim \frac{1}{\e^r},\ r=1,2,\ldots$, and the finite part. One can then add the so-called counterterms to the Lagrangian that are of higher order in the coupling constant(s) and that are $\sim \frac{1}{\e^r}$ with exactly the right coefficients, so that they produce a tree-level vertex that  cancels the diverging piece of the corresponding loop diagram. 
Allowed counterterms must be local expressions in the fields involving finitely many derivatives, corresponding to counterterm vertices that are polynomial in the momenta. It is thus important that the divergences from the loop diagrams have a corresponding structure.
Instead of adding these counterterms ``by hand", one can also consider that the original Lagrangian was made from so-called bare fields and bare masses and bare coupling constants, and then split this Lagrangian into a finite Lagrangian plus these same counterterms. We will not describe this further as we will not need it in these notes.

\subsection{An example : vacuum polarization in QED}

\begin{figure}[h]
\centering
\includegraphics[width=0.30\textwidth]{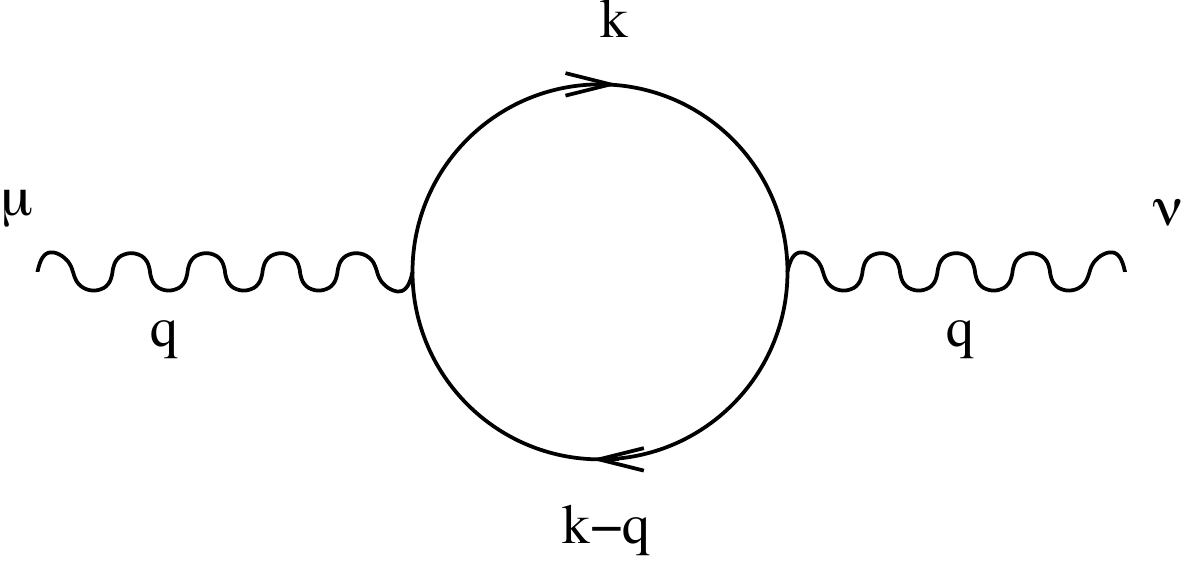}\\
\caption[]{The one-loop vacuum polarization diagram in QED}\label{vacpol}
\end{figure}

It is interesting to briefly discuss the simplest one-loop diagram in QED which is the vacuum polarization diagram. This is the 1PI two photon function $\Pi_{\m\n}(q)$. At one loop, there is a single Feynman diagram that contributes at order $e^2$ as shown in Fig.~\ref{vacpol}.  
The Feynman rule for the Dirac propagator was already given in \eqref{Diracprop}. The vertex follows from the expression of the covariant derivative in $\cL_{\rm matter}=-\pb (\Dsl+m)\p$ with $D_\m=\del_\m +ie A_\m$, so that $\cL_{\rm int}= -ie \pb\g^\m \p A_\m$. Thus, the electron-electron-photon vertex contributes $i (2\pi)^4\times (-ie\g^\m)=(2\pi)^4 e\g^\m$. Then  we get for this diagram
\be\label{vacpolexpr}
i(2\pi)^4 \Pi^{\m\n}_{{\rm loop},e^2}(q)
= - \int \d^4 k \tr \left\{ 
{(-i)\over (2\pi)^4} {-i\ksl+m\over k^2+m^2-i\e} (2\pi)^4 e \g^\m 
{(-i)\over (2\pi)^4} {-i(\ksl-\qsl)+m\over (k-q)^2+m^2-i\e}
(2\pi)^4 e \g^\n \right\} \ ,
\ee
where the overall minus sign is due to the fermion loop. 

We will not carry out the computation of this diagram which can be found in most textbooks. Let us only comment on a few properties. First of all, one sees that for large loop-momentum $k$ the integrand behaves as $\d^4 k\, k^{-2}$, so that one expects this diagram to diverge quadratically, i.e.~as $\L^2$, if one cuts off the loop-momentum at $|k|\le\L$. One also says it has superficial degree of divergence 2. Suppose now that one regularizes this expression  e.g. by dimensional regularization, letting $d=4-\e$, so that for a finite regulator ($\e\ne 0$) the integral is actually finite.  One can then split this expression into a diverging part which is $\sim\frac{1}{\e}$ and a finite part, which has a finite limit when $\e\to 0$. One can then take derivatives of this regularized integral \eqref{vacpolexpr} with respect to the external momentum $q$ and see that this  changes the large-momentum behavior of the integrand into $\d^4 k\, k^{-3}$, so that the superficial degree of divergence has been lowered from 2 to 1. Taking 2 more derivatives yields a degree of divergence $-1$, i.e.~a convergent integral. But this means that  3 derivatives have annihilated the diverging part, so the latter must be a polynomial of degree 2 in the external momentum. Indeed, the explicit computation yields \cite{ABM3, Weinberg-book}
\be
 \Pi^{\m\n}_{{\rm loop},e^2}(q)=(q^2 \eta^{\m\n}-q^\m q^\n) \Big(
-\frac{e^2}{6\pi^2}\frac{1}{\e} + f(q^2) \Big) \ ,
\ee
where $f(q^2)$ is some finite, non-polynomial function of $q^2$ (and $m^2$ and some additional scale one needs to introduce in dimensional regularization to keep the coupling constant $e$ dimensionless). The overall factor $(q^2 \eta^{\m\n}-q^\m q^\n)$ actually is dictated by gauge-invariance.
There is also a corresponding (polynomial) counterterm contribution $ \Pi^{\m\n}_{{\rm counter},e^2}(q)=(q^2 \eta^{\m\n}-q^\m q^\n) (Z_3-1)$ which is generated from the counterterm Lagrangian
$-\frac{1}{4}(Z_3-1) F_{\m\n}F^{\m\n}$, treated as an additional interaction. The a priori unknown constant $Z_3$ is fixed by requiring that the factor multiplying the $(q^2 \eta^{\m\n}-q^\m q^\n)$ vanishes at $q^2=0$. This yields 
\be
 \Pi^{\m\n}_{{\rm loop},e^2}(q)+ \Pi^{\m\n}_{{\rm counter},e^2}(q)=(q^2 \eta^{\m\n}-q^\m q^\n) \Big(
 f(q^2) -f(0)\Big) \ ,
\ee
which is now a well-defined finite function of the momentum.


\section{Various manifestations of an anomaly}
\setcounter{equation}{0}

In this section, we will give a brief overview of how an anomaly can arise and how it can manifest itself.  
We begin by recalling Noether's theorem that asserts the existence of  a classically conserved current for any symmetry of the classical action, and derive the  conservation of the corresponding quantum current in the functional integral formalism. This conservation is seen to hold provided the functional integration measure also is invariant. Said differently, this shows that an anomaly
can arise from a non-trivial  transformation of the integration measure. We will explicitly compute this anomaly for chiral transformations (involving $\g_5$) of Dirac fermions, and more specifically see how it arises from the regularization process of the determinant of the Jacobian. Secondly, we will derive the anomalous Ward identities (also called Slavnov-Taylor identities) that are identities between $n$-point functions and the anomaly. This will allow us below, in Sect.~4, to do an explicit Feynman diagram computation of different anomalies. We will also look at anomalies under gauge transformations and how the (covariant) non-conservation of the gauge current is related to the non-invariance of an appropriately defined effective action, with this non-invariance again related to the anomalous transformation of the functional integration measure. We will mostly follow \cite{ABanomaly, Weinberg-book}.
 In the next section, we will also  explicitly see how the anomaly arises in the computation of Feynman diagrams  when one cannot regularize a divergent Feynman diagram in a way that respects all the  symmetries.

\subsection{Noether's theorem and a Ward  identity}

Noether's theorem states that to any global symmetry of a classical field theory corresponds a classically conserved current. Let us give the proof for a relativistically invariant field theory described by an action $S=\int \d^4 x\, \cL(x)$ with $\cL(x)$ a scalar density. Recall that the classical field equations are the Euler-Lagrange equations obtained from requiring stationarity of the action~: under arbitrary variations $\dd\p$ of the fields $\p$ (subject to the boundary condition of vanishing at infinity) one has $\dd S=0$ if the fields satisfy the Euler-Lagrange equations. Assume now that the action is invariant, i.e.~$\dd S=0$, under some infinitesimal variation $\p(x)\to \p(x) +\D\p(x)$ where $\D\p(x)=\e\, \cF(\p(x))$, for constant infinitesimal $\e$. Then, for {\it non-constant} $\e(x)$ there will be a non-vanishing variation of $S$ that must involve the derivatives of $\e$ and thus be of the form $\dd S=-\int \d^4 x\, j^\m(x) \del_\m \e(x)$ with some four-vector $j^\m(x)$. Upon integrating by parts, this is 
\be\label{Sepsvar}
\dd S=\int \d^4 x\, \big(\del_\m j^\m(x)\big) \e(x) \ .
\ee
In general, this has no reason to vanish. But if the fields satisfy the Euler-Lagrange equations then, as just explained, one must have $\dd S=0$. Since $\e(x)$ is arbitrary, this implies that $j^\m$ is a conserved current~:
\be
\del_\m j^\m(x) \Big\vert_{\rm field\ equations}=0 \ .
\ee

In the quantum theory, there will be a corresponding (quantum) current operator $J^\m(x)$. (This is not to be confused with the source fields of the generating functionals $Z[J]$.) The question then is whether this quantum current is still conserved. More precisely, one needs to check whether $\del_\m J^\m(x)$ vanishes as an operator, which means that all time-ordered vacuum expectation values of products of operators $O_j(\P(y_j))$ and a factor $\del_\m J^\m(x)$ should vanish~:
\be\label{qcurrent1}
\bra{\rm vac,\ out} T\{\del_\m J^\m(x) \prod_j O_j(\P(y_j)) \} \ket{\rm vac,\ in} = 0 \ .
\ee
If this is not the case, one has an anomaly. Note that \eqref{qcurrent1} is not the same as the statement  $\frac{\del}{\del x^\m} \bra{\rm vac,\ out} T\{ J^\m(x) \prod O_j(y_j)\} \ket{\rm vac,\ in} = 0$, since the time-ordering involves various $\theta(x^0-y_j^0)$ and $\theta(y_j^0-x^0)$ and their time derivatives generate $\dd(x^0-y_j^0)$ times (equal-time) commutators of $J^0$ and the corresponding $O_j$. Thus generically, if \eqref{qcurrent1} holds, one has
\ba\label{Ward}
&&\hskip-2.cm\frac{\del}{\del x^\m} \bra{\rm vac,\ out} T\big\{ J^\m(x) \prod_j O_j(\P(y_j))\big\} \ket{\rm vac,\ in}\nonumber\\
&&=\sum_j \dd(x^0-y_j^0) \bra{\rm vac,\ out} T\big\{ \, [J^0(x), O_j(\P(y_j))] \, \prod_{k\ne j} O_k(\P(y_k))\big\} \ket{\rm vac,\ in} \ ,
\ea
which is called a Ward identity. If there is an additional anomalous term on the right-hand side of this equation, i.e.~if $\del_\m J^\m\ne 0$ and \eqref{qcurrent1} does not hold, then the analogue of \eqref{Ward} is called an anomalous Ward identity.

Let us show how the non-conservation of the quantum current is related in the functional integral formalism to the non-invariance of the functional integral measure.
Consider the same transformation of the fields as before, such that $\dd S$ is given by \eqref{Sepsvar}, and assume furthermore that the operators $O_j(\P(y_j))$, or rather their classical counterparts $O_j(\p(y_j))$, are invariant. We then assume that the measure transforms as
\be
\cD \p'=\cD \p\, e^{i\wh\cA} \ ,
\ee
with some possibly non-trivial Jacobian factor $e^{i\wh\cA}$. Since $\wh\cA$  vanishes for $\e=0$, and since $\e$ is infinitesimal, $\wh\cA$ must be of the form $\wh\cA=\int \d^4 x\, \e(x) \cA(x)$. We then proceed by first renaming the functional integration variables from $\p$ to $\p'$, and then identifying the $\p'$ with the transformed fields $\p+\dd\p$~:
\ba
\int\cD \p\,  \prod_j O_j(\p(y_j)) e^{iS(\p)}
\hskip-1.mm&=&\hskip-1.mm\int\cD \p'\,  \prod_j O_j(\p'(y_j)) e^{iS(\p')}
=\int\cD \p\, e^{i\int \e \cA}\, \prod_j O_j(\p(y_j)) e^{iS(\p)+i\int \del_\m j^\m \e}
\nonumber\\
&=&\hskip-1.mm\int\cD \p\, \Big( 1 + i\int  (\cA + \del_\m j^\m)\e\Big)\, \prod_j O_j(\p(y_j)) e^{iS(\p)} \ .
\ea
In the last expression, the 1 in the parenthesis reproduces the initial expression and, hence, the terms linear in $\e(x)$ must vanish~:
\be\label{anwardFI}
\int\cD \p\, \Big(\cA(x) + \del_\m j^\m(x)\Big)\, \prod_j O_j(\p(y_j)) e^{iS(\p)} = 0 \ .
\ee
But as we have extensively discussed, the functional integral computes time-ordered vacuum expectation values of the corresponding products of operators, and this translates into the operator statement
\be\label{Wardanom}
 \bra{\rm vac,\ out} T\big\{ \del_\m J^\m(x) \prod_j O_j(\P(y_j))\big\} \ket{\rm vac,\ in} 
 = -  \bra{\rm vac,\ out} T\big\{ \cA(x) \prod_j O_j(\P(y_j))\big\} \ket{\rm vac,\ in}.
 \ee
 The non-vanishing term on the right-hand side will appear as the anomalous contribution on the right-hand side of the Ward identities \eqref{Ward}. In any case, it relates the non-conservation of the quantum current to the anomalous transformation of the functional integral measure.
 
There is an important subtlety when translating between the functional integral relation \eqref{anwardFI} and the operator statement \eqref{Wardanom}. Indeed, we could have written \eqref{anwardFI} as
\be\label{anwardFI-2}
\frac{\del}{\del x^\m}\int\cD \p\,   j^\m(x)\, \prod_j O_j(\p(y_j)) e^{iS(\p)} = -\int\cD \p\, \cA(x) \, \prod_j O_j(\p(y_j)) e^{iS(\p)} \ ,
\ee
with the space-time derivative pulled outside the functional integral. Then the corresponding operator statement would be
\be\label{Wardanom-2}
 \frac{\del}{\del x^\m}\bra{\rm vac,\ out} T\big\{ J^\m(x) \prod_j O_j(\P(y_j))\big\} \ket{\rm vac,\ in} 
 = -  \bra{\rm vac,\ out} T\big\{ \cA(x) \prod_j O_j(\P(y_j))\big\} \ket{\rm vac,\ in} \ .
 \ee
But as discussed above, the difference between \eqref{Wardanom} and \eqref{Wardanom-2} are the equal-time commutator terms $\sim [J^0(x), O_j(\P(y_j))]$ which are $\sim \dd^{(4)}(x-y_j)$ and are sometimes refered to as contact terms. But the space-integral of $J^0$ is the corresponding charge operator which generates the transformations of the $O_j(\P(y_j))$ under the symmetry, and $[J^0(x), O_j(\P(y_j))]$ is a local version of this. Our assumption that the $O_j(\P(y_j))$ are invariant under the symmetry amounts to the vanishing of these commutators and, at present, \eqref{Wardanom} and \eqref{Wardanom-2} are equivalent. However, if the $O_j$ transform non-trivially, additional terms will be present and one has to be more careful.\footnote{
A related subtlety is that the functional integral representation was derived in Hamiltonian form, and only for operators not involving the momentum fields could one easily pass to the Lagrangian version of it. One can show that the Lagrangian version is still valid without modification if there is only one momentum field, but it receives additional contact term contributions if there are more momentum fields involved. Typically, $J^0$ involves a momentum field, so one needs to be careful.
}

\subsection{Non-invariance of the fermionic measure under chiral transformations (abelian anomaly)}

The first example of anomalies we will study is the so-called abelian anomaly. 
This anomaly arises from chiral transformations (i.e.~involving $\g_5$) of massless Dirac fermions. More specifically, we consider the usual  matter Lagrangian for a Dirac field $\p$ coupled to a gauge field $A_\m=A_\m^\a t_\a$ through the covariant derivative \footnote{
Note that no gauge coupling constant $g$ explicitly appears. Our convention is that it is included in the gauge-group generators $t_\a$, so that in particular $\tr t_\a t_\b \sim g^2$, and the structure constants are $C_{\a\b\g}\sim g$.
}
\be
\cL_{\rm matter}[\p,\pb,A]=-\pb (\Dsl+m)\p = -\pb \Big(\g^\m (\del_\m - i A^\a_\m t_\a)+m\Big) \p \ .
\ee
Recall that our signature is  $(-+++)$ and the $\g^\m$ matrices obey $(\g^0)^\dag=-\g^0,\ (\g^i)^\dag=\g^i$ and $\g_5^\dag=\g_5$.
 The Dirac field $\p$ is taken to be in some representation $\cR$ of some Lie group with (hermitian) generators $t_\a$. We want to study quite generally
the chiral transformations $\p(x) \to \p'(x)=e^{i \e^\a(x) t_\a \g_5}\p(x)$. Since $\pb=\p^\dag i\g^0$, and since $\{\g_5,\g^\m\}=0$, one also has $\pb(x) \to \pb'(x)= \p^\dag(x) e^{-i \e^\a(x) t_\a \g_5} i\g^0  =\pb(x)e^{i \e^\a(x) t_\a \g_5}$. These transformations are {\it not} symmetries of the matter Lagrangian unless (i) $\e$ is constant, (ii) we replace  $\e^\a t_\a$ by $\e\, t$ where $[t,t_\a]=0$, and (iii) take $m=0$. Then the chiral transformation $\p\to e^{i\e t\g_5}\p$ is a symmetry of the {\it massless} matter Lagrangian \footnote{
As a matter of fact, in order for this transformation to be a classical symmetry, the mass term has to vanish. Indeed, the invariance comes from the fact that $\{\g_5,\g^\m\}=0$ so that the transformation of $\p$ and $\pb$ compensate each other in the $\pb\Dsl\, \p\ $  term. Since the mass term doesn't involve the $\g^\m$, the transformation of this term doesn't cancel out.
}
with corresponding chiral Noether current 
\be\label{jmu5}
j_5^\m=i \pb \g^\m \g_5 t\p \ .
\ee
The anomaly we will soon compute means that this chiral current is not conserved in the quantum theory~: $\langle \del_\m J_5^\m\rangle_A \ne 0$, where the notation $\langle \ldots \rangle_A$ is shorthand for $\bra{\rm vac,\, out} (\ldots) \ket{\rm vac,\, out}$ computed in the presence of the interaction Lagrangian density  $i \pb A^\a_\m t_\a \p$. We will be more precise shortly.

Let us now  use the functional integral approach to compute vacuum expectation values of time-ordered products of operators $\cO_i$ that involve the matter fields.  
The computation can then be divided into two parts: the integration that interests us most is the integration over matter fields. In fact, this is enough to study anomalies. This is due to the fact that the only matter field - gauge field coupling we consider here is through the covariant derivative, thus the currents $j^\m_\a= \del \cL / \del A_\m^\a$ do not involve the gauge fields $A_\m^\a$. As such, all current correlation functions only involve the matter fields and one can do the functional integral over these matter fields, keeping the gauge fields fixed. This is indicated by writing $\langle \ldots \rangle_A$, as we have already done above. The functional integration over the gauge fields can be done in a second step, and only then one has to deal with all the subtleties of gauge-fixing and adding the Faddeev-Popov ghost fields - but this will not be our focus here.

In the rest of this subsection, we will solely be concerned with the transformation of the functional integral measure for the (Dirac) matter fields $\p$ and $\pb$. Quite generally, we consider a local transformation:
\be
\p(x)\to \p'(x)=U(x) \p(x)
\quad , \quad
\pb(x) \to \pb'(x) =\pb(x) \Ub(x) 
\quad , \quad
\Ub(x)=i\g^0 U^\dag(x) i\g^0\ ,
\ee 
where $U(x)$ is a unitary matrix acting on the indices of the representation of the gauge group and on the Dirac indices.\footnote{Other complications arise if $\p$ and $\pb$ transform under additional representations like flavor symmetry, but we will not tackle these additional complexities here.}
Of course, in order for the matter action to also be invariant,  $U$ has to be constant.

Let us examine how this transformation affects the fermionic measures $\cD\p$ and $\cD\pb$. Since these are measures for anticommuting fields, they transform with the inverse Jacobian:
\be
\cD\p\to \cD\p'= (\Det \cU)^{-1}\, \cD\p 
\quad , \quad \cD\pb\to \cD\pb'= (\Det \overline\cU)^{-1}\, \cD\pb \ ,
\ee
where the operators $\cU$ and $\cUb$ are defined by
\be
\bra{x} \cU \ket{y} = U(x)\, \dd^{(4)}(x-y) \quad , \quad
\bra{x} \cUb \ket{y} = \Ub(x)\, \dd^{(4)}(x-y) \ .
\ee
Thus, the overall Jacobian is
\be
\cD\p'\, \cD\pb'= e^{i\wh \cA}  \cD\p\, \cD\pb\quad , \quad
e^{i\wh \cA} = (\Det \cU)^{-1}(\Det \overline\cU)^{-1} \ .
\ee

\vskip5.mm

\subsubsection{Transformation of the measure}

\textbf{a) Non-chiral transformation}
\vskip2.mm
\no
Let us first study what happens if the transformation we consider is not chiral, i.e.~if $U$ does not involve $\g_5$. It easy to see that in this case the transformations of $\cD \p $ and $\cD \pb$ cancel each other out. 
Indeed, if $U$ is a unitary non-chiral transformation of the form: 
\be
U(x)=e^{i \e^\a(x) t_\a} \quad , \quad {\rm with}\quad t_\a^\dag=t_\a \quad {\rm and}\quad [\g^\m,t_\a]=0 \ ,
\ee
one has
\be
\Ub(x)=i\g^0 e^{-i \e^\a(x) t_\a} i\g^0=e^{-i \e^\a(x) t_\a} (i\g^0)^2=e^{-i \e^\a(x) t_\a} =U^{-1}(x) 
\ee
Thus, $\cUb = \cU^{-1}$ which means that the overall measure is invariant: 
\be
e^{i\cA}= (\Det \cU)^{-1} (\Det \cUb)^{-1} =(\Det \cU)^{-1} (\Det \cU ^{-1} )^{-1}=1 \ .
\ee
This remains true even after regularization; since both determinants are regularized in the same way, they  still cancel each other.
Hence,  there is no anomaly.

\vskip2.mm
\no
\textbf{b) Chiral transformations}

\vskip2.mm
\no
 On the other hand, if the transformation we consider is chiral, i.e.~involves a $\g_5$ matrix, the transformations of $\cD \p $ and $\cD \pb$ do not compensate each other, as we now proceed to show. 
Consider a unitary chiral transformation of the form:
\be\label{chiraltra}
U(x)=e^{i \e^\a(x) t_\a \g_5} \quad , \quad {\rm with}\quad t_\a^\dag=t_\a \quad {\rm and}\quad [\g^\m,t_\a]=0 \ .
\ee
Then, due to the anti-commutativity of $\g_5$ with  $\g^0$ one has: 
\be
\Ub(x)=i\g^0 e^{-i \e^\a(x) t_\a \g_5} i\g^0=e^{+i \e^\a(x) t_\a \g_5} (i\g^0)^2=e^{+i \e^\a(x) t_\a\g_5} =U(x) \ ,
\ee
Then also $\Det \cUb=\Det \cU$, and the overall fermionic measure transforms as follows: 
\be
e^{i\wh\cA}=  (\Det \cU)^{-1} (\Det \cUb)^{-1} = (\Det \cU)^{-2} =e^{-2 \Tr \log \cU}\ ,
\ee
where we used the operator analogue of the relation $\det M =e^{\tr\log M}$ which is easily proven for any diagonalizable matrix $M$.
Since the determinant of $\cU$ no longer cancels out, one now has to compute it. This is where complexities leading to anomalies arise. 

\subsubsection{Defining and computing the determinant}\label{detsec}

Defining the determinant of an operator in an infinite dimensional space comes with a set of challenges. 
Let us begin by identifying what the matrix elements of the powers of $\cU$ are, which will allow us to generalize this relation to all Taylor expandable functions. For a local kernel, we have $\bra{x}\cU^2\ket{y}= \int \d^4 z \bra{x}\cU\ket{z} \bra{z}\cU \ket{y}=\int \d^4 z\, U(x) \dd^{(4)}(x-z) U(z) \dd^{(4)}(z-y)=U^2(x) \dd^{(4)}(x-y) =U^2(x) \langle x\ket{y}$, and similarly for all powers of $\cU$.
As such, we obtain the generic relation $\bra{x} f(\cU)\ket{y}= f(U(x))\ \langle x\ket{y}$, which we can now apply to $\Tr\log\cU$:  
\be
\Tr\log\cU = \int \d^4 x \bra{x} \tr\log(\cU) \ket{x}
=\int \d^4 x\, \dd^{(4)}(x-x) \, \tr \log(U(x))
=\int \d^4 x\, \dd^{(4)}(0) \ i\e^\a(x)\, \tr  t_\a \g_5 \ ,
\ee
where $\Tr$ is a functional and matrix trace, while $\tr$ is only a matrix trace (with respect to the $\g$ matrices and the gauge representation matrices). This allows us, in turn to obtain the following formula for the determinant: 
\be
(\Det \cU)^{-2}=e^{-2\Tr \log\cU}= e^{i \int\d^4 x\, \e^\a(x) \cA_\a(x)}
\quad {\rm with}\quad
\cA_\a(x)=-2  \dd^{(4)}(0) \tr  t_\a \g_5 \ ,
\ee
where $\cA_\a$ is called the anomaly function or simply the anomaly.

However, this anomaly function needs to be regularized as it is ill-defined at this stage given that $\dd^{(4)}(0)$ is infinite and that $\tr \g_5 t_\a$ vanishes.\footnote{
This is due to the fact that actually the trace over the matrices of the representation and over the Dirac matrices are taken separately, i.e.~$\tr \g_5 t_\a \equiv (\trR t_\a ^\cR)\, ({\rm tr}_D \g_5)$ and ${\rm tr}_D \g_5 = 0$.
}
The $\dd^{(4)}(0)=\dd^{(4)}(x-x)$ factor is to be interpreted through its Fourier transform $\int{\d^4 p\over (2\pi)^4} e^{ip(x-y)}\Big\vert_{x=y}$ as a large momentum (UV) divergence.
We will then proceed by using the Fourier transform and regularize the anomaly through a  cut-off function which limits the contribution of larger momenta. 
To this end, we introduce a smooth cut-off function $f$ such that $f(0)=1$, $f(\infty)=0$, as well as $xf'(x)=0$ at $x=0$ and at $x=\infty$, like e.g.~$f(x)=e^{-x}$. The precise reason for these conditions will become clear in the computation of eq.~(\ref{intsaddle}). 
We then rewrite the integral of the anomaly as the limit of a quantity dependent on the regulator $\L$: 
\vskip-5.mm
\be \label{anomalyaslim}
\int\d^4 x\, \e^\a(x) \cA_\a(x)=-2 \lim_{\L\to\infty}\Tr \cE_\L
\quad , \quad {\rm where}\quad
\cE_\L=\e^\a(\hat X) \g_5 t_\a\ f\Big( -{\hat\Dsl^2\over \L^2}\Big) \ ,
\ee
\vskip-2.mm
\no
where $\hat X$ is the quantum mechanical position operator and $\hat \Dsl$ denotes the quantum mechanical operator that acts as: $\bra{x} \hat\Dsl \ket{\chi}=\Dsl \ \langle x\ket{\chi}$. We choose to regulate the integral through $\hat \Dsl \ ^2$ which is both gauge and Lorentz invariant. Below, we will further discuss the importance of this choice. 
Let us now compute this quantity: 
\vskip-7.mm
\ba\label{regtracecomp}
\Tr \cE_\L &\hskip-2.mm \equiv \hskip-2.mm& 
\int \d^4 x\, \tr \bra{x} \e^\a(\hat X) \g_5 t_\a f\Big( -{\Dsl^2\over \L^2}\Big)\ket{x} 
=\int \d^4 x\, \e^\a(x) \int\d^4 p\, \tr \g_5 t_\a \bra{x} f\Big( -{\hat\Dsl^2\over \L^2}\Big)\ket{p} \langle p\ket{x} 
\nonumber\\
&=\hskip-2.mm&\int \d^4 x\, \e^\a(x) \int{\d^4 p\over (2\pi)^4} e^{-ipx} \tr \g_5 t_\a\ f\Big( -{1\over \L^2}\Big[ \g^\m \Big( {\del\over \del x^\m} - i A_\m^\cR(x)\Big) \Big]^2\Big) e^{ipx}
\nonumber\\
&=\hskip-2.mm&\int \d^4 x \e^\a(x) \int{\d^4 p\over (2\pi)^4} e^{-ipx} \tr \g_5 t_\a \ \e^{ipx} f\Big( -{1\over \L^2}\Big[ \g^\m \Big( {\del\over \del x^\m} +i p_\m - i A_\m^\cR(x)\Big) \Big]^2\Big) 
\nonumber\\
&&\hskip5.6cm
\nonumber
\\
&=\hskip-2.mm&\int \d^4 x\ \e^\a(x) \int{\d^4 p\over (2\pi)^4} \tr \g_5 t_\a\ f\Big( -{1\over \L^2}\Big[  i \psl+\Dsl\,\Big]^2\Big) 
\nonumber\\
&=\hskip-2.mm&
\int \d^4 x\ \e^\a(x)\, \L^4 \int{\d^4 q\over (2\pi)^4} \tr \g_5 t_\a\ f\Big( -\Big[  i \qsl+{\Dsl\over \L}\Big]^2\Big) 
\nonumber\\
&=\hskip-2.mm&
\int \d^4 x\ \e^\a(x)\, \L^4 \int{\d^4 q\over (2\pi)^4} \tr \g_5 t_\a\ f\Big( q^2 -2i {q^\m D_\m\over \L}-{\Dsl^2\over \L^2} \Big) \ .
\ea
When going from the second to the third line, we have moved the $e^{ipx}$ past the differential operator $f(-\Dsl^2/\Lambda^2)$ which should be thought of as a power series in $\Dsl^2/\Lambda^2$. In doing so, each $\del_\m-iA_\m(x)$ gets replaced by a $\del_\m +i p_\m-i A_\m(x)$, with the right-most differential operator acting on 1 (which we did not write explicitly). Also, in the last steps, we set $p^\m=\L q^\m$ and use that 
$-\left[  i \qsl+{\Dsl\over \L}\right]^2= \qsl^2 -{i \over\L}\big(\qsl\Dsl+\Dsl\,\qsl\big) -{1\over \L^2}\Dsl^2 = q^2 - {2i\over \L} q^\m D_\m -{1\over \L^2} \Dsl^2 $.  We then Taylor expand $\L^4 f$ around $q^2$. This yields the following terms 
\be\label{Lambdaexp}
\L^4 \left( f(q^2) +f'(q^2) \big(- {2i\over \L} q^\m D_\m -{1\over \L^2} \Dsl^2\big) + {1\over 2} f''(q^2) \big( -{2i\over \L} q^\m D_\m -{1\over \L^2} \Dsl^2\big)^2 + \ldots\right) \ ,
\ee
acting on 1.
Note that the only term involving Dirac matrices are the  powers of $\Dsl^2$.
Two elements then come into play to select the non-vanishing term. First, in order for the Dirac trace involving $\g_5$ to be non vanishing, the term has to involve at least four $\g^\m$, i.e.~two factors of $\Dsl^2$, which means expanding at least to second order. On the other hand, for the term to be non vanishing as we take the limit $\L \to + \infty$, it has to contain no power of the inverse of $\L$ greater than four. 
Thus, the only term that contributes to $\lim_{\L\to\infty}\Tr \cE_\L$ is the term of second order: ${1\over 2} f''(q^2) (-\Dsl^2/\L^2)^2$.
i.e. 
\be\label{TLI}
\lim_{\L\to\infty} \Tr \cE_\L
=\int \d^4 x\ \e^\a(x)  \int{\d^4 q\over (2\pi)^4}  
{1\over 2} f''(q^2)\tr \g_5 t_\a\, (-\Dsl^2)^2\ .
\ee
We can easily evaluate this integral after performing a Wick rotation:
\ba \label{intsaddle}
\int{\d^4 q\over (2\pi)^4} {1\over 2} f''(q^2) 
&=&  {i\over 2(2\pi)^4} {\rm vol}(S^3) \int_0^\infty \d q\, q^3 f''(q^2)
={i\over 2(2\pi)^4} 2\pi^2 {1\over 2} \int_0^\infty \d\xi\, \xi f''(\xi)  \nonumber\\
&=& 
{i\over 32\pi^2}\left( \xi f'(\xi)\Big\vert^\infty_0 - \int_0^\infty \d\xi\, f'(\xi) \right) 
= {i\over 32\pi^2} \ ,
\ea
where we used the special properties of the function $f$.
Furthermore, 
\ba
\Dsl \ ^2&=&\g^\m \g^\n D_\m D_\n={1\over 2} \{\g^\m,\g^\n\} D_\m D_\n + {1\over 2} [\g^\m,\g^\n] D_\m D_\n = \eta^{\m\n} D_\m D_\n + {1\over 4} [\g^\m,\g^\n] [D_\m, D_\n] \nonumber\\
&=& D^\m D_\m -{i\over 4} [\g^\m,\g^\n] F_{\m\n} \ ,
\ea
where we used that the gauge field strength is defined as $F_{\m\n}=i[D_\m,D_\n]$.
Using furthermore
\be\label{g5trace}
\trD \g_5 \g^\m\g^\n\g^\r\g^\s=4i\e^{\m\n\r\s}
\quad , \quad {\rm with} \ \e^{0123}=+1 \ ,
\ee
we get
\be
\quad \tr \g_5 t_\a\, (-\Dsl \ ^2)^2 
= \left({i\over 4} \right)^2 
\trD \g_5 [\g^\m,\g^\n] [\g^\r,\g^\s]\ 
\trR t_\a F_{\m\n}F_{\r\s}
=-i \e^{\m\n\r\s} \trR t_\a F_{\m\n}F_{\r\s} \ .
\ee
Putting everything together we finally obtain $\lim_{\L\to\infty} \Tr\cE_\L = {1\over 32\pi^2}\e^{\m\n\r\s} \trR t_\a F_{\m\n}F_{\r\s}$ and hence for
the anomaly function
\be\label{anfunc}
\cA_\a(x)= -{1\over 16\pi^2}\e^{\m\n\r\s} \trR t_\a F_{\m\n}(x)F_{\r\s}(x) \ .
\ee

A remark is in order~: The trace over the Hilbert space should be taken with respect to some basis of square-normalizable, gauge-covariant functions which one could chose to be eigenfunctions of $i\Dsl$. Of course, we don't know this basis explicitly and, instead, we have taken the trace with respect to the improper, non-normalisable basis of position eigenstates $\ket{x}$. This is not a problem though, as long as we are only interested in the leading terms for large $\Lambda$. However, this might not be correct for the sub-leading terms.

Let us now specify to the case where $t_\a$ is replaced by an abelian generator $t$ so that we consider abelian chiral transformations, in which case the chiral current was classically conserved.  Then the anomaly in the quantum theory is immediately obtained from \eqref{anfunc} by this replacement~:
\be\label{abeliananfunc}
\cA(x)= -{1\over 16\pi^2}\e^{\m\n\r\s} \trR t F_{\m\n}(x)F_{\r\s}(x) \ .
\ee
In view of the Ward identity \eqref{Wardanom-2}, we can conclude that the chiral current non-conservation is given exactly by (minus) the anomaly function~:
\be\label{delmuj5isminusA}
\del_\m \langle  J_5^\m(x)\rangle_A 
=-\cA(x)\ .
\ee

At first sight, it may seem a bit strange that the divergence of the chiral current $J_5^\m$ - which does not depend on the gauge fields -  involves the gauge field strength. Of course, this gauge-field dependence is a quantum effect. Below, we will relate it to a one-loop diagram, and the gauge-field dependence appears naturally through the matter-gauge field vertices. In our previous computation the gauge-field dependence is due to our regulator $f(-\Dsl^2/\L^2)$ which is made from the gauge-covariant derivative. Had we used a regulator $f(-\del_\m \del^\m/\L^2)$ there would be no chiral anomaly, but this regulator would break the gauge invariance of the theory. The lesson is that an anomaly arises if the regulator cannot maintain {\it all} the classical symmetries, and then one needs to make a choice which symmetries are more important. Since the gauge symmetry is fundamental for the consistency of the theory the regulator function must be taken gauge-invariant (if possible).

The computation of the abelian anomaly and the result \eqref{abeliananfunc} bear a strong resemblance with the index of the Dirac operator. It is instructive to spell out this relation in more detail, as we will do in the next subsection. Later-on we will come upon another occurrence of index theorems which will relate the chiral anomaly in $d$ dimensions to the index of a Dirac operator in $d+2$ dimensions.

\subsubsection{Relation of the abelian anomaly with the index theorem}\label{abelianindextheorem}

Let us then briefly explore the relation between the abelian anomaly and the index theorem. 
Let us begin by recalling the definition of the abelian anomaly as the limit of the regularized quantity \eqref{anomalyaslim} but now with $\e^\a\to \e$ and $ t_\a\to t$~:
\vskip-5.mm
\be
\int\d^4 x\, \e(x) \cA(x)=-2 \lim_{\L\to\infty}\Tr \cE_\L
\quad , \quad {\rm where}\quad
\cE_\L=\e(\hat X) \g_5 t\ f\Big( -{\hat\Dsl^2\over \L^2}\Big) \ .
\ee 
Note that if we set $\e=t=1$, this relation becomes
\be
\int\d^4 x\, \cA(x)=-2 \lim_{\L\to\infty}\Tr \g_5 f( -{\hat\Dsl^2/\L^2}) \ .
\ee 
As a matter of fact, we will show that the trace $\Tr \g_5 f(-\hat\Dsl^2/\L^2)$ counts the number of positive chirality zero-eigenmodes of $i\Dsl$ minus the number of its negative chirality zero-eigenmodes. This is called the index of $i\Dsl$.

This argument is best done in Euclidean signature where $i\Dsl$ is hermitian. Indeed, in Euclidean signature, all $\g^\m$ can be chosen to be hermitian and then, with appropriate boundary conditions, $\int \d^4 x_E\, (i\Dsl\, \chi)^\dag \psi=\int\d^4 x_E\, \chi^\dag i\Dsl\,\p$. Let us now compute the trace in the eigenbasis of $i\Dsl$. Let then $\p_n$ be an eigenmode of $i\Dsl$ with eigenvalue $\l_n$~:
\be
i\Dsl \, \p_n=\l_n \p_n 
 .
 \ee 
Since $\Dsl \, \g_5 = -\g_5 \Dsl$ \ ,
we have 
\be
i\Dsl (\g_5 \p_n)=-\g_5 i\Dsl\,\p_n = -\g_5 \l_n \p_n =(-\l_n) (\g_5\p_n) \ ,
\ee
so that $\g_5\p_n$ is an eigenmode of $i\Dsl$ with eigenvalue $-\l_n$. 

Then, if $\l_n\ne 0$, $\p_n$ and $\g_5\p_n$ are both eigenmodes of the hermitian $i\Dsl$ with different eigenvalues and, as such they are orthogonal, and in particular linearly independent. It follows that the two combinations $\p_n^\pm=\p_n\pm\g_5\p_n$ are both non-vanishing. Clearly, $\p_n^\pm$ are eigenstates of $\g_5$ and have chirality $\pm 1$. However, they are no longer eigenmodes of $i\Dsl$. Nevertheless, they are still eigenmodes of $(i\Dsl)^2=-\Dsl^2$ with eigenvalues $\l_n^2$. One concludes that $\p_n^+$ contributes $+f(\l_n^2/\L^2)$  and $\p_n^-$ contributes $-f(\l_n^2/\L^2)$ to $\Tr \g_5 f( -{\hat\Dsl^2/\L^2})$, and the sum of their contributions vanishes. Since this holds for all non-zero modes ($\l_n\ne 0$), only zero-modes possibly give a non-vanishing contribution to this trace.

Consider now the zero-modes, with $\l_n=0$. Then $\p_n$ and $\g_5\p_n$ have the same eigenvalue of $i\Dsl$ (namely 0) and one no longer can conclude that they are linearly independent. If they are linearly dependent, i.e.~if $\g_5\p_n \sim \p_n$, one necessarily has $\g_5\p_n=\pm \p_n$ and then $\p_n$ has definite chirality. If they are linearly independent, one can still form the $\p_n^\pm=\p_n\pm \g_5\p_n$ which have definite and opposite chirality. This shows that within the zero-eigenspace of $i\Dsl$, we can choose a basis of zero-modes with definite chirality. However, there is no reason any more to have (only) pairs of opposite chirality, 
and the number of positive chirality zero-modes may differ from the number of negative chirality zero-modes. As already said, this difference is called the index of $i\Dsl$. Since $\l_n=0$, all these zero-modes just contribute a $\pm f(0/\L^2)=\pm 1$ to the trace $\Tr \g_5 f( -{\hat\Dsl^2/\L^2})$. Thus (writing $\Dsl_E$ to insist that this is in Euclidean signature)
\be\label{index1}
{\rm index} (i\Dsl_E) = \Tr \g_5 f(-\Dsl_E^2/\L^2) \ ,
\ee
independently of the value of $\Lambda$. Above we have computed this trace in the large $\Lambda$ limit. As discussed below \eqref{anfunc}, our computation is reliable only in this limit, and we conclude that 
 the integral of the abelian anomaly $\int\d^4 x\, \cA(x) \big\vert_{t=1}$ is just $(-2)$ times the quantity \eqref{index1} (in the large $\Lambda$ limit). In view of \eqref{abeliananfunc} we find that
\be
{\rm index} (i\Dsl_E) = {1\over 32\pi^2}\int \d^4 x_E\, \e_E^{\m\n\r\s} \trR F^E_{\m\n}(x)F^E_{\r\s}(x) \ .
\ee
This is a special case of the famous Atiyah-Singer index theorem \cite{indextheoremAS,EGH,indextheorem}.

\subsection{Matter effective action and more Ward identities}

In the next section we will do a Feynman diagram computation of one-loop triangle diagrams. In order to relate these computations to the non-conservation of the currents, we need to  derive some further identities which usually go under the name of (generalized) Ward identities.

As mentioned above, it is useful to consider the partial functional integral done over the matter fields only~:
\be\label{Gammaeff}
e^{i\wt\G[A]}=\int \cD\p \cD\pb\, e^{i\int \cL_{\rm matter}[\p,\pb,A]} \ .
\ee
Integrating over the matter fields only, and not over the gauge fields, avoids all the complications related to the gauge fixing, Faddeev-Popov determinant and introduction of ghost fields. More importantly, for the discussion of anomalies under gauge transformations, the $\wt \G[A]$ is the appropriate object to consider.
This $\wt\G[A]$ is some sort of quantum effective action. It is different from - but related to - the quantum effective action $\G$ discussed  in subsection \ref{queffacsec}. Indeed, the matter Lagrangian $\cL_{\rm matter}[\p,\pb,A]=-\pb (\Dsl + m)\p$ is quadratic in $\p$ and $\pb$ and, at least formally, the integral is given by the determinant 
\ba
\hskip-1.cm\Det (\Dsl + m)&=&\Det ( \dsl + m -i \Asl)=\Det \big(( \dsl + m)( {\bf 1}-i(\dsl+m)^{-1}\, \Asl)\big)\nonumber\\
&=&\Det ( \dsl + m)\Det( {\bf 1}-i(\dsl+m)^{-1}\, \Asl)= \cN \times \Det( {\bf 1}-i(\dsl+m)^{-1}\, \Asl) \ ,
\ea
so that
\be
i\wt \G[A]=\log \Det (\Dsl + m)= i\wt\G[0]+ \Tr \log ( {\bf 1}-i(\dsl+m)^{-1}\, \Asl)
=i\wt\G[0] +\sum_{r\ge 1} \frac{1}{r} \Tr \big(i(\dsl+m)^{-1}\, \Asl\big)^r\ .
\ee
This sum is to be interpreted as a sum of one-loop diagrams with $r$ Dirac propagators $-i(\dsl+m)^{-1}$ and $r$ vertices $-\g^\m t_\a$ where the $A^\a_\m$ are attached. These are all  1PI diagrams that contribute to the full quantum effective action $\G$. The full $\G$ also contains higher-loop diagrams with internal gauge field propagators that arise when also doing the functional integral over the gauge fields. For our present purpose, it is enough to work with $\wt\G[A]$ and we simply refer to it as an effective action.

Taking functional derivatives of $e^{i\wt \G[A]}$ with respect to $A_\m^\a$ will insert into the functional integral a factor $i\frac{\del \cL}{\del A_\m^\a}= i j^\m_\a$ where $j_\a^\m=i\pb \g^\m t_\a\p$ is the ``gauge current". Taking multiple functional derivatives produces multiple insertions of the corresponding currents and then the functional integral computes the vacuum expectation value of the corresponding product of quantum currents (up to a normalization constant that we drop)~:
\be
\frac{\dd}{\dd A_{\m_1}^{\a_1}(x_1)} \ldots \frac{\dd}{\dd A_{\m_r}^{\a_r}(x_r)} e^{i\wt\G[A]}
= i^r \bra{\rm vac,\ out} T\big\{ J^{\m_1}_{\a_1}(x_1) \ldots J^{\m_r}_{\a_r}(x_r)\big\} \ket{\rm vac,\ in}_A
\ee
This same result can also be obtained directly from Dyson's formula derived in subsection \ref{Dysonsec}, without the need to rely on the functional integral.
As discussed in subsection \ref{sec-conngreen}, we get the {\it connected} vacuum expectation value  if we take functional derivatives of $i\wt \G[A]$ instead. We may furthermore set $A=0$ in the end~:
\be
\frac{\dd}{\dd A_{\m_1}^{\a_1}(x_1)} \ldots \frac{\dd}{\dd A_{\m_r}^{\a_r}(x_r)} i\wt\G[A] \Big\vert_{A=0}
= i^r \bra{\rm vac,\ out} T\big\{ J^{\m_1}_{\a_1}(x_1) \ldots J^{\m_r}_{\a_r}(x_r)\big\} \ket{\rm vac,\ in}_{\rm connected}\ .
\ee
This corresponds exactly to a one-loop diagram with $r$ currents (times $i$)  inserted at the vertices and Dirac propagators in between. We will compute such diagrams in the next section.

\subsubsection{Anomalous Ward identity for the abelian anomaly}

We can apply similar manipulations to
\be
\frac{\del}{\del x^\m}\langle J^\m_5(x)\rangle_A 
= \int \cD\p \cD\pb\, \del_\m j^\m_5(x)\, e^{i\int \cL_{\rm matter}[\p,\pb,A]} \ ,
\ee
which yields \footnote{
One might  worry whether this is also justified if $x$ coincides with one of the $x_j$. This corresponds exactly to the possible occurrence of the contact terms that originate from $[J_5^0,J^{\mu_r}_{\a_r}]$. But these commutators vanish again, and one does not need to worry.
}
\be\label{aWW}
\frac{\dd}{\dd A_{\m_1}^{\a_1}(x_1)} \ldots \frac{\dd}{\dd A_{\m_r}^{\a_r}(x_r)} \frac{\del}{\del x^\m}\langle  J^\m_5(x)\rangle_A
= i^r \frac{\del}{\del x^\m}\bra{\rm vac,\ out} T\big\{ J^\m_5(x)\, J^{\m_1}_{\a_1}(x_1) \ldots J^{\m_r}_{\a_r}(x_r) \big\} \ket{\rm vac,\ in}_A^{\rm connected} \ .
\ee
Now use \eqref{delmuj5isminusA} to replace $  \del_\m\langle J^\m_5(x)\rangle_A$ by $-\cA(x)$ and, after taking the functional derivatives, set $A=0$ to get
\be\label{anWard}
\frac{\dd}{\dd A_{\m_1}^{\a_1}(x_1)} \ldots \frac{\dd}{\dd A_{\m_r}^{\a_r}(x_r)} \cA(x) \Big\vert_{A=0}
= -i^r \frac{\del}{\del x^\m}\bra{\rm vac,\ out} T\big\{ J^\m_5(x)\, J^{\m_1}_{\a_1}(x_1) \ldots J^{\m_r}_{\a_r}(x_r) \big\} \ket{\rm vac,\ in}_{\rm connected} \ .
\ee
The abelian anomaly  \eqref{abeliananfunc} contains terms that are quadratic and higher in the gauge fields.  Thus the first non-trivial relation \eqref{anWard} is for $r=2$ and reads
\be\label{anWard2}
\frac{\dd}{\dd A_{\n}^{\b}(y)} \frac{\dd}{\dd A_{\r}^{\g}(z)} \cA(x) \Big\vert_{A=0}
=  \frac{\del}{\del x^\m}\bra{\rm vac,\ out} T\big\{ J^\m_5(x)\, J^{\n}_{\b}(y) J^{\r}_{\g}(z) \big\} \ket{\rm vac,\ in}_{\rm connected}\equiv - \frac{\del}{\del x^\m}\G^{\m\n\r}_{5\b\g} \ ,
\ee
where $i \G^{\m\n\r}_{5\b\g}$ is exactly the 1PI one-loop triangle diagram with one (abelian) chiral current and two (non-abelian) non-chiral currents at the vertices.\footnote{
More precisely, there is an insertion of $i\times J$ at each vertex, in agreement with the factors $i^r$ in \eqref{aWW} or \eqref{anWard}.
} 
This makes the precise relation with the Feynman diagram computation given in the next section. 
Using the explicit form  \eqref{abeliananfunc} of $\cA$ and Fourier transforming from $(x,\, y,\, z)$ to $(k,p,q)$ and factoring the usual $(2\pi)^4\dd^{(4)}(p+k+q)$, one can show that the previous relation is equivalent to
\be\label{abeliananomcharac}
-i(p_\m+q_\m) \G^{\m\n\r}_{5\b\g}(-p-q,p,q) = -\frac{1}{2\pi^2}\e^{\n\r\l\s}p_\l q_\s \trR t\, t_{(\b} t_{\g)} \ .
\ee

\subsubsection{Anomalous Ward identity for the chiral anomaly}

Somewhat misleadingly, the name chiral anomaly refers to an anomaly under gauge transformations. Let us first consider whether $\wt\G[A]$ is gauge invariant. Let $A'_\m=A_\m+D_\m\e$ be the gauge transformed gauge field. Then 
\be
e^{i\wt\G[A']}=\int \cD \p \cD\pb e^{i\int\cL_{\rm matter}[\p,\pb,A']}
=\int \cD \p' \cD\pb' e^{i\int\cL_{\rm matter}[\p',\pb',A']}
=\int \cD \p \cD\pb\, e^{i\wh\cA} \,e^{i\int\cL_{\rm matter}[\p,\pb,A]} \ ,
\ee
where we first renamed ($\p$, $\pb$) into ($\p'$, $\pb'$), and then identified them with the gauge transformed fields (under the same transformation as $A_\m'$). Lastly, we used the gauge invariance of the matter Lagrangian and the possibly anomalous transformation of the integration measure. It turns out that $\wh\cA=\int \d^4 x\, \e^\a(x)\cA_\a(x)$ only depends on the gauge fields $A_\m$, and not on $\p$ or $\pb$, and can be taken out of the functional integral. Thus
$\wt\G[A']=\wh\cA +\wt\G[A]$, or
\ba\label{noninvGamma}
\int \d^4 x\, \e^\a(x)\cA_\a(x)&=&\dd\wt\G[A]=\wt\G[A']-\wt\G[A]=\wt\G[A+\dd A]-\wt\G[A]
=\int \d^4 x\, \dd A_\m^\a(x) \frac{\dd\wt\G[A]}{\dd A_\m^\a(x)}
\nonumber\\
&=&\int \d^4 x\, (D_\m \e(x))^\a \langle J^\m_\a(x)\rangle_A \ .
\ea
Thus the anomaly here exactly equals the non-invariance of the effective action $\wt\G$.
Upon integrating the gauge covariant derivative by parts, we relate again the anomaly to the, now covariant, current non-conservation~:
\be
(D_\m \langle J^\m(x)\rangle_A)_\a = - \cA_\a(x) \ ,
\ee
much analogous to the relation \eqref{delmuj5isminusA} for the abelian anomaly. As before, one can take successive functional derivatives of this relation with respect to the gauge fields. Since now the covariant derivative also involves the gauge field, there are some extra contributions involving the structure constants. In particular, taking two functional derivatives with respect to the gauge fields gives 
\cite{ABanomaly}
\be\label{anomWardgauge}
-\frac{\dd}{\dd A_\n^\b(y)} \frac{\dd}{\dd A_\r^\g(z)} \cA_\a(x) \Big\vert_{A=0}
=\frac{\del}{\del x^\m} \G^{\m\n\r}_{\a\b\g}(x,y,z) + C_{\a\g\e} \dd^{(4)}(x-z) \G^{\r\n}_{\e\b}(x,y) + C_{\a\b\e} \dd^{(4)}(x-y) \G^{\n\r}_{\e\g}(x,z) \ .
\ee
Note that $ -\G^{\m\n\r}_{\a\b\g}(x,y,z)=\langle T\{J_\a^\m(x) J_\b^\n(y) J_\g^\r(z)\}\rangle_{\rm connected}$ and that, upon taking $\frac{\del}{\del x^\m}$ one gets\break $\langle T\{\del_\m J_\a^\m(x) J_\b^\n(y) J_\g^\r(z)\}\rangle_{\rm connected}$, plus two equal-time commutator terms of the form $\dd(x^0-y^0)$ $\times \langle T\{ [ J_\a^0(x) ,J_\b^\n(y)] J_\g^\r(z)\}\rangle_{\rm connected}$. The latter are non-vanishing at present, and  will exactly cancel the two terms in \eqref{anomWardgauge} involving the structure constants, thus showing that the anomaly ``is  contained" in $\del_\m J_\a^\m$.

In any case, \eqref{anomWardgauge} shows that the computation of the triangle diagram $ \G^{\m\n\r}_{\a\b\g}$ with a gauge current at each vertex will allow us to obtain  the part of the anomaly that is bilinear  in the gauge fields.  (The $ \G^{\r\n}_{\e\b}$ are vacuum-polarization type 2-point functions.) Again, we will present this computation in the next section.


\section{Computing the triangle diagrams}
\setcounter{equation}{0}

This section focuses on the Feynman diagram computation of the two anomalies we have discussed~: the abelian anomaly under chiral transformations and the chiral anomaly which is an anomaly under gauge transformations in the presence of chiral fermions. As we have shown in the previous section via the anomalous Ward identities, these anomalies manifest themselves in one-loop triangle Feynman diagrams.  

Firstly, we will compute the triangle diagram with one chiral current and two non-chiral gauge currents at the vertices. This will allow us to confirm exactly the relation \eqref{abeliananomcharac}, which in turn confirms the form and precise coefficient of the abelian anomaly \eqref{anfunc} (or at least of the part that is bilinear in the gauge fields).  Secondly, we will tackle the triangle diagram for chiral fermions coupled to non-abelian gauge fields. This corresponds to computing the terms on the right-hand side of the anomalous Ward identity \eqref{anomWardgauge}, and hence will establish the form and coefficient of the part of the chiral (gauge) anomaly that is bilinear in the gauge fields. We will explain later-on how the higher order terms are determined by the Wess-Zumino consistency condition. 

We will see throughout this section how anomalies originate in the impossibility to regulate the one-loop triangle diagrams  in a way that manifestly preserves all the symmetries we would like to preserve.
In particular, due to the prominent presence of the $\g_5$ matrix, dimensional regularization is not available. 
We will instead rely on Pauli-Villars regularization and will pay special attention to where the anomalies arise.  The computations follow closely those in \cite{ABanomaly}, see also \cite{Weinberg-book}.

\subsection{Computations for the abelian anomaly \label{AVVFeynman}}

Let us then compute the so-called AVV triangle diagram, with one axial current and two vector currents and rederive the non-conservation of the axial current. 
We will carry out our computations in a slightly more general, non-abelian setting, thus replacing $j_5^\m=i\pb\g^\m\g_5t\p$ with
\be
j_{5\a}^\m=i\pb\g^\m\g_5 t_\a\p \ ,
\ee
and compute the vertex function $\G_{5\a\b\g}^{\ \m\n\r}$. In the end, we substitute back $t_\a \to t$ to get the abelian anomaly.

\begin{figure}[h]
\centering
\includegraphics[width=0.5\textwidth]{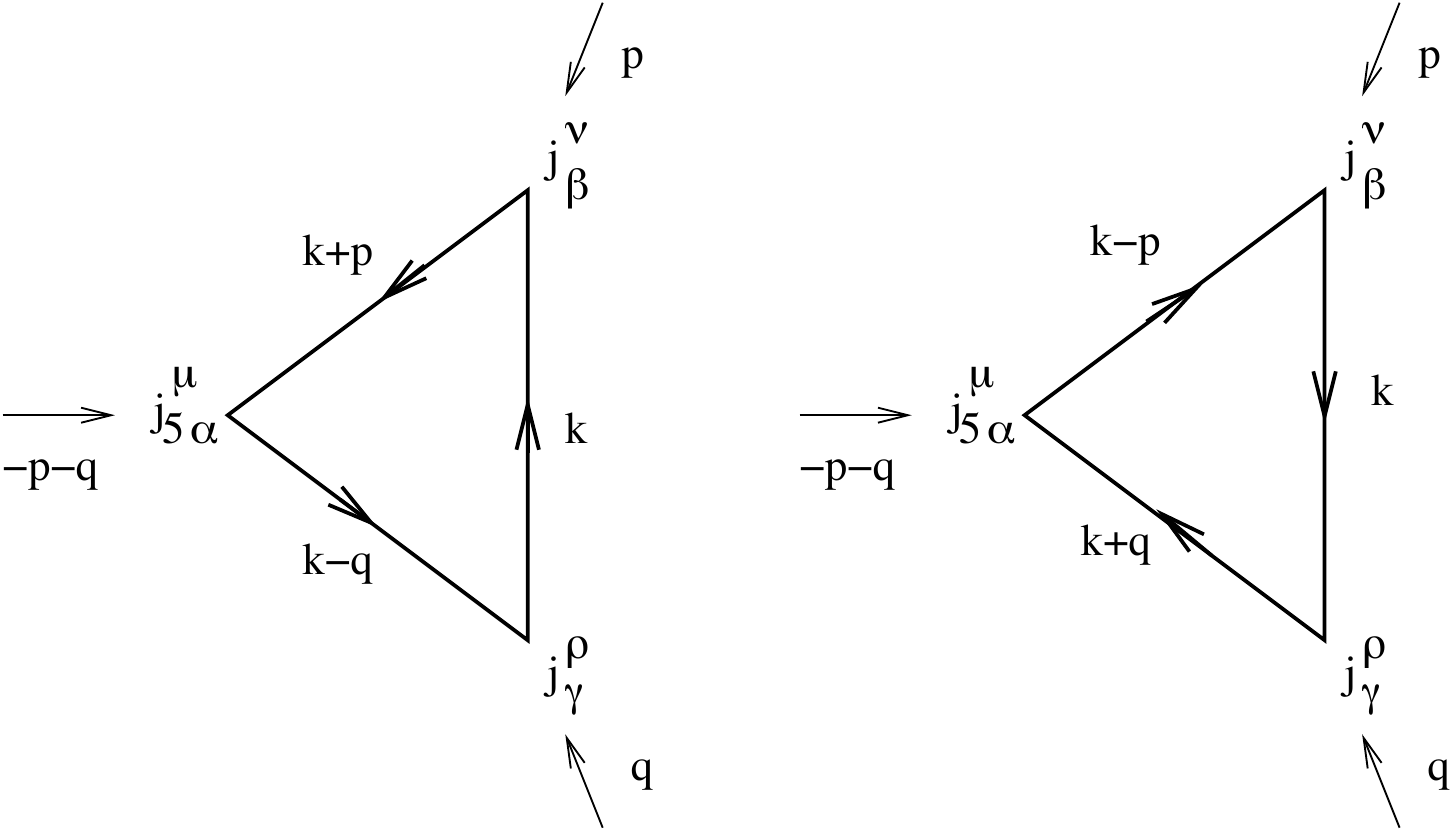}\\
\caption[]{The two triangle diagrams  contributing to the abelian anomaly} \label{triangle}
\end{figure}
As shown in Fig.~\ref{triangle}, 
there are two diagrams contributing to  $\G_{5\a\b\g}^{\ \m\n\r}$, corresponding to the two ways to contract the fermion fields contained in the currents. 
The vertices contribute  $i(2\pi)^4\times i\g^\m\g_5 t_\a=-(2\pi)^4\g^\m\g_5 t_\a$ and $-(2\pi)^4\g^\n t_\b$, resp. $-(2\pi)^4\g^\r t_\g$, while the fermion loop contributes an extra minus sign. Recall also that a propagator for a fermion of momentum $k$ and mass $m$ is ${-i\over(2\pi)^4}{1\over i\footksl+m-i\e} = \frac{-i}{(2\pi)^4} {-i\footksl +m\over k^2+m^2-i\e}$.
Of course, here the fermions are massless. The different factors of $(2\pi)^4$ cancel between the vertices and propagators, but there is one factor $(2\pi)^4$ left since the Feynman diagrams compute \footnote{
More precisely, we should take external momenta $p,\, q$ and $k$ and then the Feynman diagrams compute $i(2\pi)^4 \dd^{(4)}(k+p+q) \G_{5\a\b\g}^{\ \m\n\r}(k,p,q)$. The $(2\pi)^4 \dd^{(4)}(k+p+q)$ arises upon Fourier transforming $\G_{5\a\b\g}^{\ \m\n\r}(x,y,z)$. But as usual, we ``factorize" and drop the overall $\dd^{(4)}(k+p+q)$ on both sides of the equation \eqref{G51}. We could have made similar remarks when we computed the vacuum polarization in \eqref{vacpolexpr}.
} 
$i(2\pi)^4 \G_{5\a\b\g}^{\ \m\n\r}$. Then
\ba\label{G51}
&&\hskip-1.cm i \G_{5\a\b\g}^{\ \, \m\n\r}(-p-q,p,q)\nonumber\\
&&= - \int{\d^4 k\over (2\pi)^4}\, \tr \Big[ 
(-\g^\m\g_5 t_\a)(-i) {-i(\ksl+\psl)\over (k+p)^2 -i\e}
(-\g^\n t_\b) (-i){-i\ksl\over k^2 -i\e}
(-\g^\r t_\g) (-i){-i(\ksl-\qsl)\over (k-q)^2 -i\e}\Big]
\nonumber\\
&&\hskip3.cm + \{( p, \n, \b) \leftrightarrow (q, \r, \g)\}\ .
\ea 
Let us  focus on the first term, which we denote  $i \G_{5\a\b\g}^{ (I) \m\n\r}(-p-q,p,q)$ and which corresponds to the left diagram. The second term then is simply obtained through the permutations $\{( p,\n, \b) \leftrightarrow (q,\r, \g)\}$.
Simplifying the numerical factors and signs,\footnote{
Note that we have anticommuted the $\g_5$ with the $\g^\m$, resulting in another minus sign.
} 
and specifying the trace, this yields: 
\be\label{G52}
\G_{5\a\b\g}^{(I) \m\n\r}(-p-q,p,q)
= -i \int\hskip-1.5mm {\d^4 k\over (2\pi)^4}\, \tr_D \Big[ \g_5\g^\m {(\ksl+\psl)\over (k+p)^2 -i\e}\g^\n {\ksl\over k^2 -i\e}\g^\r {(\ksl-\qsl)\over (k-q)^2 -i\e}\Big] \, \trR t_\a t_\b t_\g \ .
\ee
For large loop-momentum $k$ the integrand behaves as $\sim \d^4 k\, k^{-3}$ and the integral is linearly divergent. Again, the presence of $\g_5$ prevents us from using dimensional regularization. As such, we revert to Pauli-Villars regularization, which consists of adding an artificial  term with ``fermions" of mass $M$ and opposite statistics. Since it follows opposite statistics, the minus sign due to the fermion loop is not there and this term appears with an opposite sign. 
It should be noted that if the integral (\ref{G52}) had been convergent, the contribution of the regulator fields would have vanished as $M\to\infty$, as expected. However, since the integral is divergent, this translates into an $M$-dependence of the regularized integral. 
The regularized $\G_{5\a\b\g}^{ (I) \m\n\r}$ can be expressed as: 
\be\label{gamma5reg}
\G_{5\a\b\g}^{ (I)\m\n\r}(-p-q,p,q)\Big\vert_{\rm reg}
= i \hskip-1.mm\int\hskip-1.mm{\d^4 k\over(2\pi)^4}\, \Big( J^{\m\n\r}_M(k,p,q) - J^{\m\n\r}_0(k,p,q)\Big)\ \trR t_\a t_\b t_\g ,
\ee
where ($m=0$ or $m=M$)
\be
J^{\m\n\r}_m(k,p,q)
=\tr_D \Big[ \g_5\g^\m {\ksl+\psl+im\over (k+p)^2 +m^2-i\e}
\g^\n {\ksl+im\over k^2 +m^2-i\e}
\g^\r {\ksl-\qsl+im\over (k-q)^2 +m^2-i\e}\Big] \ .
\ee
For a finite value of $M$, (\ref{gamma5reg}) is a convergent well-defined integral. As previously discussed, the Lagrangian for massive regulator Dirac fields lacks the chiral symmetry, and this is why an anomaly is found in the end.
We are only interested in
$(p_\m+q_\m) \G_{5\a\b\g}^{\ \m\n\r}(-p-q,p,q)\Big\vert_{\rm reg}$ and, hence, we  only need to compute $(p_\m+q_\m) J^{\m\n\r}_m(k,p,q)$. 
Rewriting:
\be
\g_5(\psl+\qsl)=\g_5 (\psl+\ksl-i m) + (\ksl-\qsl-im)\g_5 +2i m\g_5\ ,
\ee
and using the cyclicity of the trace, certain terms cancel out between the numerator and denominator, and we end up with: 
\ba\label{divIM}
(p_\m+q_\m) J^{\m\n\r}_m(k,p,q)&=& {\tr_D \g_5 \g^\n  (\ksl+im)
\g^\r (\ksl-\qsl+im)\over [ k^2 +m^2-i\e][(k-q)^2 +m^2-i\e]}
+ {\tr_D \g_5 (\ksl+\psl+im)
\g^\n (\ksl+im) \g^\r \over [(k+p)^2 +m^2-i\e] [k^2 +m^2-i\e]}
\nonumber\\
&+&2i\,m\ {\tr_D  \g_5 (\ksl+\psl+im)
\g^\n (\ksl+im)\g^\r (\ksl-\qsl+im)
\over  [(k+p)^2 +m^2-i\e] [k^2 +m^2-i\e] [(k-q)^2 +m^2-i\e]} \ .
\ea
For the Dirac trace involving $\g_5$ to be non vanishing, it should involve at least four $\g^\m$. Thus, the only non-vanishing contribution to the first term comes from $\tr_D \g_5\g^\n\ksl\g^\r(\ksl-\qsl)=4i\e^{\n\l\r\s}k_\l (-q_\s)$. 
Then, when computing the regularized integral $\int\d^4 k$, the $k_\l$ necessarily gets replaced by a four-vector proportional to $q_\l$, which is the only available one in this expression.
When contracted with $\e^{\n\l\r\s} q_\s$ this vanishes. Similarly, the second term in (\ref{divIM}) only involves $k$ and $p$ and, after integration over $\d^4 k$, yields a contribution $\sim\e^{\l\n\s\r}p_\l p_\s=0$. Hence, only the third term in (\ref{divIM}) will contribute and, since
$\tr_D  \g_5 (\ksl+\psl+im)
\g^\n (\ksl+im)\g^\r (\ksl-\qsl+im)=4m\, \e^{\n\r\l\s} p_\l q_\s$, we have
\be\label{divIM2}
\int d^4k\, (p_\m+q_\m) J^{\m\n\r}_m(k,p,q) =
\int d^4k {8i\, m^2\ \e^{\n\r\l\s} p_\l q_\s
\over  [(k+p)^2 +m^2-i\e] [k^2 +m^2-i\e] [(k-q)^2 +m^2-i\e]}  \ .
\ee
This term vanishes for $m=0$, i.e.~in $(p_\m+q_\m) J_0^{\m\n\r}$, making it seem that the anomalous contribution solely comes from the regulator term. However, it should be noted that the first  two terms in (\ref{divIM}) only vanish after integration if we consider convergent integrals, and thus the combination $J_0-J_M$. This is reassuring as it shows that the anomaly is not an artifact. 
Using (\ref{divIM2}) and (\ref{gamma5reg}) we arrive at
\be
-i (p_\m+q_\m) \left[\G_{5\a\b\g}^{ (I) \m\n\r}(-p-q,p,q)\right]_{\rm reg}
=8i M^2\, \e^{\n\r\l\s} p_\l q_\s\, J(p,q,M)\, \trR t_\a t_\b t_\g \ ,
\ee
with
\be
J(p,q,M)= \int{\d^4 k\over (2\pi)^4} 
{1\over [(k+p)^2 +M^2-i\e] [k^2 +M^2-i\e] [(k-q)^2 +M^2-i\e]}  \ .
\ee
The regulator $M$ should be taken to $\infty$ in the end, so we only need the asymptotics of this integral for large $M$ which is easily obtained by letting $k=M\, l$:
\be\label{asympt}
J(p,q,M)\sim {1\over M^2} 
\int{\d^4 l\over (2\pi)^4} {1\over [l^2 +1-i\e]^3} 
={i\over 32\pi^2 M^2} \ ,
\ee
where the $i$ comes from the Wick rotation.\footnote{
One has 
$\int{\d^4 l\over (2\pi)^4} {1\over [l^2 +1-i\e]^3} =i \int{\d^4 l_E\over (2\pi)^4} {1\over [l_E^2 +1]^3} =\frac{i}{(2\pi)^4} 2 \pi^2 \int_0^\infty l^3 \d l \frac{1}{[l^2+1]^3} =\frac{i}{8 \pi^2} \frac{1}{2} \int_0^\infty \d\xi \frac{\xi}{(\xi+1)^3} =\frac{i}{32 \pi^2}$.
} 
Hence, we get a finite limit for $M^2 J(p,q,M)$ as we remove the regulator ($M\to\infty$). Adding the contribution from the second diagram with the appropriate permutations results in
\ba\label{abeliantriangleresult}
-i (p_\m+q_\m) \G_{5\a\b\g}^{\ \m\n\r}(-p-q,p,q)
&=&-{1\over 2\pi^2}\,\e^{\n\r\l\s} p_\l q_\s\,
\trR t_\a t_{(\b} t_{\g)} \ .
\ea
Upon setting $t_\a=t$, this exactly reproduces the result \eqref{abeliananomcharac}, confirming once more that in the quantum theory the axial current $J_5^\m$ is not conserved.

At this stage, it is important to assess if this truly is a problem. In the present case, the anomaly is a mere indicator that the (anomalous) global chiral symmetry $\p\to e^{i\e \g_5}\p$ is broken by a quantum effect. This means that certain selection rules no longer hold and otherwise forbidden (or suppressed) processes may occur. As previously mentioned, this precisely explains the observed disintegration rate $\pi^0\to\g\g$.  On the other hand, anomalies become problematic if they break gauge-invariance. In non-abelian gauge theories, vector currents couple to gauge fields and their non-conservation thus becomes a serious issue. Let us check that the vector currents are conserved. Because of the symmetry under exchange of $(p,\n,\b)$ and $(q,\r,\g)$, it is enough to probe one of them and compute $p_\n \Big[\G_{5\a\b\g}^{\ \m\n\r}(-p-q,p,q)\big]_{\rm reg}$. But now $p_\n J_m^{\m\n\r}(k,p,q)$ contains
\ba
\hskip-1.cm\big(\ksl+\psl+im\big) \psl \big(\ksl+im)
&=&\big(\ksl+\psl+im\big)
\Big[ \big(\psl+\ksl-im\big) -\big(\ksl-im\big) \Big]\big(\ksl+im\big)
\nonumber\\
&=&\big[ (k+p)^2+m^2\big]\big(\ksl+im\big)
\ - \ \big(\psl+\ksl+im\big)\big[ k^2+m^2\big] \ ,
\ea
which now leads to
\be\label{divIM3}
p_\n J^{\m\n\r}_M(k,p,q)= {\trD \g_5 \g^\m  (\ksl+im)
\g^\r (\ksl-\qsl+im)\over [ k^2 +m^2-i\e][(k-q)^2 +m^2-i\e]}
- {\trD \g_5 \g^\m (\ksl+\psl+im)
 \g^\r (\ksl-\qsl+im)\over [(k+p)^2 +m^2-i\e] [(k-q)^2 +m^2-i\e]} \ .
\ee
As compared to (\ref{divIM}) the third term is now absent. This difference is due to the absence of $\g_5$ in the vector current and hence in \eqref{divIM3}. As before, the two terms in (\ref{divIM3}) vanish after integration over $\d^4 k$, provided we always consider the convergent combination \footnote{
For the first term the argument is exactly as before. For the second term, since the integral is convergent, we may shift $k\to k+q$ and then the integral can only depend on $p+q$ and there is no way to get a non-vanishing result when contracting with the $\e$-tensor.
}
 $p_\n \big(J^{\m\n\r}_M(k,p,q)- J^{\m\n\r}_0(k,p,q)\big)$.
We conclude that
\be
p_\n\, \G_{5\a\b\g}^{\ \m\n\r}(-p-q,p,q)
=q_\r\, \G_{5\a\b\g}^{\ \m\n\r}(-p-q,p,q)=0 \ .
\ee

\vskip5.mm

\subsection{Computations for the chiral anomaly\label{trichir}}

\subsubsection{Chiral fermions\label{chirprelim}}

Let us begin by going over some facts about chiral fermions.
Chirality projectors \footnote{
There does not seem to be a unique choice of conventions used in the literature about what is called left-handed and right-handed. It would be better to talk about positive and negative chirality. But even then, the definitions used for $\g_5$ often differ by a sign. However, as long as we stick to one set of conventions, observable quantities will not depend on which one is chosen. Here, we follow the same conventions as \cite{Weinberg-book,ABanomaly}.
} 
$P_L$ and $P_R$ are defined as: 
\be
P_L={1+\g_5\over 2} \quad , \quad P_R={1-\g_5\over 2} \ .
\ee
Denoting $\cH$ the Hilbert space of all fermions, these projectors define the supplementary subspaces of left- and right-handed fermions, respectively $\cH_L$ and $\cH_R$. Left-handed and right-handed fermions are thus defined as follows: 
\be
\p_L=P_L\,\p \quad , \quad \p_R=P_R\, \p \quad ,\quad \p=\p_L+\p_R \ . 
\ee
The projected fields $\p_L$ and $\p_R$ are eigenstates of the chirality matrix $\g_5$:
\ba 
\g_5 \p_L=\p_L 
 \quad &, \quad P_L\p_L=\p_L  \quad &, \quad P_R\p_L=0
\nonumber\\
\g_5\p_R=-\p_R
 \quad &, \quad P_R\p_R=\p_R  \quad &, \quad P_L\p_R=0 \ .
\ea
$\p_L$ has positive chirality and $\p_R$ has negative chirality. 

Since $\g_5^\dag=\g_5$ and $\{\g_5, \g^\m\}=0$, the projectors are hermitian and satisfy the following relations: 
\be
P_L \g^\m=\g^\m P_R \quad , \quad  P_R\g^\m=\g^\m P_L \quad  , \quad   P_L P_R=P_R P_L=0 \quad , \quad  P_L \g^\m P_L=0 \ 
\ee
\be
\overline{\p_L} P_L=0 \ , \quad \overline{\p_L} P_R=\overline{\p_L}
\quad , \quad 
\overline{\p_R} P_R=0 \ , \quad \overline{\p_R} P_L=\overline{\p_R}
\ .
\ee
As such,
\be
\overline{\p_L} \p_L=\overline{\p_L} \left(P_L \p_L\right)
=\left(\overline{\p_L}P_L\right) \p_L = 0 \quad , \quad
\overline{\p_R} \p_R =\left(\overline{\p_R}P_R\right) \p_R = 0 \ .
\ee
It follows that a mass term only couples opposite-chirality parts of a fermion,
\be
\overline{\p} \p = \overline{\p_L} \p_R + \overline{\p_R} \p_L \ ,
\ee
and a chiral fermion (with either $\p_L=0$ or $\p_R=0$) cannot have a mass term.\footnote{
One can still write a so-called Majorana mass term for chiral fermions, of the form  $\overline{(\p_L^c)} \p_L$  using the charge conjugate spinor $\p_L^c$ (which is right-handed). But such terms violate fermion number conservation. Moreover, they are only compatible with gauge invariance if $\p_L$ is in a real representation of the gauge group, and, as we will see, for real representations the anomaly vanishes. 
} 
On the other hand,
\be
P_R \Dsl\, \p_L=\Dsl P_L \p_L= \Dsl\, \p_L
\quad , \quad
P_L \Dsl\, \p_R=\Dsl P_R \p_R= \Dsl\, \p_R \ ,
\ee
so that $\Dsl\,\p_L$ is right-handed and $\Dsl\,\p_R$ is left-handed, and $\overline{\p_L} \Dsl\,\p_L$
and $\overline{\p_R}\Dsl\,\p_R$ are non-vanishing.

\subsubsection{Matter Lagrangian for chiral fermions\label{chidd}} 

For a massless right-handed fermion, one can write a gauge-invariant kinetic term in the Lagrangian as
\be\label{chiralmatlagr}
\cL_{\rm matter}^R = - \overline{\p_R} \Dsl\, \p_R = - \pb \Dsl P_R\p \ .
\ee
The latter formulation, relying on non-chiral fermion fields and chirality projectors, allows us to continue using the usual Feynman rules. It follows that the propagators are now simply 
$\frac{-i}{(2\pi)^4}\times P_R\, {-i\ksl\over k^2-i\e}$ and that the vertices are $i(2\pi)^4 \times (i \g^\m t_\a) P_R$. 

Taking a closer look at the computation of the triangle diagram, one notices that there is always a projector $P_R$ from a vertex next to a projector $P_R$ from a propagator,
\be
\ldots (i \g^\m t_\a) P_R \   P_R\, {-i(\ksl+\psl)\over (k+p)^2-i\e} \ldots 
=\ldots (i \g^\m t_\a) P_R  \, {-i(\ksl+\psl)\over (k+p)^2-i\e}
\ee 
so that this expression coincides exactly with the expression that would have been obtained if the vertices had been the same but the propagators were $\frac{-i}{(2\pi)^4}\times\, {-i\ksl\over k^2-i\e}$ i.e.~without the chirality projector. 
These Feynman rules would be obtained from a matter Lagrangian
\be\label{chiralmatlagr3}
\cL_{\rm matter}^{R\,'} = - \pb \dsl \p - \pb (-i\Asl) P_R \p
= - \pb \dsl P_L \p - \pb \Dsl P_R \p\ ,
\ee
containing a left- and a right-handed (i.e.~a non-chiral) propagating fermion, but only the right-handed part couples to the gauge field. This asymmetry causes the Lagrangian $\cL_{\rm matter}^{R\, '}$ to not be gauge-invariant. 

In the case of the gauge-invariant Lagrangian (\ref{chiralmatlagr}), the origin of the anomaly manifests itself through the definition of the following functional integral: 
\be\label{chiralFI}
e^{i\wt\G[A]} = \int \cD\p_R \cD \overline{\p_R}\, e^{-i \int  \overline{\p_R} D\hskip-2.0mm/ \,\p_R} \sim {\rm Det}_R ( \Dsl P_R) \ .
\ee
Since the functional integral is only over right-handed fermions, the corresponding determinant should be that of the restriction of $\Dsl  P_R$ to $\cH_R$. Since $\Dsl P_R$ is an operator that maps right-handed fermions to left-handed fermions, which live in supplementary subspaces of the Hilbert space, this restriction is not well defined and thus neither is the determinant. As we had seen in subsection \ref{detsec}, in the case of the abelian anomaly, the presence of an anomaly can again be related to the ill-definition of a determinant. 

One may instead consider, in Euclidean signature, $(\Dsl_E  P_R)^\dag \Dsl_E  P_R$ which maps $\cH_R$ to itself and does have a well-defined and positive determinant. Taking its square root allows to define the modulus of what we would like to be the determinant of  $\Dsl  P_R$, but there remains an ambiguity in its phase. The phase corresponds to the imaginary part of the Euclidean effective action. If this ambiguity cannot be fixed in a gauge-invariant way one precisely has an anomaly. Thus the anomaly resides in the imaginary part of the Euclidean effective action. As we will discuss later-on, such an imaginary part necessarily involves the $\e$-tensor~!

\subsubsection{Feynman diagram computation of the triangle for chiral fermions\label{Feynanpart}}

\begin{figure}[h]
\centering
\includegraphics[width=0.5\textwidth]{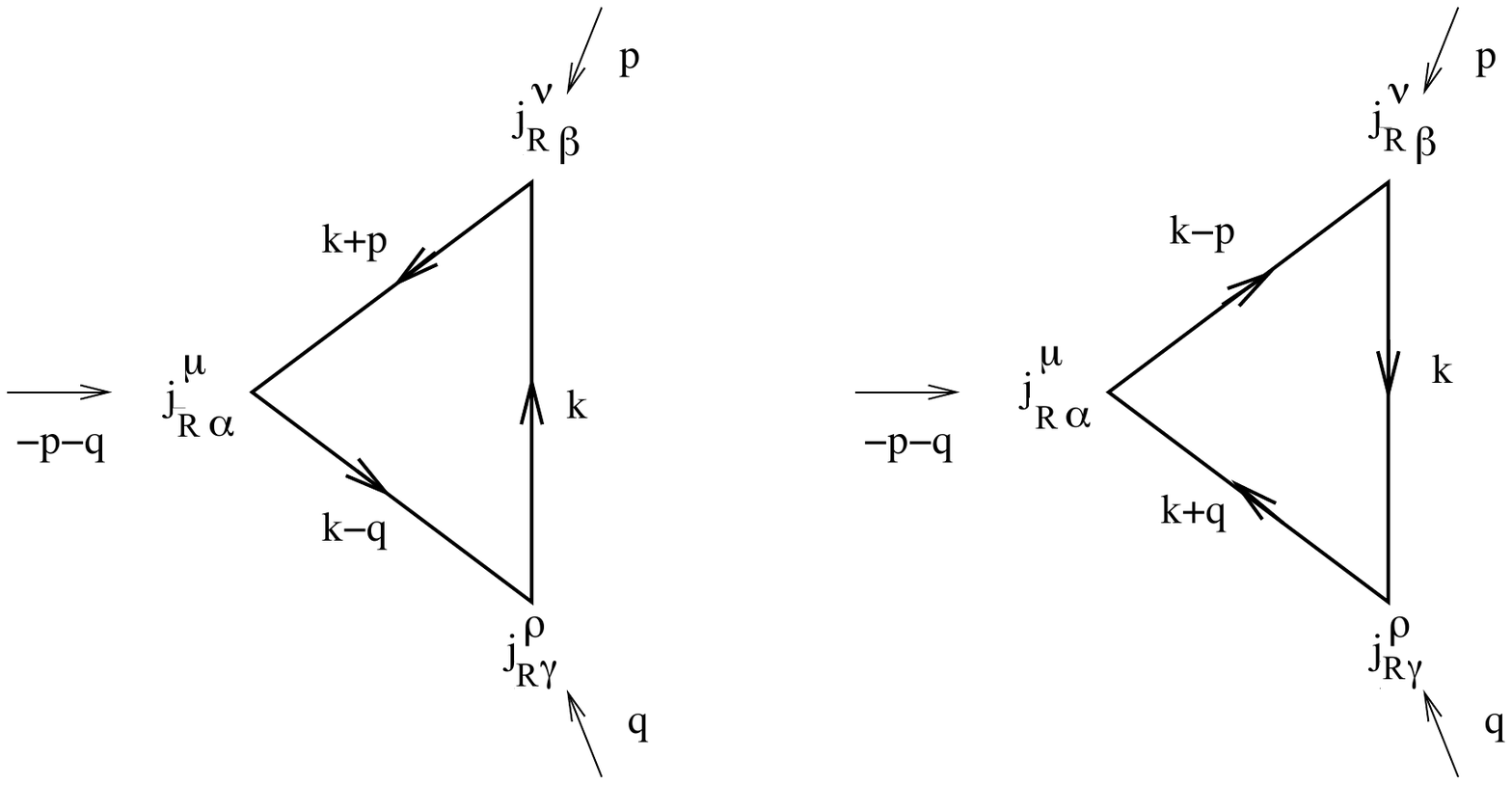}\\
\vskip-10.mm
\caption[]{The two triangle diagrams  contributing to the non-abelian chiral anomaly} \label{trianglechir}
\end{figure}

Let us  now compute the three-point vertex function $\G^{\ \,\m\n\r}_{R,\a\b\g}$ for a chiral right-handed fermion using the second formalism, i.e.~using ordinary propagators and inserting a chirality projector $P_R$ at each vertex.
This is equivalent to computing the expectation value of three right-handed currents, $\G^{\ \,\m\n\r}_{R,\a\b\g}=-\langle T\big( J_{R\a}^\m J_{R\b}^\n J_{R\g}^\r\big)\rangle$ where $J_{R\a}^\m=i\pb \g^\m t_\a P_R\p$. There are again two Feynman diagrams as shown in Fig.~\ref{trianglechir}.
\no
The same computations as in the case of the abelian anomaly lead us to the following expression \footnote{
The sign difference with respect to \eqref{G52} is only due to the fact that in \eqref{G52} we have anti-commuted the $\g_5$ to the left of the $\g^\m$.
}: 
\ba\label{Gmnr}
\G_{R,\a\b\g}^{\ \,\m\n\r}(-p-q,p,q)\hskip-1.mm 
&=& \hskip-1.mm i \int\hskip-1.mm 
{\d^4 k\over (2\pi)^4}\, \trD \Big\{\g^\m P_R{(\ksl+\psl)\over (k+p)^2 -i\e}\g^\n P_R {\ksl\over k^2 -i\e}\g^\r P_R{(\ksl-\qsl)\over (k-q)^2 -i\e}\Big\}\, \trR t_\a t_\b t_\g
\nonumber\\
&&\hskip2.6cm+\ \{( p, \n, \b) \leftrightarrow (q, \r, \g)\}
\ .
\ea
We will, once more, regularize this integral using the Pauli-Villars method. This seems quite counter intuitive as we had established that it does not make sense to consider a mass term in the case of chiral fermions. This  obstacle is overcome through the alternative formulation \eqref{chiralmatlagr3} where we consider non-chiral fermions and chiral interactions. This allows us to add a mass term to the propagator, and thus to regularize using the Pauli-Villars method. 
The origin of the anomaly can then be traced back to the fact that the Lagrangian \eqref{chiralmatlagr3} used is not gauge invariant.

Since $\G_{R,\a\b\g}^{\ \,\m\n\r}(-p-q,p,q)$  has  a degree of divergence equal to one it is enough to add one regulator field of opposite statistics. However, at intermediate steps of the computation, we will encounter integrals of degree of divergence equal to two and, for this reason, we add one more pair of regulator fields (one fermionic and one bosonic). We use the notation $\eta_1=\eta_3=1$ for the fermionic fields, and $\ \eta_2=\eta_4=-1$ for the bosonic ones, and denote their masses by $M_n$ where, of course, the original field has $M_1=0$. Then we have
\be\label{gammareg}
\left[\G_{R,\a\b\g}^{\ \,\m\n\r}(-p-q,p,q)\right]_{\rm reg}
= i \int{\d^4 k\over(2\pi)^4}\,  \left(\sum_{n=1}^4 \eta_n 
K^{\m\n\r}_{M_n}(k,p,q) \right) \trR t_\a t_\b t_\g
\ + \ \{( p, \n, \b) \leftrightarrow (q, \r, \g)\}\ ,
\ee
with
\be
K^{\m\n\r}_M(k,p,q)
=\trD \Big[ \g^\m P_R {\ksl+\psl+iM\over (k+p)^2 +M^2-i\e}
\g^\n P_R {\ksl+iM\over k^2 +M^2-i\e}
\g^\r P_R {\ksl-\qsl+iM\over (k-q)^2 +M^2-i\e}\Big] \ .
\ee

Let us now simplify this expression  using the properties of the chiral projectors. Firstly, due to the cyclicity of the trace, all mass terms of the numerator appear as
\be
\ldots \g^\l P_R iM \g^\s P_R \ldots = \ldots \g^\l  iM P_R P_L \g^\s \ldots = 0 \ .
\ee
Thus, they do not contribute. We can then further simplify this expression using the fact that $P_R^2 = P_R$:
\be
K^{\m\n\r}_M(k,p,q)
={\trD  \big(\ksl-\qsl\big) \g^\m  \big(\ksl+\psl\big)\g^\n  \ksl
\g^\r P_R 
\over [(k+p)^2 +M^2-i\e] [k^2 +M^2-i\e]  [(k-q)^2 +M^2-i\e] }
\ .
\ee
Again,  we are only interested in $(p_\m+q_\m) K^{\m\n\r}_M(k,p,q)$. Thus we need 
\ba
\hskip-0.0cm\trD  \big(\ksl-\qsl\big) \big(\psl+\qsl) \big(\ksl+\psl\big)\g^\n  \ksl\g^\r P_R
&=&\trD  \big(\ksl-\qsl\big) \big(\psl+\ksl-\ksl+\qsl) \big(\ksl+\psl\big)\g^\n  \ksl\g^\r P_R
\nonumber\\
&&\hskip-3.7cm
=(k+p)^2 \trD  \big(\ksl-\qsl\big) \g^\n  \ksl\g^\r P_R
-(k-q)^2\trD \big(\ksl+\psl\big)\g^\n  \ksl\g^\r P_R
\nonumber\\
&&\hskip-3.7cm
= \big[(k+p)^2 +M^2\big]\trD  \big(\ksl-\qsl\big) \g^\n  \ksl\g^\r P_R
-\big[(k-q)^2+M^2]\trD \big(\ksl+\psl\big)\g^\n  \ksl\g^\r P_R
\nonumber\\
&&\hskip-3.3cm
+M^2\, \trD \big(\psl+\qsl\big) \g^\n \ksl\g^\r P_R \ ,
\ea
which, through cancellations between the numerators and denominators,  leads us to: 
\ba
(p_\m+q_\m) K^{\m\n\r}_M(k,p,q)
&=&{\trD  \big(\ksl-\qsl\big) \g^\n  \ksl\g^\r P_R \over
[k^2 +M^2-i\e]  [(k-q)^2 +M^2-i\e] }
- {\trD \big(\ksl+\psl\big)\g^\n  \ksl\g^\r P_R \over
[(k+p)^2 +M^2-i\e] [k^2 +M^2-i\e] }
\nonumber\\
&+& M^2 {\trD \big(\psl+\qsl\big) \g^\n \ksl\g^\r P_R \over
[(k+p)^2 +M^2-i\e] [k^2 +M^2-i\e]  [(k-q)^2 +M^2-i\e] } \ .
\ea
Again, in the first two terms, the $\g_5$ in the projectors $P_R$ give vanishing contributions after doing the integration (since they only involve a single independent momentum), but now the $1$ in $P_R$ does lead to a non-vanishing integral for each of these to terms.
Hence,
\ba\label{GammaLdiv}
- i(p_\m+q_\m)\left[\G_{R,\a\b\g}^{\ \,\m\n\r}(-p-q,p,q)\right]_{\rm reg}
&=&\Big( L^{\n\r}(-q) - L^{\n\r}(p) 
+ \sum_{n=2}^4 \eta_n M_n^2 J_{M_n}^{\n\r}(p,q) \Big) 
\trR t_\a t_\b t_\g \nonumber\\
&&
+\ \{( p, \n, \b) \leftrightarrow (q, \r, \g)\} \ ,
\ea
where $L^{\n\r}(p)$ and $J_{M_n}^{\n\r}(p,q)$ are {\it convergent} integrals (at least as long as we keep the regulator masses $M^2_N$ finite), allowing us in particular to shift the integration variables:
\be\label{Inr}
L^{\n\r}(p)= \frac{1}{2}\int{\d^4 k\over(2\pi)^4}\, 
\sum_{n=1}^4  \eta_n \
{\trD \big(\ksl+\psl\big)\g^\n  \ksl\g^\r \over
[(k+p)^2 +M_n^2-i\e] [k^2 +M_n^2-i\e] } \ ,
\ee
and, introducing Feynman parameters,
\ba\label{JMs}
J_{M_n}^{\n\r}(p,q)\hskip-2.mm&=&\hskip-2.mm\int{\d^4 k\over(2\pi)^4}\,
{\trD \big(\psl+\qsl\big) \g^\n \ksl\g^\r P_R \over
[(k+p)^2 +M_n^2-i\e] [k^2 +M_n^2-i\e]  [(k-q)^2 +M_n^2-i\e] }
\nonumber\\
&=&\hskip-2.mm 2\int_0^1\d x \int_0^{1-x}\hskip-1.mm \d y \int{\d^4 k\over(2\pi)^4}\,
{\trD \big(\psl+\qsl\big) \g^\n \ksl\g^\r P_R \over
[(k+xp-yq)^2 +M_n^2+ r(p,q,x,y)-i\e]^3}
\nonumber\\
&=&\hskip-2.mm 2\int_0^1\d x \int_0^{1-x} \hskip-2.mm\d y\ 
\trD \big(\psl+\qsl\big) \g^\n \big( y\qsl-x\psl\big)\g^\r P_R 
\int{\d^4 k'\over(2\pi)^4}\,
{1 \over [{k'}^2 +M_n^2+ r(p,q,x,y)-i\e]^3} \ ,\nonumber\\
\ea
where we used that the odd power of $k'$ in the numerator does not contribute, and where $r(p,q,x,y)= x(1-x)p^2+y(1-y)q^2 +2xy\,pq$.
We only need the large $M_n$ limits of $M_n^2 J_{M_n}^{\n\r}(p,q)$. They are all the same, (cf~(\ref{asympt}))
\ba\label{JMlimit}
\lim_{M_n\to\infty} M_n^2 J_{M_n}^{\n\r}(p,q) &=& 2\int_0^1\d x \int_0^{1-x} \d y\, 
\trD \big(\psl+\qsl\big) \g^\n \big( y\qsl-x\psl\big)\g^\r P_R
\times {i\over 32\pi^2}
\nonumber\\
&=&{i\over 16\pi^2} {1\over 6}\,\trD \big(\psl+\qsl\big) \g^\n \big( \qsl-\psl\big)\g^\r P_R \nonumber\\
&=&{i\over 24\pi^2} \left[ q^\n q^\r-p^\n p^\r + {p^2-q^2\over 2} \eta^{\n\r} - i \e^{\l\n\s\r} p_\l q_\s \right]\ .
\ea

We have mentioned above in subsection \ref{chidd} (and explicitly observed for the abelian anomaly) that the anomaly is given by the terms involving $\e^{\n\r\l\s}$. The $L^{\n\r}(p)$ and $L^{\n\r}(-q)$ do not give rise to any such terms. 
But we know from the (anomalous) Ward identity \eqref{anomWardgauge} that these additional pieces without $\e$-tensor are present. They correspond to the two terms in  \eqref{anomWardgauge} that involve the structure constants \footnote{
One easily sees that $L^{\n\r}(p)=L^{\n\r}(-p)=L^{\r\n}(p)$, so that the combination $L^{\n\r}(-q)-L^{\n\r}(p)$ is odd under exchange of $(p,\n)$ and $(q,\r)$. The same is true for the terms in \eqref{JMlimit} that do not involve the $\e$-tensor. This means that upon adding the second diagram, one has $\trR t_\a t_\b t_\g \to \trR t_\a (t_\b t_\g - t_\g t_\b) = i C_{\b\g\dd}\, \trR t_\a t_\dd \sim C_{\b\g\a}$, and this is how the structure constants arise from these terms.
} 
$C_{\a\g\e}$ and $C_{\a\b\e}$. They are essentially vacuum-polarization diagrams, similar to the one computed above in QED. Just as the latter, they require the addition of a counterterm to yield a finite limit as the regulator masses are taken to infinity. We will not discuss this any further here.

Hence, inserting the $\e^{\l\n\s\r}$-terms of (\ref{JMlimit}) into (\ref{GammaLdiv}), 
we can already state the main result of this section
\ba\label{GammaLdiv2}
-i(p_\m+q_\m)\G_{R,\a\b\g}^{\ \,\m\n\r}(-p-q,p,q)\Big\vert_{\e -{\rm terms}}
\hskip-4.mm&=&\hskip-2.mm \sum_{n=2}^4 \eta_n\, {i\over 24\pi^2}\,  (-i) \e^{\l\n\s\r} p_\l q_\s\,
\trR t_\a t_\b t_\g
+\ \{( p, \n, \b) \leftrightarrow (q, \r, \g)\}
\nonumber\\
&=& \hskip-2.mm
{1\over 12 \pi^2}\ \e^{\n\r\l\s} p_\l q_\s\, D^\cR_{\a\b\g}
\ ,
\ea
where we used $\sum_{n=2}^4\eta_n=-\eta_0=-1$ and $\trR t_\a t_{(\b} t_{\g)} \equiv D^\cR_{\a\b\g}$, (and another sign comes from rearranging the indices on the $\e$-tensor). 
As compared to the analogous formula \eqref{abeliananomcharac} for the abelian anomaly where the numerical factor was  $-\frac{1}{2\pi^2}$, we now have a factor $\frac{1}{12\pi^2}$. The sign difference and one factor $\frac{1}{2}$ are easily understood as coming from the replacement of $\g_5$ by $P_R=\frac{1}{2}(1-\g_5)$, but the presence of the 3 projectors also eliminated all mass terms in the numerator, implying further differences in the computation. 

Using a reasoning similar to what we did for the abelian anomaly, equation \eqref{GammaLdiv2} allows us to infer the part of the anomaly that is bilinear in the gauge fields~:
\be\label{chiralanom1}
\cA^R_\a(x) 
={1\over 24\pi^2} \e^{\m\n\r\s} \trR t_\a \del_\m A_\n \del_\r A_\s + \cO(A^3) \ ,
\ee
where we added a superscript $R$ on $\cA_\a(x)$ to remind us that this anomaly is computed for right-handed fermions. Note that our computation did not yield the $\cO(A^3)$ contributions to the anomaly. To determine them one would need to evaluate similarly a  four-point function which is given by the sum of 6 box diagrams. Of course, one only needs to compute one of them, the others correspond to permutations of the cyclic order of the vertices. Actually, this computation is not needed as the so-called Wess-Zumino consistency condition, to be discussed in the next section, completely fixes these higher-order contributions in terms of  (\ref{chiralanom1}). The result is
\be\label{fullnaanom}
\cA^{R}_{\a}(x)={1\over 24\pi^2} \e^{\m\n\r\s} \trR t_\a\, \del_\m \Big(A_\n \del_\r A_\s -{i\over 4} A_\n [ A_\r , A_\s]\Big)\ .
\ee
Note the absence of terms quartic in $A$, which means that the 5-point function (pentagon diagrams) is not anomalous. Finally the $A^3$-term  can also be rewritten in terms of the symmetrized trace of 3 generators,\footnote{
Note that $\e^{\m\n\r\s} [A_\n,[A_\r,A_\s]]$ vanishes by the Jacobi identity. This allows us to rewrite
$
\e^{\m\n\r\s} \trR t_\a A_\n[A_\r,A_\s]= {1\over 2}\e^{\m\n\r\s} \trR t_\a \big( A_\n[A_\r,A_\s]+ [A_\r,A_\s] A_\n\big) = {1\over 2}\e^{\m\n\r\s} A_\n^\b i C^{\g}_{\ \dd\e} A_\r^\dd A_\s^\e\,\trR t_\a ( t_\b t_\g+t_\g t_\b) = \e^{\m\n\r\s} A_\n^\b i C^\g_{\ \dd\e} A_\r^\dd A_\s^\e\, D_{\a\b\g}^\cR \ .$
}
 i.e.~the $D^\cR_{abc}$-symbol~:
\be\label{fullnaanom2}
\cA^R_\a(x)={1\over 24\pi^2} \e^{\m\n\r\s}  \del_\m \Big(A_\n^\b \del_\r A_\s^\g -{i\over 4} A_\n^\b [ A_\r , A_\s]^\g\Big) D^\cR_{\a\b\g}\ .
\ee

If one does the same computations for a left-handed fermion, the only difference being $\g_5\to -\g_5$, and then obviously the anomaly obtained is exactly the opposite~:
\be\label{anLanR}
\cA ^L_\a(x)=-\cA_\a^R(x) \ .
\ee
Obviously also, since a non-chiral fermion is composed of a left-handed and a right-handed fermion, both in the same representation of the gauge group, there is no anomaly  for a non-chiral fermion.



\section{General characterization and properties of the anomalies}
\setcounter{equation}{0}

\subsection{Group structure}

The possibility of having an anomaly is intrinsically linked to the structure of the gauge group. 
In fact, if $D^\cR_{\a\b\g}= \trR t_\a t_{(\b}t_{\g)} = \str_\cR t_\a t_{\b}t_{\g} $ vanishes, then so does the chiral anomaly \eqref{fullnaanom2} .
This is the case when we consider  real or pseudo-real representations. Indeed, if the representation $\cR$ is real, it is equal to its complex conjugate representation $\cR^*$ which is given by
$\exp(i\th^\a t_\a^{\cR^*}) = \big(\exp(i\th^\a t_\a^\cR)\big)^* = \exp(-i\th^\a(t_\a^\cR)^*) =\exp(-i\th^\a(t_\a^\cR)^T)$. \footnote{We take the generators to be hermitian.} This translates into the following condition on the generators $t_\a^{\cR^*}=-(t_\a^\cR)^T$ and for a real representation $t_\a^\cR=t_\a^{\cR^*}=-(t_\a^\cR)^T$, while for a pseudo-real one, this relation reads: 
$t_\a^\cR=Pt_\a^{\cR^*}P^{-1}= -P(t_\a^\cR)^T P^{-1}$, with some fixed matrix $P$.
As such, for a real or pseudo-real representation
\ba
D^\cR_{\a\b\g} &=& \strR t_\a t_{\b}t_{\g} = (-)^3 \strR P (t_\a)^T P^{-1} P (t_{\b})^T P^{-1} P (t_{\g})^T P^{-1} \nonumber\\
&=& - \strR t_\a^T t_{\b}^T t_{\g}^T=  - \strR (t_\g t_{\b} t_{\a})^T =  - \strR t_\g t_{\b} t_{\a}= - \strR t_\a t_{\b}t_{\g} = - D^\cR_{\a\b\g} \ ,
\ea
(where $P={\bf 1}$ if the representation is real). This implies that $D^\cR_{\a\b\g}=0$. 

Thus, we cannot have anomalies in (four-dimensional) gauge theories with a gauge group which only has real and pseudo-real representations. This is the case for many well-known groups such as $SO(4n),\ n\ge 2$ or $SO(2n+1),\ n\ge 1$ as well as several others. Yet other groups do have non (pseudo) real representations, but still with all $D^\cR_{\a\b\g}$ vanishing.
In fact, the only groups which have non-vanishing $D^\cR_{\a\b\g}$ and can lead to anomalies are $SU(n), n\geq 3$, $U(1)$ and product groups involving these as factors. Note that this argument is specific to 4 mod 4 dimensions where the relevant $D$-symbol is the trace of an odd number of  generators. 

In particular, in the low-energy QFT limit of ten-dimensional superstring theories anomalies will involve the 6-index $D_{\a_1\ldots a_6}$-symbol, which has no reason to vanish for the $SO(n)$ groups.

\subsection{Particles or antiparticles}

It should be clear that if different chiral particles (fields) are present, one has to add the contributions of the different one-loop diagrams. As a result, all chiral particles contribute additively to the anomaly. Since the chiral fermions are massless, their contributions to the anomaly cannot depend on a mass parameter, but only on the group representation $\cR$ they are in. As a result, their contribution to the anomaly is fully captured by the corresponding $D^\cR$-symbol. Since the fermionic propagator in the loop is the propagator of the particle {\it and} its anti-particle, it is clear that each particle-anti-particle pair only contributes once.
But in four dimensions the antiparticle of a left-handed particle is right-handed and vice versa, so one might ask whether it is important which one of the pair has been designated as the particle. We will now show that it does not matter.

Suppose we have a left-handed fermionic particle in some representation $\cR_L$ of the gauge group, with associated quantum field $\p$. Then its antiparticle is described by the charge-conjugated $\p^c= i\g^0 \cC \p^*$. The charge conjugation matrix $\cC$ satisfies $\cC \g_5^*=\g_5 \cC$ so that
\be
\g_5 \p^c =\g_5  i\g^0 \cC \p^*=-i\g^0 \cC (\g_5 \p)^* 
=\mp i\g^0 \cC \p^*
=\mp \p^c 
\quad {\rm if}\ \ \g_5\p =\pm\p \ .
\ee
We see that indeed, the charge conjugated field has opposite chirality. On the other hand, due to the complex conjugation, $\p^c$ transforms in the complex conjugate representation $(\cR_L)^*$. As we have just seen, the $D$-symbol for the complex conjugate representation
has the opposite sign. So if one starts out with the anti-particle, one gets one minus sign from having the opposite chirality and one gets another minus sign from being in the complex conjugate representation, so that overall the contribution to the anomaly is the same.
Again, this argument is correct in 4 mod 4 space-time dimensions. In 2 mod 4 dimensions the complex conjugate representation has the same $D$-symbol. But one also has $\cC \g^* = - \g\cC$ (where $\g$ is the chirality matrix $\sim \g^0\ldots \g^{d-1}$ in $d$ dimensions), so that particles and anti-particles have the same chirality. Then switching from particles to anti-particles results in two plus signs rather than two minus signs, and again it does not matter. 

Note that in 2, 6, 10, etc dimensions one can have a chiral particle that is its own anti-particle, usually called a Majorana-Weyl fermion.

\subsection{Locality} \label{local}
Let us show that the anomaly is local i.e.~that it can be written as $\int d^4 x\, f(x)$ where $f(x)$ only depends on the values of the fields and their derivatives at $x$. In Fourier transform terms, this means that the anomaly is polynomial in the momenta. 

Firstly, it should be underlined that anomalies arise from the regularization of divergent one-loop diagrams, and that convergent diagrams do not need to be regularized. 
Secondly, taking a derivative with respect to external momenta decreases the degree of divergence by one.\footnote{In fact, the derivation yields a sum of terms where one of the internal propagators is replaced by $
{\del\over\del p_\m}\left(-{1\over \footksl+\footpsl+\ldots}\right)
={(\footksl+\footpsl+\ldots)\g^\m(\footksl+\footpsl+\ldots)\over [(k+p+\ldots)^2]^2}
$. For a one-loop diagram it is easy to see that all terms have their degree of divergence reduced by one.}

The three-point vertex function has a degree of divergence $\cD=1$, and two derivatives result in a convergent integral. Then three derivatives of $ (p_1)_{\m_1} \G^{\m_1\ldots\m_3}(p_1, p_2, p_3)$ result in a convergent integral which doesn't need regularization and which then cannot have any anomalous part. As such, three  derivatives kill the anomaly meaning that it is a polynomial of degree at most two in the momenta. This is indeed what we have observed. This argument is general for all anomalous one-loop diagrams in an arbitrary (even) space-time dimension~: If the diagram has degree of divergence $\cD$, then $\cD+1$ derivatives with respect to the external momenta result in a convergent integral that needs no regularization. As such, $\cD+2$ derivative kill the anomalous contribution to $ (p_1)_{\m_1} \G^{\m_1\ldots\m_n}$ which must be a polynomial of degree $\cD+1$ in the $p_j$.

\subsection{How many points necessary for an anomalous diagram?}

This observation leads to an interesting constraint on the number of points (vertices) needed to observe an anomaly for a given space-time dimension $d$. Firstly, in odd dimensions, one cannot define a $\g_5$ matrix satisfying the ordinary properties. Thus, there are no anomalies.\footnote{More precisely, there are no anomalies of the kind we discuss here : abelian anomaly, anomaly under gauge transformations, or anomalies under diffeomorphims / local Lorentz transformations. There are however anomalies under discrete symmetries like parity \cite{AGDPM}.
} 
Secondly, let us consider an even dimensional space-time with $d=2n$. The quantity we are interested in is the contraction of  the $(m+1)$-point vertex function with one of the external momenta, $ p^{(1)}_{\m_1}\G^{\m_1...\m_{m+1}}$, and we expect to find a quantity proportional to the epsilon tensor $\e^{\n_1\ldots\n_d}$. As just seen, we also expect the anomaly to be polynomial in the momenta. Thus, we need $d-m$ independent external momenta that we can contract this quantity with. This means that we need to have $d-m \leq m$, i.e. $m \geq n$. Thus, the minimal number of points that can lead to an anomalous one loop diagram is $n+1$. In the case of 4 dimensional spacetime, this is in fact the triangle, as we have extensively discussed. In 6 dimensions the anomaly first shows up in a box diagram, in 8 dimension in the pentagon diagram, and in 10 dimensions one must compute the hexagon diagram. This requires way more work and quite a few tricks \cite{Alvarez-Gaume-Witten} !

\subsection{Finiteness}
Both in the case of the abelian and chiral anomaly, the anomalies we computed had finite, i.e.~regulator independent, coefficients. (This was not the case for the non-anomalous part of the triangle amplitude which required addition of counterterms to cancel their divergent pieces.)

Let us now show why this in fact \textit{has} to be the case because of locality and dimensionality. 
Firstly, the three-point vertex function has a degree of divergence one, which means, per Section \ref{local}, that the anomalous part is a polynomial of degree at most two in the external momenta. The most general form of such a polynomial is the following: 
\be\label{genanomform}
(p_\m+q_\m)\G^{\m\n\r}_{\a\b\g}(-p-q,p,q)\Big\vert_{\rm anom\ part}
= {D_1}^{\n\r\l\s}_{\a\b\g} p_\l p_\s + 
{D_2}^{\n\r\l\s}_{\a\b\g} q_\l q_\s + 
{D_3}^{\n\r\l\s}_{\a\b\g} p_\l q_\s + 
{C_1}^{\n\r\l}_{\a\b\g} p_\l +{C_2}^{\n\r\l}_{\a\b\g} q_\l
+B^{\n\r}_{\a\b\g} \ ,
\ee
where $D_1, D_2, D_3, C_1, C_2, B$ are constant Lorentz tensors independent of momenta. First of all, there are no constant Lorentz tensors independent of momenta with three indices, so that $C_1$ and $C_2$ must vanish.  Next, we expect the anomaly to be proportional to the epsilon tensor. Thus, $D_1, D_2$ and $B$ have to vanish also, as the associated terms do not respect the required anti-symmetry. 
All that we are left with is the term ${D_3}^{\n\r\l\s}_{\a\b\g} p_\l q_\s$. 

Secondly, let us look at how the scaling dimension constrains the tensor ${D_3}^{\n\r\l\s}_{\a\b\g}$. 
The three-point vertex function $\G^{\m\n\r}_{\a\b\g}(-p-q,p,q)$ has a scaling dimension 1 since each fermion propagator contributes a $-1$ and the integration contributes a $+4$. Thus,  $(p_\m+q_\m)\G^{\m\n\r}_{\a\b\g}(-p-q,p,q)$ has a scaling dimension 2, which implies that ${D_3}^{\n\r\l\s}_{\a\b\g}$ has a scaling dimension 0. As it does not involve the momenta (and the fields are massless), by dimensional considerations, it cannot involve the Pauli-Villars regulator mass $M$. 
This means that the anomaly is necessarily finite, i.e.~does not involve regularization-dependent coefficients.

The same type of argument can also be applied to the box-diagram responsible for the $A^3$-part of the anomaly $\cA_{\rm cubic} \sim  (p_\m+q_\m+k_\m)\G^{\m\n\r\s}_{\a\b\g\dd}(-p-q-k,p,q,k)\Big\vert_{\rm anom\ part}$. This box diagram has one more fermion propagator and a degree of divergence 0, so that $\cA_{\rm cubic}$  is a polynomial of degree one.  On the other hand the box diagram has  scaling dimension 0, so that $\cA_{\rm cubic}$ has scaling dimension 1. Again, the only possible structure is ${E_3}^{\n\r\s\l}_{\a\b\g\dd} (p_\l+q_\l+k_\l)$ with a dimensionless $E_3$ proportional to the $\e$-tensor. So this term is also finite.

It is easy to see that these arguments generalize to arbitrary even dimensions~: in $d=2n$ dimensions, the $n+1$ point one-loop diagram has degree of divergence $d-(n+1)=n-1$ so that $k_\m \G^{\m\ldots}$ is annihilated by $n+1$ derivatives and must be a polynomial of degree $n$ in the external momenta. But this expression has scaling dimension $n$ and contains $n$ momenta, hence the tensor-coefficient is dimensionless and independent of any momentum and, again, cannot depend on the Pauli-Villars regulator mass $M$. The anomalous higher-point functions are treated similarly. This proves finiteness of the anomaly in any (even) dimension.

\subsection{One-loop exactness}

The Adler-Bardeen theorem \cite{Adler-Bardeen} states that higher loop diagrams do not contribute to the anomaly, meaning that the one-loop computation we have done here is actually exact to all orders in perturbation theory. The original proof is complicated and well beyond the scope of these notes. Part of the reasoning is that higher loops necessarily contain bosonic (gauge field) propagators and these can be regulated in a gauge invariant way, thus effectively regulating the full diagram  \cite{Alvarez-Gaume-Witten, Weinberg-book}. The subtlety here is to show that 
regulating the gauge boson propagators sufficiently regulates all loop integrals and that this does not generate any non-gauge invariant (anomalous) terms.

Alternatively, we have argued that  anomalies originate from a non-trivial transformation of the functional integral measure, giving rise to a non-trivial (functional) Jacobian determinant.  But determinants always correspond to one-loop effects. This can be seen e.g.~by introducing a loop counting parameter $\l$, as we did in subsection 2.5. The determinant cannot contain any $\l$, so it is of order $\l^{L-1}=\l^0$, confirming indeed $L=1$, i.e.~it contributes at the one-loop order only.

Below, we will sketch how the chiral anomaly can be related to the index of an appropriate extension of the Dirac operator. This shows again that the anomaly is independent of any loop-counting parameter and must correspond to  a one-loop contribution only.


\subsection{Relevant and irrelevant anomalies}

We have shown that the anomaly is related to the non-invariance of the quantum effective action, see e.g.~\eqref{noninvGamma}, $\int\d^4 x\, \e^\a(x) \cA_\a(x) = \dd_\e \wt\G[A]$.
Obviously also, since the anomaly arises as a one-loop effect, it is of ``higher order" than the classical action. If we re-introduced $\hbar$ as a loop counting parameter, one would see that the anomaly has an extra $\hbar$ as compared to the classical action. Also, as obvious from the triangle, it involves 3 gauge coupling constants $g$ (which we have hidden in the generators and hence in the $D$-symbol). As such the anomaly is of order $g^3$ where the tree-level couplings are of order $g$. We see that the anomaly is of the same order as the counterterms one adds to the action to cancel possible divergences (in other diagrams). One may then wonder whether it is possible to add some non gauge-invariant higher-order ``counterterm" $\hat S_{ct}[A]$ to the classical action such that $\dd_\e  \hat S_{ct} = - \dd_\e \wt \G[A]$. More precisely, $\hat S_{ct}[A]$ would be an order $g^3$ term to be added to the classical action and then would also appear as an additional term in $\wt \G[A]$~:
\be
\wt \G[A]\to \wt \G[A]+\hat S_{ct}[A] \quad , \quad \dd_\e(\wt \G[A]+\hat S_{ct}[A])=0 \ .
\ee
It looks as if one can simply take $\hat S_{ct}[A]=-\wt \G[A]$ and get rid of the anomaly. This is of course not true. In any local quantum field theory, we are only allowed to add local counterterms to the action, i.e.~a space-time integral of a local expression involving only the fields at the point $x$ as well as a finite number of derivatives acting on these fields. While we have shown that the anomaly $\dd_\e \wt \G[A]$ is such a local functional, the effective action $\wt \G[A]$ itself is not. To understand this, note that the anomaly corresponds to $(p_1)_\m \G^{\m\n\r}_{\a\b\g}(p_1,p_2,p_3)$ which upon Fourier transforming is $ \frac{\del}{\del x^\m} \G^{\m\n\r}_{\a\b\g}(x,y,z)$. On the other hand, the effective action contains the $\G^{\m\n\r}_{\a\b\g}$ itself, and  ``removing" the derivative with respect to $x^\m$ is not a local operation.

Thus the question is whether or not we can find a {\it local functional} of the gauge fields $\hat S_{ct}[A]$ such that its gauge variation cancels the anomaly $\dd_\e\wt\G[A]$. If we can, the anomaly is called irrelevant because it can be removed by adding such a counterterm to the classical action. If we cannot, the anomaly is called relevant~:
\be
\text{The anomaly is relevant} \quad \Longleftrightarrow\quad 
-\int\d^4 x\, \e^\a(x) \cA_\a(x)\ne \dd_\e \hat S_{ct}[A]\quad \text{with $\hat S_{ct}[A]$ a {\it local} functional.}
\ee
\noindent
It is relatively easy to see that the chiral anomaly is a relevant anomaly. The $\hat S_{ct}[A]$ would need to give an additional contribution $\D \G^{\m\n\r}_{\a\b\g}(k,p,q)$ to the 3-point function that must be of scaling dimension 1, i.e. linear in the external momenta and totally symmetric under exchange of the external legs (i.e.~$(k,\m,\a)\leftrightarrow (p,\n,\b)\leftrightarrow (q,\r,\g)$). This is usually referred to as the Bose symmetry, since this $\D \G^{\m\n\r}_{\a\b\g}$ arises from three functional derivatives with respect to the gauge fields which are bosonic and commute. On the other hand, $\D \G^{\m\n\r}_{\a\b\g}$ must involve the $\e$-tensor in order to possibly cancel the anomaly. To cancel the anomaly due to a right-handed fermion, the requirement is that 
\be
-i(p_\m+q_\m) \D \G^{\m\n\r}_{\a\b\g}(-p-q,p,q) = -\frac{1}{12\pi^2} \e^{\n\r\l\s} p_\l q_\s D_{\a\b\g} \ ,
\ee
and similarly (with the opposite sign) for a left-handed fermion.
It is easy to see that no such $\D \G^{\m\n\r}_{\a\b\g}(k,p,q)$ with the required Bose symmetry can exist. We conclude that the chiral anomaly is indeed relevant. 

\subsection{Anomaly cancellation}

As already mentioned, the contribution of fermions to the gauge anomaly is additive: the total anomaly is obtained by summing over all chiral species. Furthermore, the structure of the anomaly is universal and the only model dependence factors in through the group--theoretical factor
\(
D^{\mathcal R}_{\alpha\beta\gamma}
\),
which depends only on the gauge group and representations \(\mathcal R\) of the fermions.

For a general chiral gauge theory with left-handed fermions in representations $\cR_i^L$ and right-handed ones in representations $\cR_j^R$, the anomaly reads: 
\ba
\hskip5.mm \mathcal A_\alpha =
-\frac{1}{24\pi^2}\,\epsilon^{\mu\nu\rho\sigma}\,
\del_\mu\!\Big(
A_\nu^\beta \partial_\rho A_\sigma^\gamma
-\frac{i}{4}A_\nu^\beta [A_\rho,A_\sigma]^\gamma
\Big)
\Big(
\sum_i D^{\cR_i^L}_{\a\b\g}
-
\sum_j D^{\cR_j^R}_{\a\b\g}
\Big).
\ea
\no
Breaking of the gauge invariance by anomalies ruins the consistency of the quantum theory. Hence, gauge invariance at the quantum level requires that the expression 
$(\sum_i D^{\cR_i^L}_{\a\b\g} - \sum_j D^{\cR_j^R}_{\a\b\g})$ vanishes, which is the condition for anomaly cancellation.  

In the standard model the gauge group is the product $SU(3)\times SU(2) \times U(1)$. While we have seen that $SU(2)$ alone has a vanishing $D$-symbol, for product groups one can have different generators appear in the trace of the $D$-symbol. Since the generators for the different groups act on different representation spaces, the trace factorizes accordingly. If we call $T_a$ the 8 generators of the $SU(3)$, $t_\a$ the 3 generators of $SU(2)$ and $q$ the generator of the $U(1)$ then the different non-vanishing traces are (within any irreducible reprsentation)
\be\label{mixed}
\tr T_aT_bT_c \quad , \quad \tr t_\a t_\b q= q \tr t_\a t_\b \sim q \dd_{\a\b}
\quad \quad \tr T_a T_b q=q \tr T_a T_b \sim q \dd_{ab}
\quad , \quad \tr q q q=q^3 \ .
\ee
For example, $\tr t_\a t_\b q$ will result form a triangle diagram with two vertices where one could attach $SU(2)$ gauge bosons and one vertex where a $U(1)$ gauge boson could be attached. $q$ is the $U(1)$ (hyper) charge of the chiral fermion ``circulating" in the loop.
The different anomalies corresponding to the terms in \eqref{mixed} are referred to as $SU(3)^3$, $SU(2)^2 \times U(1)$, $SU(3)^2\times U(1)$ and $U(1)^3$-anomalies. For each of these anomalies one must check that they vanish  when summed over all chiral fermion representations (with the signs according to the chirality).
Remarkably, in the Standard Model this condition is satisfied separately within each fermion generation, leading to a non-anomalous gauge theory. Historically the requirement of anomaly cancellation  had let to the prediction of the existence of the (then still unobserved) charm quark \cite{BIM}. 

\subsection{Wess-Zumino consistency condition}

 We have defined the anomaly $\cA_\a$ as the gauge variation of the effective action after integration over the matter fields: 
\be\label{andef33}
\dd_\e \wt\G[A] = \int \e^\a(x) \cA_\a(x)
\quad \Leftrightarrow \quad
\cA_\a(x) = - \Big( D_\m {\dd\over \dd A_\m(x)}\Big)_\a \wt\G[A] \ .
\ee
\no
This definition constrains the expression of $\cA_\a$ and leads to a linear equation called the Wess-Zumino consistency condition. 
In fact, one can rewrite \eqref{andef33} using a functional derivative operator $\cF$ defined as: 
\be\label{cfop}
\cF_\a(x)= - \left( D_\m {\dd\over \dd A_\m(x)}\right)_\a
= - {\del\over \del x^\m} {\dd\over \dd A_\m^\a(x)}
- C_{\a\dd\g} A_\m^\dd(x) {\dd\over \dd A_\m^\g(x)} \ ,
\ee
which yields $\cA_\a(x)=\cF_\a(x) \wt\G[A]$. 
Using the Jacobi identity, one can show that this functional derivative satisfies the following algebra:  
\be\label{Galgebra2}
[\cF_\a(x), \cF_\b(y)]
=\dd^{(4)}(x-y) C_{\a\b\g}  \cF_\g(x)\ ,
\ee
\no
Applying this to the effective action then yields the following relation for the anomaly~: 
\be\label{WZcond}
\cF_\a(x) \cA_\b(y) - \cF_\b(y) \cA_\a(x) = C_{\a\b\g} \dd^{(4)}(x-y) \cA_\g(x)  .
\ee
\no
This is Wess-Zumino's consistency condition \cite{Wess-Zumino}. It constrains the form of the anomaly and, since $\cF$ mixes terms of different degrees in the gauge fields,  it allows us to determine the higher order ($\cO(A^3)$) terms that contribute to the anomaly  using only the term found in \eqref{chiralanom1} from the triangle computation. This will be done in the next section after a  reformulation of the Wess-Zumino consistency condition in terms of the nilpotency of the BRST transformation.

\subsection{Reformulation in terms of differential forms}\label{diffformsec}

We have seen that the anomalies are intrinsically related to the presence of the $\g_5$-matrix which in turn implies that the expression of the anomaly involves the $\e^{\m\n\r\s}$-tensor. This will allow us to rewrite the anomaly $\int \d^4 x\, \e^a \cA_a$ as the integral of a differential 4-form. Differential $p$-forms can be naturally integrated over $p$-manifolds, and here the 4-manifold is the (flat) space-time. But the formulation in terms of differential forms immediately generalizes to curved manifolds. On a curved manifold we need to distinguish the antisymmetric symbol $\wh\e^{\m\n\r\s}$ which equals $\pm 1$ or 0 and  the ``true" $\e^{\m\n\r\s}$-tensor. They are related as $\wh\e^{\m\n\r\s} =\sqrt{-g}\,\e^{\m\n\r\s} $ where $\sqrt{-g}$ is the square-root of (minus) the determinant of the space-time metric $g_{\m\n}$. Then
\be
\d x^\m \wedge \d x^\n\wedge \d x^\r\wedge\d x^\s
= \wh\e^{\m\n\r\s} \d^4 x = \e^{\m\n\r\s} \sqrt{-g}\, \d^4 x \ .
\ee
The gauge field potentials $A_\m=A_\m^\a t_\a$ are combined into the Lie algebra-valued one forms
\be\label{Aoneform}
A=A_\m \d x^\m = A_\m^\a \d x^\m t_\a \ .
\ee
Recall that the exterior derivative is $\d = \d x^\m \del_\m$.
Using these relations, we can  rewrite the anomaly for a left-handed (i.e.~positive chirality) fermion, which is the opposite of \eqref{fullnaanom}, now on a curved space-time manifold, as
\ba\label{an4form}
\int \d^4 x\, \sqrt{-g}\, \e^a \cA^L_a 
&=& -\int \d^4 x\, \sqrt{-g}\, \frac{1}{24\pi^2}  \e^{\m\n\r\s} \tr \e\, \del_\m \Big( A_\n\del_\r A_\s-\frac{i}{4} A_\n A_\r A_\s + \frac{i}{4} A_\n A_\s A_\r \Big)
\nonumber\\
&&\hskip-2.cm = -\frac{1}{24\pi^2}   \int \d x^\m \wedge \d x^\n\wedge \d x^\r\wedge\d x^\s
\tr \e\, \del_\m \Big( A_\n\del_\r A_\s-\frac{i}{4} A_\n A_\r A_\s + \frac{i}{4} A_\n A_\s A_\r \Big)
\nonumber\\
&&\hskip-2.cm = -\frac{1}{24\pi^2}   \int \tr \e \,\d \Big( A \wedge \d A -\frac{i}{2} A\wedge A\wedge A\Big) \ .
\ea
The final expression as the integral of a 4-form does not involve the metric, as usual with differential forms. 
Note that in the language of differential forms the field strength is given by 
\be
F=\frac{1}{2} F_{\m\n} \d x^\m \wedge \d x^\n 
= (\del_\m A_\n- i A_\m A_\n )  \d x^\m \wedge \d x^\n
=\d A - i A^2 \ .
\ee
We can similarly rewrite the abelian anomaly \eqref{abeliananfunc} in terms of the differential forms resulting in \footnote{
Since $t$ commutes with the $A$, it follows from the cyclicity of the trace that $\tr t A\wedge A\wedge \d A=\tr t \d A \wedge A\wedge A$.  For the same reason $\tr t A\wedge A\wedge A\wedge A=0$.  
}
\be\label{abelianform}
\int \d^4 x\, \sqrt{-g}\, \e \cA=-\frac{1}{4\pi^2} \int  \e\tr  t\, F\wedge F
=-\frac{1}{4\pi^2} \int  \e \tr t\, \big(\d A \wedge \d A -2i\, \d A \wedge A\wedge A \big) \ .
\ee
It is worth noting that the chiral anomaly \eqref{an4form} is {\it not} $\sim \int\tr\e F\wedge F$.



\section{BRST cohomology and descent equations}
\setcounter{equation}{0}

\subsection{BRST transformations}

We have discussed the anomalies under gauge transformations in terms of the effective action $\wt\G[A]$ obtained by doing the functional integral over the matter fields  of $e^{iS_{\rm matter}[\p,\pb,A]}$, cf \eqref{Gammaeff}. Since the gauge anomalies originated in the anomalous transformation of the functional integral measure for the chiral matter fermions, this was indeed the appropriate object to consider for studying the anomalies. However, in the full quantum theory one needs to go on and add the classical Yang-Mills action \footnote{
Here we assume that the Lie algebra is compact, i.e.~that $\tr t_\a t_\b$ is a positive definite bilinear form. One can then adopt a basis where $\trR t_\a t_\b=C_\cR\, \dd_{ab}$ with the constant $C_\cR$ depending on the representation. One can then use the ``metric" $\dd_{\a\b}$ and its inverse $\dd^{\a\b}$ to raise and lower the indices $\a, \b, \g, \ldots$ which means that we can write them without distinction as upper or lower indices. Also, the structure constants $C_{\a\b\g}$ then are totally antisymmetric - as we already implicitly assumed earlier in these notes.
}
 $\int (-\frac{1}{4} F_{\m\n}^\a F^{\a\m\n})$ and also do the functional integral over the gauge fields, possibly with the relevant operator insertions, as described in Sect.~2. As is well known, one then needs to fix the gauge by some appropriate gauge-fixing action and compensate for the arbitrariness of this gauge fixing by inserting the Faddeev-Popov determinant. The latter can be rewritten in terms of  a Gaussian functional integral over a so-called ghost and anti-ghost fields. This results in the full action being
\be
S_{\rm full}=\int\d^4 x\left( \underbrace{\cL_{\rm matter}[\p,\pb,A] - \frac{1}{4} F_{\m\n}^\a F^{\a\m\n}}_{\cL_{\rm class}}
+\underbrace{\frac{\xi}{2} h_\a h_\a + h_\a f^\a[A]}_{\cL_{\rm gauge\, fixing} }+ \underbrace{\o_\a^* \r^\a[A,\o]}_{\cL_{\rm ghost}} \right)
\ee
The first two terms are referred to as the classical Lagrangian.
$f^\a[A]$ is the gauge-fixing condition, e.g.~$f^\a=\del_\m A^{\a\m}$, and $h_\a$ is an (adjoint) auxiliary field. Since it appears quadratically, it can be trivially integrated out, resulting in an equivalent gauge-fixing Lagrangian  $\frac{1}{2\xi} f^\a f^\a$. The parameter $\xi$ is positive but otherwise arbitrary. The quantity $\r_\a$ is given by
\be
\r^\a(x)= \int\d^4 y \frac{\dd f^\a[A(x)]}{\dd \e^\b(y)} \o^\b(y) \ ,
\ee
and $\o^\a$ and $\o_\a^*$ are the anti-commuting ghost and anti-ghost fields. Integrating them out would give back the Faddeev-Popov determinant $\Det \frac{\dd f^\a[A(x)]}{\dd \e^\b(y)}$. Note that the quantity $\r_\a$ can be interpreted as the gauge transformed gauge-fixing function $f^\a$ with the parameter $\e^\a$ replaced by the ghost field $\o^\a$. Despite its non-local appearance, for the standard choices of gauge-fixing functions $f^\a$, the quantity $\r^\a(x)$ only depends on $A^\b_\m(x)$ and $\o^\b(x)$ and their derivatives.

Of course, $S_{\rm full}$ is the gauge-fixed action and no longer is gauge invariant. But it is still invariant under some ``remnant" of the gauge invariance which is  invariance under BRST transformations. These BRST transformations  are denoted by $s$ and defined to act on the gauge and matter fields as ordinary gauge transformations but with $\o^\a$ replacing $\e^\a$. Then the gauge-invariance of the classical Lagrangian implies its BRST invariance.
Defining $\o=\o^\a t_\a$, $\o^*=\o^*_\a t_\a$ as well as $h=h_\a t_\a$, the full set of BRST transformations is defined as
\ba\label{BRST1}
s A_\m= \del_\m \o -i [A_\m,\o] \quad &, \quad s\p=i\o \p \quad ,& \quad s \pb = i \pb \o \nonumber\\
s\o = i \o \o  \quad \quad &, \quad  s\o^*=-h  \quad ,& \quad s h = 0 \ ,
\ea
together with the fact that $s(F G)=(sF) G \pm F sG$ with a minus sign if $F$ is anti-commuting. Note for later use that this implies $s\o^\a=-\frac{1}{2}C_{\a\b\g}\o^\b\o^\g$.
The action of $s$ on $\o, \ \o^*$ and $h$ is chosen such that this BRST transformation $s$ is nilpotent, i.e.~when $s$ acts twice on any of the fields one gets zero~:
\be\label{nilpotency}
s^2=0 \ .
\ee
It is now easy to see that also the remaining part of the Lagrangian is BRST invariant. First note that, by its definition, one has $\r^\a = s f^\a$, so that by the nilpotency of $s$ one has $s \r^\a=0$. Then
\be\label{sLgfLgh}
s\left(\cL_{\rm gauge\, fixing}+\cL_{\rm ghost}\right)
=h_\a (s f^\a) + (s \o^*_\a) \r^\a - \o_\a^* (s \r^\a) = h_\a \r^\a - h_\a \r^\a=0 \ .
\ee
One can actually write
\be
\cL_{\rm gauge\, fixing}+\cL_{\rm ghost}= s\left(-\o^*_\a \left(\frac{\xi}{2} h^\a +f^\a
\right)\right) 
\equiv s\, G_{-1} \ ,
\ee
and then \eqref{sLgfLgh} trivially follows from \eqref{nilpotency}.

Since the ghost fields are anti-commuting, as are the differential forms of odd degree, it makes some sense to adopt the convention that  the ghost (and anti-ghost) fields also anticommute with $\d x^\m$. Accordingly then,  one has $s\, \d = - \d\, s$. However, here we find it more straighforward {\it not} to adopt this convention, and rather have the ghost fields {\it commute} with $\d x^\m$ so that
\be
s\, \d = \d\, s\ .
\ee

\subsection{Reformulating relevant anomalies as BRST cohomologies}

Since $s^2=0$, one introduces the notion of BSRT-closed functionals $F$ obeying $sF=0$, as well as of BRST-exact functionals $H$ obeying $H=s G$ for some $G$. This is much as for the exterior derivative $\d$ acting on differential forms. The role of the degree of the forms is played here by the ghost number, with $\o$ and $\o^*$ having ghost numbers $+1$ and $-1$. Indeed, acting on any field, $s$ increases the ghost number by one unit. We can now say that $\cL_{\rm class}$ is BRST-closed and that $\cL_{\rm gauge\, fixing}+\cL_{\rm ghost}$ is  BRST-exact.
Since gauge-invariant physical observables should not depend on the gauge-fixing function $f_a[A]$, changing it amounts to changing this BRST-exact term. The physics is then determined by the BRST-closed $\cL_{\rm class}$ up to addition of BRST-exact terms., i.e.~by the BRST cohomology class at ghost number zero.

We now show that relevant anomalies can be understood as BRST-cohomology classes at ghost number one. First, since $\wt\G[A]$ only depends on $A$, its BRST variation simply is its gauge variation with $\e$ replaced by $\o$, i.e.
\be\label{sanom}
\cA[\o,A]\equiv \int\d^4 x\, \o^\a(x) \cA_\a(x) = s \wt\G[A] \ .
\ee
From now on, we refer to the left-hand side as the anomaly. It clearly has ghost number one and is BRST-closed since $s \cA[\o,A]=s^2 \wt\G[A]=0$. Actually it looks as if the anomaly $\cA[\o,A]$ is also BRST-exact. But as with the discussion of relevant anomalies, we will only allow local functionals and, although the anomaly is a local functional, $\wt\G$ is not. Hence, the anomaly is BRST-closed, but not BRST-exact. As such it constitutes a BRST-cohomology class at ghost number one. 

We have seen before that the Wess-Zumino consistency condition simply was a consequence of the anomaly being defined as the gauge variation of some (non-local) functional. At present this condition is a simple consequence of $s^2=0$. Indeed, applying $s$ to \eqref{sanom},  we get
\ba\label{WZs}
0 &=&s^2 \,\wt\G[A]= s \int\d^4 x\, \o^\a(x) \cA_\a(x) = \int\d^4 x \Big((s\o^\a(x)) \cA_\a(x) - \o^\a(x) s \cA_\a(x)\Big)
\nonumber\\
&=&-\frac{1}{2}\int\d^4 x\, C_{\a\b\g}\o^\b(x)\o^\g(x) \cA_\a(x)- \int\d^4 x\, \d^4 y\, \o^\a(x) \o^\b(y) \cF_\b(y) \cA_a(x) \ ,
\ea
where we used the functional differential operator $\cF$ introduced in \eqref{cfop} to express the BRST variation of $\cA_\a$. Indeed, $\dd_\e \cA_\a(x)=\int\d^4 y\, \e^\b(y)\cF_\b(y) \cA_\a(x)$ translates into $s \cA(x)=\int\d^4 y\, \o^\b(y)\cF_\b(y) \cA_\a(x)$.
Since the ghost fields anti-commute, we can anti-symmetrize the last term in $(\a,x)$ and $(\b,y)$ and then \eqref{WZs} can be written as
\be\label{WZs2}
0=\int \d^4 x \d^4 y \, \o^\a(x) \o^\b(y) \Big( \dd^{(4)}(x-y)  C_{\a\b\g} \cA_\g(x) + \cF_\b(y)\cA_\a(x) -\cF_\a(x) \cA_\b(y) \Big) \ ,
\ee
which vanishes by the Wess-Zumino condition \eqref{WZcond}. Hence, the fact that the anomaly is BRST closed is equivalent to satisfying the Wess-Zumino condition.\footnote{
At first sight it looks as if the Wess-Zumino condition only implies BRST-closedness. But the vanishing of the double integral in \eqref{WZs2} for all ghost fields $\o$ implies the vanishing of the term in parenthesis. Indeed the latter has exactly the same antisymmetry properties as $\o^\a(x)\o^\b(y)$. Upon setting $\o^a=\o^a_{(1)}\pm \o^a_{(2)}$ with arbitrary (anticommuting) $\o^a_{(1,2)}$, a standard argument then implies the vanishing of the parenthesis. But this is exactly the Wess-Zumino consistency condition.
}

\subsection{Solving the WZ-condition}

The next task is to solve $s\cA[\o,A]=0$ which will determine the terms cubic in $A$ in terms of the quadratic terms.   One way to do this is by ``brute force", i.e.~one writes all possible terms compatible with the general constraints on the anomaly, i.e.~$\cA[\o,A]$ must be the integral of a 4-form since, as discussed in subsection \ref{diffformsec}, it must necessarily involve the $\e$-tensor, and it must be made from $\d$'s and $A$'s. Thus we may start with
\be\label{anomansatz}
\cA^L[\o,A]= -\frac{1}{24\pi^2} \int \tr \o \Big( \d A \d A + \b_1 \d A  A^2+ \b_2 A \d A A + \b_3 A^2 \d A +\g A^4\Big)\ ,
\ee
where the wedge product is understood, i.e.~$\d A A^2=\d A \wedge A\wedge A$, etc. We have already adjusted the coefficient of the $\d A\d A$-term as appropriate for the anomaly due to a left-handed (positive chirality) fermion. In principle one could also add double-trace terms such as $\tr\o A \tr A\d A$. But there is no way to achieve a cancellation between such double-trace terms and the single trace terms written in \eqref{anomansatz}. 

Note that since $s$ changes the scaling dimensions of all field uniformly, we can only achieve cancellation between the different terms if we start out with all terms in \eqref{anomansatz} having the same scaling dimension.
One can then compute $s \cA[\o,A]$ and impose that it vanishes. Note that one assumes that $\o=\o^a(x) t_a$ vanishes at spatial infinity as well as at $t\to\pm\infty$ so that one is allowed to freely integrate by parts.\footnote{
For topologically non-trivial space-times requiring several charts, one would need to further assume that the transition functions for the gauge-field are trivial or else that $\o^\a(x)$ is non-vanishing only inside a given chart.
}
Said differently $\int \d (\ldots)=0$. This means that $s$ of the integrand must be an exact 4-form. 
A straightforward computation then yields $\g=0$ and $\b_1=-\b_2=\b_3 =-\frac{i}{2}$, so that $ \b_1 \d A  A^2+ \b_2 A \d A A + \b_3 A^2 \d A=-\frac{i}{2}\d A^3$, and one gets exactly the expression  \eqref{an4form} we already anticipated.

The previous method to solve the Wess-Zumino consistency condition is certainly sufficient in 4 dimensions, but when one wants to study anomalies in higher dimensions $d=2n$, things quickly become painful. There is a much more elegant way to solve for the BRST-cohomology at ghost-number one on the space of local functionals that are integrals of local $d$-forms made from the gauge-field one-forms, exterior derivatives and one ghost field. It is based on the so-called descent equations which relate this $d$-form to a charcteristic class which is a (formal) $(d+2)$-form on a $(d+2)$-dimensional space-time. This higher-dimensional space can be constructed as a product manifold $M=\cM\times \cD$ with $\cM$ the space-time manifold and $\cD$ a (two-dimensional) disc. The geometry of this $(d+2)$-dimensional space-time, as well as the descent equations play quite an important role in the study of \cite{AGG} which relates the anomaly on the $d$-dimensional $\cM$ to the index of the Dirac operator $\Dsl$ on the $(d+2)$-dimensional space-time $M$, as we will briefly explain in the next section.

\subsection{Descent equations  : characteristic classes and Chern-Simons forms}

\subsubsection{Characteristic classes}

Within the present context, we define a  characteristic class $P_r$ as a local $2r$-form on a (compact) $2r$-dimensional manifold $M$ that is constructed from the gauge field strength $F$ (or curvature 2-form $R$) such that its integral over the manifold is sensitive only to non-trivial transition functions of the gauge bundle. Well-known examples are instantons and magnetic monopoles. For the (four-dimensional) instantons $P=\tr F\wedge F$ and the manifold is compactified ${\mathbb R}^4\simeq S^4$. For the monopole $P=F$ and the manifold is $S^2$. Since this is the simplest example let us exhibit the construction of the monopole bundle in some more detail.

For the \underline{(abelian) magnetic monopole} one looks at static configurations with space being ordinary space ${\mathbb R}^3$ with a little ball of radius $r_0$ at the origin removed, so that one deals with $[r_0,\infty)\times S^2$. Since the gauge field strength 2-form here does not depend on the radial coordinate we may view it as defined on the sphere $S^2$. The integral $\int_{S^2} F$ computes the flux of the magnetic field through the sphere and, by Gauss' theorem, this measures the total magnetic charge ``inside" the sphere. Obviously, if $F=\d A$ with a globally well-defined gauge-field one-form $A$, then by Stoke's theorem one has $\int_{S^2} F=\int_{S^2} \d A=\int_{\del S^2} A=0$ since the boundary $\del S^2$ of the sphere is empty, and there is no magnetic charge inside. If instead $F=g_m \sin\theta \d\theta \d\vf$, then $\int_{S^2} F = 4\pi g_m$. Since this is independent of the radius of the sphere, we interpret $g_m$ as the magnetic charge of a monopole sitting at the origin (or rather inside the little ball of radius $r_0$). Since $F$ is closed, it must be locally exact, i.e.~given locally by $F=\d A$, but $A$ can no longer be globally defined. One needs to introduce two (topologically trivial) charts which are the upper half and lower half spheres $S_\pm^2$ (or the product of these half-spheres with $[r_0,\infty)$), and then
\be
A_+=g_m (1-\cos\theta) \d\vf \ {\rm on}\ S^2_+ 
\quad , \quad
A_-=g_m (-1-\cos\theta) \d\vf \ {\rm on}\ S^2_- \ \  .
\ee
Each $A_\pm$ is well-defined and non-singular on its domain of definition, but would become ill-defined when one tried to extend it to the full sphere.\footnote{
The point is that $\d\vf$ is well-defined, except at the north pole ($\theta=0$) and at the south pole ($\theta=\pi$). Now, in $A_+$ the factor $1-\cos\theta$ vanishes at the north pole, and in $A_-$ the factor $-1-\cos\theta$ vanishes at the south pole. So, $A_+$ would be singular at the south pole and $A_-$ at the north pole. If one tries to just use a single gauge potential, 
its singularity corresponds to the ``Dirac string" in the earlier descriptions of the Dirac monopole.
}
Both $A_\pm$ yield the same $F$ as $F=d A_\pm$ and are related by a U(1) gauge transformation on the overlap $S^2_+\cap S^2_-=S^1$~:
\be
A_+ - A_-=2 g_m \d\vf =i g_{+-}^{-1} \d g_{+-}\quad , \quad g_{+-}=\exp(-2ig_m \vf) \ .
\ee
For $g_{+-}$ to be well-defined on $S^1$ one must have $g_m=\frac{N}{2}$ with integer $N$, so that $g_{+-}=\exp(-iN \vf)$. We see that the magnetic charge is quantized, and equivalently 
\be\label{Chern1}
\frac{1}{2\pi} \int_{S^2} F = N \ .
\ee
One can change $F$ by adding an exact term $F\to  F+\d \wt A$ (with a globally well-defined $\wt A$) and this does not change the integral of $F$ over the manifold $S^2$. The value off this integral only depends on the transition function $g_{+-}$, i.e.~on $N$. Thus $F$ is the characteristic class $P_1$ for the U(1) gauge field strength on $S^2$. It is instructive to compute $\int F$  again, in a way which clearly shows why the value of the integral is only sensitive to the transition function $g_{+-}$~
\ba\label{Chern1-2}
\int_{S^2}  F &=& \int_{S^2_+}  F \ +\int_{S^2_-}  F = \int_{S^2_+}  \d A_++\int_{S^2_-}  \d A_-
=  \int_{\del S^2_+}  A_+ +\int_{\del S^2_-}  A_- \nonumber\\
&=&\int_{S_1} A_+ - \int_{S_1} A_-= \int_{S_1} i g_{+-}^{-1} \d g_{+-} = 2\pi N\ ,
\ea
where the minus sign in the second line is due to the fact that the boundaries $S^1$ of the upper and lower half spheres have opposite orientation.

The general case proceeds exactly along the same line of argument. For an arbitrary gauge group, define
\be\label{Prdef}
P_r(F) = \tr F^r \equiv \tr \underbrace{F\wedge \ldots \wedge F}_{r\, {\rm times}} 
=F^{\a_1}\wedge F^{\a_r}\, \tr t_{\a_1}\ldots t_{\a_r}
=F^{\a_1}\wedge F^{\a_r}\ {\rm str}\,  t_{\a_1}\ldots t_{\a_r}\ ,
\ee
since the 2-forms $F^{\a_i}$ and $F^{\a_j}$ all commute and effectively symmetrise the trace of the product of the $t_\a$. The $P_r(F)$ are gauge invariant, since under
a gauge transformation $F\to g F g^{-1}$,  and by the cyclicity of the trace all $g$ and $g^{-1}$ cancel. Hence, $P_r(F)$ has trivial transition functions and is a globally well-defined $2r$-form. Recall the Bianchi identity, $D F \equiv \d F-i [A,F]$=0, from which follows~\footnote{
Note that for Lie-algebra-valued $p$ and $q$-forms $\zeta$ and $\xi$ one has $\tr \zeta\xi=(-)^{pq} \tr\xi\zeta$.}
\be
\d P_r(F)= \d \tr F^r = r \tr (\d F) F^{r-1} = i r \tr A F^{r} - i r \tr F A F^{r-1} =0\ ,
\ee
so that $P_r(F)$ is closed, and hence locally (i.e.~on each topologically trivial chart) exact. Moreover, as we now show, if $A_1$ and $A_2$ have the same transition functions, then $P_r(F_1)-P_r(F_2)= \d Q$ with some {\it globally well-defined} $2r-1$ form $Q$. Since we assume the manifold $M$ to be compact it then follows from Stoke's theorem that
\be\label{F2F1int}
\int_M P_r(F_2)=\int_M P_r(F_1) \ ,
\ee
as required for a characteristic class. We now construct the $2r-1$ from $Q$. Let $A_t=A_1 + t(A_2-A_1)$, $\ t\in [0,1]$ be a gauge field interpolating between $A_1$ and $A_2$. Clearly, $A_t$  has the same transition functions as both $A_1$ and $A_2$. Let $F_t$ be the associated field strength $F_t=\d A_t -i A_t^2$. One computes
\be
\frac{\del}{\del t} F_t= \d(A_2-A_1) -i  (A_2-A_1) A_t  - i A_t(A_2-A_1) = D_t (A_2-A_1) \ ,
\ee
where $D_t$ is the covariant derivative with  gauge field $A_t$. In particular, the Binchi identity for $F_t$ is $D_t F_t=0$. Then
\be\label{Prder}
\frac{\del}{\del t} P_r(F_t) = r \tr \frac{\del F_t}{\del t} F^{r-1} = r \tr \big( D_t (A_2-A_1) \big) F_t^{r-1} = r D_t \big(\tr (A_2-A_1) F_t^{r-1} \big)\ .
\ee
Now while $A_2$ and $A_1$ each transform inhomogeneously under gauge transformations (on the overlap of two charts), their difference transforms homogeneously, just as does $F_t$. Hence $\tr (A_2-A_1) F_t^{r-1}$ is invariant and the covariant derivative $D_t$ in the last expression in \eqref{Prder} reduces to an ordinary exterior derivative. Integrating $t$ from 0 to 1, one gets
\be\label{PF2PF1diff}
P_r(F_2)-P_r(F_1)= \d Q \quad , \quad Q= r \int_0^1 d t\, \tr (A_2-A_1) F_t^{r-1} \ ,
\ee
with a globally well defined $2r-1$ form $Q$. This proves \eqref{F2F1int}.
Note that we always wrote $\tr$, but we could have just as well written everything in terms of the symmetrized trace $\str$, cf.~the remark after \eqref{Prdef}

\subsubsection{Chern-Simons forms and descent equations}\label{CSdescent}

For reasons that will become clear in the next subsection, we now set $r=n+1$.
Since the $P_{n+1}(F)$ are closed, they must be locally exact, i.e.~within each (topologically trivial) chart  one must have $P_{n+1}(F)=\d Q_{2n+1}$. The previous construction tells us how to construct these $Q_{2n+1}$. Within each chart, there is no issue of transition function, and we can simply take $A_2=A$ and $A_1=0$, and then conclude from \eqref{PF2PF1diff} that
\be\label{CSdef}
P_{n+1}(F)=\d Q_{2n+1} \quad , \quad 
Q_{2n+1}= (n+1) \int_0^1 d t\, \tr A \big( t \d A - i t^2 A^2 \big)^{n} \ .
\ee
Of course, $P_{n+1}(F)=\d Q_{2n+1}$ defines $Q_{2n+1}$ only up to adding an exact term $\d \a_{2n}$. The $Q_{2n+1}$ defined as in \eqref{CSdef} are called the Chern-Simons forms.  Explicitly one has \footnote{
One may e.g.~want to check that indeed $\d Q_3=\tr F^2$. First note that while $F^2$ contains a term $A^4$ its trace vanishes. This is fortunate since when acting with $\d$ on $Q_3$  one cannot generate a term $\sim A^4$. It is then easy to check that $\d Q_3= \tr \big( \d A \d A - 2i \d A A^2\big)=\tr F^2$. 
}
\ba
Q_3 &=& \tr \big( A \d A -\frac{2 i}{3} A^3\big) \ , 
\nonumber\\
Q_5 &=&  \tr \big( A \d A \d A -\frac{3 i}{2} A^3 \d A  - \frac{3}{5} A^5 \big) \ .
\ea
We should stress that these Chern-Simons forms are defined on each chart in terms of the gauge field on this chart.

Unlike the $P_{n+1}(F)$, the Chern-Simons forms $Q_{2n+1}$ are {\it not} gauge invariant, but one can use the gauge invariance of the $P_{n+1}(F)$ to constrain the gauge variation of the $Q_{2n+1}$ as follows~:
\be
0=\dd_\e P_{n+1}(F) = \dd_\e \d Q_{2n+1} = \d \dd_\e Q_{2n+1} \ .
\ee
We see that the (infinitesimal) gauge variation of $Q_{2n+1}$ is closed and, hence, locally exact. We conclude that on a given chart
\be\label{Q2r-21}
\dd_\e Q_{2n+1}=\d Q_{2n}^1 \ .
\ee
The ambiguity of adding $\d \a_{2n}$ to $Q_{2n+1}$ results in an ambiguity of adding $\dd_\e \a_{2n}$ to $Q_{2n}^1$. Of course, \eqref{Q2r-21} only defined $Q_{2n}^1$ up to adding an exact term. Thus, overall,  $Q_{2n}^1$  is only defined up to adding $\dd_\e \a_{2n} + \d \b_{2n-1}^1$. Altogether one has the following {\it descent equations} relating the $2n+2$ form $P_{n+1}$ to the $2n+1$ form $Q_{2n+1}$ and to the $2n$ form $Q_{2n}^1$ (locally on each chart) as~:
\be\label{descent}
\d P_{n+1}=\dd_\e P_{n+1}=0 \ \Rightarrow\ P_{n+1}=\d Q_{2n+1} \ , \ \dd_\e Q_{2n+1} =\d Q_{2n}^1
\ , \  Q_{2n}^1\simeq Q_{2n}^1 + \dd_\e \a_{2n} + \d \b_{2n-1}^1 \ .
\ee
By replacing the gauge variation parameter $\e$ by the ghost field $\o$ one immediately gets the BRST version of the descent equations~:
\be\label{BRSTdescent}
\d P_{n+1}=s P_{n+1}=0 \ \Rightarrow\ P_{n+1}=\d Q_{2n+1} \ , \  s Q_{2n+1} =\d Q_{2n}^1
\ , \  Q_{2n}^1\simeq Q_{2n}^1 + s \a_{2n} + \d \b_{2n-1}^1 \ .
\ee
The upper index now indicates the ghost number, i.e. $Q_{2n}^1$ contains one ghost field. Explicitly one has for $n=1$ and $n=2$
\be
Q_2^1 = \tr \o \d A \quad , \quad Q_4^1=\tr \o\, \d\big( A \d A -\frac{i}{2} A^3 \big) \ .
\ee
Again, we could have used the symmetrized trace throughout. Indeed, one can explicitly show that $Q_2^1$ and $Q_4^1$ as given here are equal to the same expressions but using the symmetized trace.\footnote{
Of course, for $Q_2^1$ symmetrizing the trace follows from cyclicity. For $Q_4^1$ the first term $\tr\o\d A\d A$ is symmetric under exchange of the two factors $\d A$ which again symmetrizes the trace. The second term $\sim\tr \o \d A^3$ seems to involve the trace of 4 generators, but one can easily see from the antisymmetry under exchanging two factors of $A$ that one really has the trace of two generators with the commutator of two other generators, so that one again has a trace of only 3 generators. Actually it is not difficult to show that this trace of 3 generators also is a symmetrized trace, as one expected since the whole $Q_4^1$ is obtained through descent from $P_3(F)$ which involved a symmetrized trace.
}

We have seen that the $2n$ form $Q_{2n}^1$ is determined, up to the equivalence just mentioned, by the characteristic class $P_{n+1}$.
The question then arises how many different characteristic classes there are. Gauge (or BRST) invariance implies that they must be constructed from traces of the field strengths $F$. Thus  the only possible characteristic classes are of the form 
\ba
&&P_{n+1}=\tr F^{n+1}\quad , \quad 
P_{n+1}^{(k,n+1-k)}=\tr F^k \wedge \tr F^{n+1-k} \ , \nonumber\\
&&P_{n+1}^{(k_1,k_2,n+1-k_1-k_2)}=\tr F^{k_1} \wedge \tr F^{k_2} \wedge \tr F^{n+1-k_1-k_2} \ , \ldots \ .
\ea
It is instructive to look at an example. Let $n=4$ and consider the 10-form characteristic class $P_5^{(2,3)}=\tr F^2 \wedge\tr F^3$. Then $P_5^{(2,3)}= \d Q_9^{(2,3)}$ with $Q_9^{(2,3)}=Q_3 \wedge \tr F^3$ or $Q_9^{(2,3)}=\tr F^2 \wedge Q_5$,  and  correspondingly $Q_8^{1,(2,3)}=Q_2^1 \wedge \tr F^3$  or $Q_8^{1,(2,3)}= \tr F^2 \wedge Q_4^1$. One says that one is doing the descent in the first or in the second factor. Obviously, any linear combination of the two with total coefficient one is also possible. As argued above, the difference of two solutions of the descent equations must be the sum of a BRST-exact and a $d$-exact term. Indeed, it is easy to see that the difference is given by 
\ba
&&s\big( Q_3 \wedge Q_5\big) - \d\big( Q_2^1\wedge Q_5 - Q_3 \wedge Q_4^1\big) \nonumber\\
&&=(sQ_3)\wedge Q_5 + Q_3\wedge (s Q_5) - ( \d Q_2^1)\wedge Q_5 - Q_2^1 \wedge \d Q_5 + (\d Q_3)\wedge Q_4^1 - Q_3 \wedge \d Q_4^1 \nonumber\\
&&=- Q_2^1 \wedge \d Q_5 + (\d Q_3)\wedge Q_4^1 = - Q_2^1 \wedge \tr F^2 + \tr F^3 \wedge Q_4^1 \ .
\ea

\subsection{BRST cohomology at ghost number one}

We will now consider a $d=2n$ dimensional space-time. We will show that starting with the characteristic class $P_{n+1}$ defined on the $d+2$ dimensional manifold $M=\cM\times \cD$,  the quantity $\int Q_d^1$   is a non-trivial representative of the BRST cohomology at ghost number one, where   $Q_{d}^1\equiv Q^1_{2n}$ is obtained by the descent equations \eqref{BRSTdescent}  from $P_{n+1}$. As such,  it provides a solution of the Wess-Zumino consistency condition. This allows us to identify the anomaly as
\be
\cA[\o,A]=c\,\int Q_d^1 \ ,
\ee
where the only freedom resides in the coefficient $c$. 
While we have seen that the characteristic classes in $2n+2$ dimensions can be the single-trace $P_{n+1}$ or the multiple-trace $P_{n+1}^{k_1, \ldots ,n+1-k_1-\ldots}$, only single traces can originate from the one-loop diagrams and we must keep with the $P_{n+1}$ only.\footnote{
This is true for the chiral gauge anomalies. For the gravitational anomalies to be discussed below, things are more complicated and the relevant characteristic classes are the Pontryagin classes that combine single trace and multiple trace classes in a well-defined way,  as fixed by the index theorems.
} 
A single computation of an $n+1$ point one-loop diagram is needed to determine this coefficient $c$. Moreover, as we have seen, one does not need to compute the full diagram, only the part involving the $\e$-tensor.

Let us first note that the equivalence $Q_d^1\equiv Q_d^1+s \a_d + \d \b_{d-1}^1$ is exactly what we expect for the anomaly. Indeed, the $d$-exact term disappears upon integrating and the $s$-exact term amounts to adding a local countertem $s\int \a_d$ to the anomaly, which we can always do, as discussed above. Hence, up to adding local counterterms, a relevant anomaly is  determined by a non-vanishing coefficient $c$. This in turn is corresponds to exactly one characteristic  class in $d+2$ dimensions.  

Let us now prove that $\int Q_d^1\equiv \int Q^1_{2n}$ is a non-trivial representative of the BRST cohomology at ghost number one. Let us recall that at ghost number zero the BRST-cohomology consists of the gauge-invariant functionals of the gauge field only.\footnote{
One could also add gauge invariant functionals of the matter fields, but this is irrelevant here. The point is that  any term that involves the $h$ field or an equal number of ghost and anti-ghost fields  necessarily is BRST-exact.
}
Furthermore, there are no gauge-invariant forms of odd degree, and then for these forms the BRST-cohomology at ghost number zero is empty. Lastly, still at ghost number zero, a gauge invariant even form of the gauge field strengths like the $P_{n+1}$ is a non-trivial representative of the BRST-cohomology. Indeed, the $P_{n+1}$ are BRST-invariant, and cannot be $s \g^{-1}_{2n+2}$ since this $ \g^{-1}_{2n+2}$ must contain at least one anti-ghost field $\o^*$ and $s\o^*\sim h$ which is not present in $P_{n+1}$.

Now the descent equation $s Q_{2n+1}=\d Q_{2n}^1$ and $s^2=0$ immediately give
\be
0=s(s Q_{2n+1})=s(\d Q_{2n}^1)= \d (s Q_{2n}^1) \ ,
\ee
so that $s Q_{2n}^1$ is $\d$-closed on the auxiliary $d+2$-dimensional space-time $M$. Now, even though the $d$-dimensional space-time may be topologically non-trivial and require several charts, as discussed before, we may assume that the ghost field is non-vanishing only within a single chart so that $Q_{2n}^1$, as well as $s Q_{2n}^1$ are non-vanishing only within a single chart, and being $d$-closed then is $d$-exact. We conclude
that $s Q_{2n}^1=\d \b^2_{2n-1}$ and then integrating this $d=2n$ form over the $d$-dimensional space-time $\cM$ one gets zero~:
\be
s\int_{\cM} Q_{2n}^1 = \int_{\cM} \d \b^2_{2n-1} = 0 \ ,
\ee
and we have shown that $\int_{\cM} Q_{2n}^1$ is BRST-closed, meaning that it satisfies the Wess-Zumino consistency condition.

We must still show that $\int_{\cM} Q_{2n}^1$ is not BRST-exact. We begin by showing that $Q_{2n}^1$ is not BRST-exact. Suppose on the contrary that $Q_{2n}^1= s \a_{2n}$. Then one would have $sQ_{2n+1}=\d Q_{2n}^1 = \d s \a_{2n}= s \d \a_{2n}$ and $s(Q_{2n+1}-\d\a_{2n})=0$, i.e.~$Q_{2n+1}-\d\a_{2n}$ is BRST-closed. Being a form of odd degree and ghost number 0 (where the BRST cohomology is empty), we must have $Q_{2n+1}-\d\a_{2n}= s \g^{-1}_{2n+1}$. But by the same argument as above, there is no such $\g^{-1}_{2n+1}$, and $Q_{2n+1}-\d\a_{2n}=0$. This would imply $\d Q_{2n+1}=0$ which contradicts $\d Q_{2n+1}=P_{n+1}$. Hence, $Q_{2n}^1$ is not BRST-exact. We must still show that $\int_\cM Q_{2n}^1$ is not BRST-exact. Suppose on the contrary that $\int_\cM Q_{2n}^1= s \int \b_{2n} = \int s \b_{2n}$. This implies that $Q_{2n}^1-s\b_{2n}=\d \g^1_{2n-1}$ for some $\g^1_{2n-1}$. But $Q_{2n}^1$ was only defined modulo adding such an exact term $\d \g^1_{2n-1}$ and then this redefined $Q_{2n}^1$ would be BRST-exact which  cannot be as we have just shown. Then  $\int_{\cM} Q_{2n}^1$ cannot be BRST-exact, and is a non-trivial representative of the BRST-cohomology at ghost number one, which we wanted to show.

To conclude, for each characteristic class $P_{n+1}$ there is a unique BRST-cohomology class at ghost number one, represented by $\int_{\cM} Q_{2n}^1$, and up to a multiplicative coefficient $c$, this is the anomaly in $d=2n$ dimensions. Conversely, the anomaly is unambiguously characterized by $c P_{n+1}$. 

\subsection{Gauge anomaly in $d=2n$ dimensions}

After all these formal developments let us discuss some more physical details. We are looking at the chiral anomaly, i.e.~an anomaly under gauge transformations in the presence of chiral fermions, much as before, but now in an arbitrary even dimension $d=2n$. As discussed in the previous section, this anomaly first manifests itself in the non-conservation an $n+1$-point one-loop function $ \G^{\m \n_1\ldots \n_n}_{\a_1\ldots \a_{n+1}}$. One checks for conservation at the vertex with index $\m$. This involves $n!$ individual diagrams corresponding to the $n!$ permutations of the indices of the $n$ other vertices. Each diagram gives rise to a trace ${\rm tr}_\cR\, t_{\a_1} \ldots t_{\a_{n+1}}$, and adding all other permutations (together with the cyclicity of the trace) results in a totally symmetrized trace which is the corresponding $D$-symbol
\be
D^\cR_{\a_1\ldots \a_{n+1}}={\rm str}_\cR\,  t_{\a_1} \ldots t_{\a_{n+1}} \ ,
\ee
where the symmetrized trace is taken in the representation $\cR$ of the chiral fermion.
Exactly the same symmetrized trace emerges from the characteristic class, cf.~\eqref{Prdef}
\be
P_{n+1}^\cR(F)=F^{\a_1}\wedge \ldots \wedge F^{\a_{n+1}} D^\cR_{
\a_1\ldots a_{\n+1}} \ .
\ee

More precisely, we had argued above that the anomalous contribution of the $n+1$ point one-loop diagram had to be of the form $\e^{\m_1 \ldots \m_n \n_1\ldots \n_n} p^1_{\m_1}\ldots p^n_{\m_n}\, D^\cR_{\a_1\ldots \a_{n+1}}$ which, upon multiplying with $\o^{\a_{n+1}} A^{\a_1}_{\m_1} \ldots A^{\a_n}_{\m_n}$ corresponds to a term in the anomaly proportional to
\be
{\rm str}_\cR\,  \o \underbrace{\d A \ldots \d A}_{n\, {\rm times}}\ .
\ee
But this is exactly the first term of $Q_{2n}^1$ as obtained by descent from $P^\cR_{n+1}={\rm str}_\cR\,  F^{n+1}$. More precisely, we should write $Q_{2n}^{1,\cR}$ to indicate in which representation the trace has to be taken. We then know from the previous discussion about solving the Wess-Zumino consistency condition and the BRST-cohomology at ghost number one, that the gauge anomaly in $d=2n$ dimensions is
\be\label{2nanomaly}
\cA[\o,A]=c_{2n} \int \Big(\sum_j  Q_{2n}^{1,\cR^L_j} -\sum_i  Q_{2n}^{1,\cR^R_i} \Big)\ ,
\ee
where the first sum is over the different left-handed (positive chirality)  fermions in representations $\cR_i^L$ and the second sum  over the different right-handed (negative chirality) fermions in representations $\cR_i^R$. The coefficient $c_{2n}$ can be determined by a careful generalization of our triangle diagram computation. One finds
\be\label{c2n}
c_{2n}=\frac{(-)^{n+1}}{(n+1)! (2\pi)^n} \ ,
\ee
which for $n=2$, i.e.~$d=4$ gives $c_4=-\frac{1}{24\pi^2}$, in agreement with the coefficient in \eqref{an4form}.

Above, we have discussed the issue of anomaly cancellation and the fact that this cancellation only involved the $D$-symbols  for the different representations of the different chiral classes. We now see that the issue of anomaly cancellation can be equivalently discussed at the level of the characteristic classes computed in the  representations of the different chiral fermions.

\section{Gauge and gravitational anomalies and index theorems}
\setcounter{equation}{0}

In this last section, we want to briefly touch upon some further developments. Firstly, we will mention - but not prove - the relation between anomalies in $d=2n$ dimensions and the index of an appropriate Dirac operator in $d+2$ dimensions. Then we will briefly show how our discussion of gauge anomalies can be straightforwardly generalized to  gravitational anomalies. Lastly, we discuss a prominent example of anomaly cancellation in a 10-dimensional supergravity theory.

\subsection{Anomalies and index theorems}

We have already seen in subsection \ref{abelianindextheorem} that the abelian anomaly in four dimensions was closely related to the index of the four-dimensional Dirac operator, and both were related in turn to the characteristic class $P_2(F)=\trR F\wedge F=F^\a\wedge F^\b\,  \trR t_\a t_\b=c_\cR\, F^\a\wedge F^\a$. This is very different  from what we observed for the chiral anomaly under non-abelian gauge transformations where the anomaly is given by $Q_4^1$ which is related by descent to the characteristic class $P_3(F)=\tr F^3=F^\a\wedge F^\b\wedge F^\g\, D^\cR_{\a\b\g}$ and which instead involves the $D$-symbol. There is, however, an index theorem in each even dimension, and the six-dimensional index theorem states that the index of an appropriate 6-dimensional Dirac operator equals some constant times $P_3(F)$. It is then tempting to speculate whether the chiral anomaly in 4 dimensions can be directly understood in terms of the index of some 6-dimensional extension of the 4-dimensional Dirac operator. More generally, the index theorem in $2n+2$ (euclidean) dimensions states
\be\label{generalindex}
{\rm ind}\, \big(i\Dsl_{2n+2}(A)\big) = \frac{c_{2n}}{2\pi} \int P_{n+1}(F) \ ,
\ee
with the same coefficient  $c_{2n}$ as given in \eqref{c2n}. The  overall sign depends on the precise convention used to define the chirality $\g$-matrix. One sees that the anomaly as given in \eqref{2nanomaly} is precisely obtained, up to a factor of $2\pi$, by descent from the index of the $2n+2$ dimensional Dirac operator. 

While this is a very suggestive relation, it was the achievement of Alvarez-Gaum\'e and Ginsparg \cite{AGG} to precisely construct a $2n+2$ dimensional gauge-field and Dirac operator and show that the anomaly is indeed given by descent from the corresponding index. Here we will not spell out the details of their reasoning but only give a brief sketch.
\def\t{\theta}

The starting point is the observation that in euclidean signature the anomaly resides in the imaginary part of the euclidean effective action $\G_E[A]$, i.e.~in the phase of $e^{-\G_E[A]}$. It is indeed the ambiguity in defining the phase of the ill-defined determinant $\Det \big(i \Dsl(A) P_L\big)$ which can be viewed as being at the origin of the anomaly. More precisely, one looks at the determinant of $\wh D(A)=\Dsl(A) P_L+\dsl P_R$. 
Note that the euclidean (effective) action $\G_E$ and its continuation to Minkowski signature $\G_M$ differ by a factor of $i$ as
\be\label{MEcont}
\G_E = -i \,\G_M \ .
\ee
It is not difficult to see that the only terms in a Minkowski signature action that will become imaginary upon continuing to Euclidean signature are the terms involving the $\e$-tensor.
This is very consistent with the fact that the anomaly resides in the imaginary part of the Euclidean effective action~!
Indeed, the terms involving the $\e$-tensor are integrals of $d$-forms over the $d$-dimensional manifold, $\int \d x^0\wedge\ldots \d x^{d-1} \xi_{0\ldots (d-1)}$. Upon the continuation, both $\d x^0$ and $\xi_{0\ldots (d-1)}$ get a factor of $\pm i$, so overall this term remains real. But it also gets an $i$ from  the relation \eqref{MEcont} and thus results in an imaginary contributiuon to the Euclidean action. On the other hand, in a term like $\int \d x^0 \d^{d-1}x \sqrt{-g}\, \zeta_{\m_1\ldots \m_r}\zeta^{\m_1\ldots \m_r}$, the integrand remains real, but one gets an $i$ from continuing $\d x^0$. So this term picks up an $i$, which together with the $i$ in \eqref{MEcont} results in a real contribution to the Euclidean action.

If the phase of the determinant can be unambiguously defined in a gauge-invariant way, then there is no anomaly. If not, an anomaly is present. Then \cite{AGG} introduces a family of gauge transformations $g(x,\t)$ depending on some parameter $\t\in [0,2\pi]$ such that $g(x,0)=g(x,2\pi)={\bf 1}$ and, denoting 
\be\label{Atheta}
A^\t(x,\t) = g(x,\t)^{-1} A(x) g(x,\t)- i  g(x,\t)^{-1} \d g(x,\t)
\ee
 the corresponding gauge transformed field,  they study how the phase of $e^{-\G_E[A^\t]}$ changes as $\t$ goes from $0$ to $2\pi$. Since for both $\t=0$ and $\t=2\pi$ one has the original gauge field, this phase change must be $2\pi m$ with $m$ being an integer. This integer $m$ then is related to the anomaly as \footnote{
We have $A_\m^\t=g(\t)^{-1} A_\m g(\t) - i g(\t)^{-1} \del_\m g(\t)$ and under an infinitesimal variation of $\t$ one has $\frac{\dd A^\t}{\dd \t} =\del_\m \frac{\del\e}{\del\t} - i [A_\t,\frac{\del\e}{\del\t}]=D^\t_\m \frac{\del\e}{\del\t}= D^\t_\m \wh\e$ where $\wh\e=\frac{\del\e}{\del\t}=-i g(\t)^{-1} \frac{\del g(\t)}{\del\t}$.
}
($\wh\e=-i g(\t)^{-1} \frac{\del g(\t)}{\del\t}$)
\ba
m&=&-\frac{1}{2\pi} \int_0^{2\pi} \d\t \, \frac{\del {\rm Im}\G_E[A^\t]}{\del\t}
=-\frac{1}{2\pi i} \int_0^{2\pi} \d\t \, \frac{\del\G_E[A^\t]}{\del\t}\nonumber\\
&=&
-\frac{1}{2\pi i} \int_0^{2\pi} \hskip-3.mm \d\t \int_\cM \d^{2n} x \sqrt{\det{g}}\, (D^\t_\m \wh\e(x,\t) )^a \frac{\dd \G_E[A^\t]}{\dd A^{\t,a}_\m(x)} \nonumber\\
&=& -\frac{1}{2\pi i} \int_0^{2\pi}  \hskip-3.mm \d\t \, \dd_{\wh \e} \G_E[A^\t]
= \frac{1}{2\pi} \int_0^{2\pi}  \hskip-3.mm \d\t \, \dd_{\wh\e} \G_M[A^\t]\ .
\ea
The next step is to relate the integer $m$ to the index of an appropriate Dirac operator. To this end, \cite{AGG} extends $A^\t$ to a two-parameter family of gauge fields $A^{\r,\t}(x)$ with $(\r,\t)$ parametrizing a disc ($\r\le 1$) and $A^{\r=1,\t}(x)=A^\t(x)$. The integer $m$ is then interpreted as some sort of winding number around the boundary of the disc. Then, in \cite{AGG} it is shown that this winding number is the sum of local winding numbers around points in the interior of the disc where $\Det(i \wh D(A^{\r,\t}))$ vanishes. Of course this happens whenever $\wh D(A^{\r,\t})$ has a zero-mode. The non-trivial point is that the local winding numbers are $\pm1$ according to the chirality of the zero-modes of a full $2n+2$ dimensional Dirac operator $i\Dsl_{2n+2}$, and one sees that this computes precisely the index of this operator. One concludes that
\be\label{indexanom}
{\rm ind}\, \big(i\Dsl_{2n+2}(A^{\r,\t})\big)=m= \frac{1}{2\pi} \int_0^{2\pi}  \hskip-3.mm \d\t \, \dd_{\wh\e} \G_M[A^\t]\ .
\ee
One still needs to specify how $A^{\r,\t}$ is determined, and how it is promoted to a true $2n+2$ form, and what exactly is the $2n+2$ dimensional manifold. This is done such that the manifold is $S^2 \times \cM$, with $S^2=S^2_+ \cup S^2_-$, and the previous disc is identified with $S^2_+$. Then we have $A^{\r,\t}=A_+$ on $S^2_+ \times \cM$, while on $S^2_- \times \cM$ the gauge field $A^{\r,\t}=A_-$ is not extended (i.e.~is missing the $\d\r$ and $\d\t$-pieces). Specifically, one lets
\be\label{Apm}
A_+(x,\r,\t)= f(\r) \Big( A^\t (x,\t) -i g(x,\t)^{-1} \del_\t  g(x,\t) \, \d\t \Big) \quad , \quad
 A_-(x)=A(x) \ ,
\ee
with $A^\t$ defined in \eqref{Atheta} and
 $f(\r)$ is a smooth function which is $f(\r)\sim\r$ for small $\r$ and $f(\r)=1$ for $\r$ close to 1 which corresponds to the fattened boundary of the disk.  Then, on the overlap of  $S^2_+ \times \cM$ with $S^2_- \times \cM$, the  $A_+$ and $A_-$ are related by the $d+1$-dimensional gauge transformation $g(x,\t)$. The ``true" space manifold is taken to be compactified ${\mathbb R}^{2n}\simeq S^{2n}$. Since $A_-$ is missing the $\d\t$-piece, we have $Q_{2n+1}(A_-)=0$. Then by the index theorem \footnote{
On a general curved manifold $M$ the index theorem involves an extra factor $\wh A(M)$ 
to be discussed below, but $\wh A(S^2\times S^{2n})=1$, so this is irrelevant here.
}
\ba
{\rm ind}\, \big(i\Dsl_{2n+2}(A^{\r,\t})\big)
&=&\frac{c_{2n}}{2\pi} \int _{S^2\times S^{2n}} \tr F^{n+1}
=\frac{c_{2n}}{2\pi} \left(\int _{S^2_+\times S^{2n}} \tr F^{n+1}+\int _{S^2_-\times S^{2n}} \tr F^{n+1}\right) \nonumber\\
&=&\frac{c_{2n}}{2\pi} \int _{S^2_+\times S^{2n}} \d Q_{2n+1} (A_+)
=\frac{c_{2n}}{2\pi} \int _{S^1\times S^{2n}} Q_{2n+1} (A_+) \ .
\ea
It is then shown \cite{AGG} that the last integral equals
\be
{\rm ind}\, \big(i\Dsl_{2n+2}(A^{\r,\t})\big)=\frac{c_{2n}}{2\pi}  \int_0^{2\pi}\hskip-2.mm \d\t \int_{S^{2n}} Q_{2n}^1(\wh\e,A^\t) \ .
\ee
Comparing with \eqref{indexanom} we see that
\be
\int_0^{2\pi}  \hskip-3.mm \d\t \, \dd_{\wh\e} \G_M[A^\t]=c_{2n} \int_0^{2\pi}\hskip-2.mm \d\t \int_{S^{2n}} Q_{2n}^1(\wh\e,A^\t)  \ .
\ee
The $\t$ dependence of $\wh\e(x,\t)$ is fairly arbitrary, and one concludes on the equality of the ``integrand" at any  $\t$ and any $\wh\e$. We may then fix some $\t$ and simply rename $A^\t(x)\to A(x)$ and $\wh\e(x,\t)\to \e(x)$. The final result  gives the anomalous variation of the Minkowskian effective action for a positive chirality (left-handed) chiral spin-$\frac{1}{2}$ fermion as
\be
\cA[\e,A]= \dd_\e \G_M[A^\t]=c_{2n} \int_{S^{2n}} Q_{2n}^1(\e,A)  \ ,
\ee
in perfect agreement with the above result \eqref{2nanomaly}.

In view of the generalization to include also gravitational anomalies to be discussed next, it is useful to formalize the relation just established as follows~:
\be\label{genindexdescent}
{\rm ind}\, \big(i\Dsl_{2n+2}(A^{\r,\t})\big)=\frac{1}{2\pi} \int I_{2n+2}
\quad , \quad 
I_{2n+2}= \d\, J_{2n+1} \quad , \quad \dd J_{2n+1}=\d\, K^1_{2n}
\quad , \quad \cA[\e,A]=\int_\cM K^1_{2n}\ ,
\ee
where for the pure chiral gauge anomalies $I_{2n+2}=c_{2n} \tr F^{n+1}$, $J_{2n+1} =c_{2n} Q_{2n+1}$ and $K_{2n}^1=c_{2n} Q_{2n}^1$. 

\subsection{Gravitational anomalies}

Theories involving gravity bear some similarities with non-abelian gauge theories, but are rather more complicated.  With the basic dynamical field being the space-time metric $g_{\m\n}$, the role of gauge-invariance is played by diffeomorphism invariance. Under an infinitesimal diffeomorphism $x^\m\to x^\m + \e^\m(x^\r)$ the metric transforms as $\dd g_{\m\n}=-\nabla_\m \e_\n - \nabla _\n \e_\m$, where $\nabla$ is the covariant derivative.
This diffeomorphism invariance must be ``gauge-fixed" when trying to do the gravitational functional integral. But again, one can do the functional integral in two steps, integrating over the matter fields first, resulting in an effective action $\G[g_{\m\n}]\equiv \G[g]$, with all the complications of gauge-fixing and ghosts deferred to later. The question of anomalies then is again whether this effective action $\G[g]$ is still diffeomorphism invariant~:
\be\label{Gammadiffinv}
\dd_\e \G[g] = - \int \big(\nabla_\m \e_\n + \nabla _\n \e_\m\big) \frac{\dd \G[g]}{\dd g_{\m\n}} \ .
\ee
The (matter) energy-momentum tensor is defined as 
\be\label{Temdef}
\dd_g S_{\rm matter}= \frac{1}{2}\int \sqrt{g} \, T^{\m\n} \dd g_{\m\n} 
\quad \Rightarrow \quad
T^{\m\n} = \frac{2}{\sqrt{g}} \frac{\dd S_{\rm matter}}{\dd g_{\m\n}} \ .
\ee
Since $e^{i \G[g]}=\int \cD \p\, e^{i S[g,\p]}$ where $\p$ stands collectively for all matter fields, we have, much as for the gauge currents $J_\a^\mu$ in the non-abelian gauge theories,
\be
 \frac{\dd \G[g]}{\dd g_{\m\n}(x)} = \frac{1}{2}\sqrt{g}\, \langle T^{\m\n}(x) \rangle
\ee
Inserting this into \eqref{Gammadiffinv} and integrating by parts \footnote{
Provided there are no boundary terms, covariant derivatives can be freely integrated by parts, e.g. $\int \sqrt{g}\, (\nabla_\m a_\n) b^{\m\n} = - \int \sqrt{g}\, a_\n \nabla_\m b^{\m\n}$, where, of course, $\nabla_\m a_\n=\del_\m a_\n -\G_{\m\n}^\r a_\r$ and $\nabla_\m b^{\m\n}=\del_\m b^{\m\n} +\G_{\m\r}^\m b^{\r\n}+\G_{\m\r}^\n b^{\m\r}$.
} 
results in
\be\label{Gammadiffinv2}
\cA_{\rm diff} [\e,g]\equiv \dd_\e \G[g] = \int\sqrt{g}\, \e_\n \nabla_\m \langle T^{\m\n}(x) \rangle\ ,
\ee
where we defined, much as before,  the anomaly under diffeomorphisms as the non-invariance of the effective action.
We see that diffeomorphism invariance of the matter effective action is equivalent to the covariant conservation of the expectation value of the energy-momentum tensor. Conversely, a gravitational anomaly, i.e.~an anomaly under dififfeomorphism invariance, shows up as the covariant non-conservation of this expectation value. Just as with the non-abelian gauge theories, one can take further functional derivatives with respect to the metric at different space-time points and generate, possibly anomalous, Ward identities for $m$-point functions of the energy-momentum tensor.

To couple spin-$\frac{1}{2}$ fields (or spin-$\frac{3}{2}$ fields) to gravity, one needs to introduce local orthonormal frames, or local Lorentz frames. This means that for each tangent space (at each point of the manifold) one makes a certain choice for an orthonormal basis of tangent vectors $E_a$ with components $E_a^\m$ such that $E_a^\m E_b^\n g_{\m\n}=\dd_{ab}$. (We will use  Euclidean signature here. For Minkowski signature one would of course have $E_a^\m E_b^\n g_{\m\n}=\eta_{ab}$ instead.) Equivalently, the components $e^a_\m$ of the dual forms $e^a=e^a_\m \d x^\m$ (also called the viel-bein) obey
\be
e^a_\m E^\m_b=\dd^a_b \quad , \quad e^a_\m E^\n_a=\dd^\n_\m \quad , \quad
\dd_{ab} e^a_\m e^b_\n = g_{\m\n} \ .
\ee
Obviously, at each point, there is not a unique orthonormal frame, but a whole family related by orthogonal or pseudo-orthogonal (Lorentz) transformations ${e'}^{a}=(L^{-1})^a_{\ b}e^b$. One talks about local Lorentz transformations. One imposes however, that the frames change in a smooth way as one moves along the manifold, and then the local Lorentz transformations one is allowed to do must also depend smoothly on the point. 

One can use the $E^\m_a$ and $e^a_\m$ to convert coordinate tensors like $T_\m^{\ \n}$ into frame tensors as $T_a^{\ b}=E_a^\m e^b_\n T_\m^{\ \n}$, etc. One can then show that invariance under diffeomorphisms and invariance under local Lorentz transformations are essentially equivalent. This means that, as far as anomalies are concerned, we may probe for anomalies under local Lorentz transformations \cite{BardeenZumino}. From the one-forms $e^a$ one defines the so-called spin-connection one-form $\o^{ab}=-\o^{ba}=\o^{ab}_\m \d x^\m$ by requiring the absence of torsion~:
\be
\d e^a + \o^{ab} \wedge e^b = 0 \quad , \quad \o^{ab}=-\o^{ba} \ .
\ee
Also, the curvature two-form  $R^{ab}=\frac{1}{2} R^{ab}_{\m\n} \,\d x^\m \wedge \d x^\n$, where $R^{ab}_{\m\n}=e^a_\r  e^b_\s R^{\r\s}_{\ \ \m\n}$ and $R^{\r\s}_{\ \ \m\n}$ is the Riemann curvature tensor, is obtained from the spin-connection one-form as
\be
R^{ab}=\d\o^{ab} + \o^{ac}\wedge \o^{cb} \ .
\ee
It will be useful to think of $\o$ and $R$ as matrix-valued one and two-forms and suppress the indices $a,b$, so that the previous relation simply reads $R=\d\o + \o\wedge\o\equiv \d\o + \o^2$.

This spin connection one-form and curvature 2-form behave under local Lorentz transformations $e^{'a}=(L^{-1})^a_{\ b}e^b$ very much as the gauge field one-form $A$ and gauge field strength 2-form $F$ under gauge transformations~:
\be
 \o'=L^{-1} \o L + L^{-1} \d L \quad , \quad R'=L^{-1} R L \ ,
\ee
or, for infinitesimal local Lorentz transformations $L=e^{v}$ with $v^{ab}=-v^{ba}$,
\be
\dd_v \o= \d v +[\o,v] 
\quad , \quad 
\dd_v R = [R,v]\ .
\ee
In particular, the whole discussion of characteristic classes and descent equations for the gauge theories can be transposed word by word to the gravitational case when formulated with the spin connection and curvature 2-form. While this is true, there is however an important restriction. Just as certain $D$-symbols vanish, depending on the representations, here, in $d=2n$ dimensions we are dealing with the vector representation $\cR_{\rm v}$ of $SO(2n)$. This is a real representation. Indeed,
One can choose the generators $J^{ab}=-J^{ba}$ of the vector representation to be hermitian, imaginary and antisymmetric~: 
\be\label{vectorgen}
(J^{ab}_{\rm v})_{cd}=i(\dd^a_c \dd^b_d-\dd^b_c \dd^a_d) \ .
\ee
Then the generators of the complex conjugate representation are $-(J^{ab})^T=J^{ab}$ and this is a real representation.
As discussed before, for real representations
the corresponding $D$-symbols with $n+1$  indices vanish whenever $n$ is even. Said differently, the characteristic class ${\rm tr}_{\cR_{\rm v}} \, R^{n+1}=R^{a_1 a_2}\wedge R^{a_2 a_3}\ldots \wedge R^{a_{n+1}a_1}$ vanishes for even $n$ by the antisymmetry of $R^{ab}=-R^{ba}$. 

Now, the anomaly, if any, is due to chiral fermions and these fermions are not in the vector representation but in a spin representation of $SO(2n)$ with generators $J_{\rm spin}^{ab}=\frac{i}{2} \g^{ab}=\frac{i}{4}(\g^a\g^b-\g^b\g^a)$. Then $R=\frac{1}{2}R^{ab} J^{ab}_{\rm spin}=\frac{i}{4} R^{ab}\g^{ab}=\frac{i}{4} R^{ab}\g^a\g^b$, and the characteristic class ${\rm tr}_{\rm spin} R^{n+1}$ involves the trace of $2n+2$ matrices $\g^{a_i}$. This trace  is the sum of products of $n+1$ Kronecker deltas, with  all possible pairings of the indices, times the dimension of the gamma-matrices. Those parings which contract two indices on the same $R$ drop out. For example, for $n=1$, one has
\ba
{\rm tr}_{\rm spin} \, R\wedge R
&=&-\frac{1}{16} R^{ab}\wedge R^{cd} \, {\rm tr}_{\rm spin} \, \g^a\g^b \g^c\g^d
=- \frac{1}{4} R^{ab}\wedge R^{cd} \Big(\dd^{ab}\dd^{cd}-\dd^{ac}\dd^{bd}+\dd^{ad}\dd^{bc}\Big)\nonumber\\
&=&-\frac{1}{4} \Big(-R^{ab}\wedge R^{ab} + R^{ab}\wedge R^{ba}\Big) 
=-\frac{1}{2} R^{ab}\wedge R^{ba} \ ,
\ea
which, up to the factor $-\frac{1}{2}$, is the same result as in the vector representation. This is easily seen to generalize as
\be
{\rm tr}_{\rm spin} \, R^{2k}= \wt c_k\, R^{a_1 a_2}\wedge R^{a_2 a_3}\wedge \ldots \wedge R^{a_{2k} a_1}
\quad , \quad 
{\rm tr}_{\rm spin} \, R^{2k+1}=0 \ ,
\ee
with some combinatorial factor $\wt c_k$. This being clarified we can now immediately state the gravitational descent equations~:
\be
P_{2k}(R)=\tr R^{2k} \quad , \quad P_{2k}(R)=\d Q^{\rm L}_{4k-1}(\o)
\quad , \quad \dd_v Q^{\rm L}_{4k-1}(\o)=\d Q_{4k-2}^{{\rm L},1}(v,\o) \ , \quad k=1,2,\ldots
\ee
where we added a superscript ``L" for local Lorentz. Since the only non-vanishing characteristic classes are the $P_{2k}(R)$, the $Q^{{\rm L},1}_d$ are non-vanishing only for $d=4k-2$, i.e.~$d=2, 6, 10, \ldots$. So only in these dimensions can we expect to have purely gravitational anomalies. We will also encounter characteristic classes which are (wedge) products of several traces, like e.g.$P_{2k,2l}=\tr R^{2k} \wedge \tr R^{2l}$. Then one can do the descent in one or the other factor, i.e.~$\tr R^{2k} \wedge Q^{\rm L}_{4l-1}$ or $Q^{\rm L}_{4k-1}\wedge \tr R^{2l}$. Of course, both are equally good, since their difference is an exact form, namely $\d \a_{4k+4l-2}$ with $\a_{4k+4l-2}=Q^{\rm L}_{4k-1}\wedge Q^{\rm L}_{4l-1}$. Continuing the descent, one gets either
$\tr R^{2k} \wedge Q^{{\rm L},1}_{4l-2}$ or $Q^{{\rm L},1}_{4k-2}\wedge \tr R^{2l}$. Again they only differ by a term $\d \b^{1}_{4k+4l-3} + \dd_v \g_{4k+4l-2}$, cf.~the analogous discussion at the end of section \ref{CSdescent}.

One can now continue the discussion of gravitational anomalies in two directions. On the one hand, one can compute the anomalous part of one-loop diagrams with chiral fermions coupled to gravity, i.e.~to metric perturbations $h_{\m\n}=g_{\m\n}-\dd_{\m\n}$ around flat space. Since metric perturbations couple to the energy-momentum tensor, this amounts to computing $(m+1)$-point one-loop functions with $m+1$ energy-momentum tensors at the vertices, and check for conservation, cf.~the Ward identities one could derive from \eqref{Gammadiffinv2}.
Again, one needs to saturate the $d=4k-2$-dimensional $\e$-tensor with at least $2k-1$ independent momenta and as many symmetric polarisation tensors. This means that $m\ge 2k-1$. Thus in 2 dimensions ($k=1$) we need to compute a 2-point function, in 6 dimensions 
($k=2$) a box diagram and in 10 dimensions ($k=3$) a hexagon. Of course, this is the same counting as for the gauge anomalies. These computations have been done in \cite {Alvarez-Gaume-Witten} where, among others,  the full two-point function in 2 dimensions is evaluated very explicitly, and the anomalous part of the hexagon diagram is determined - which is quite a {\it tour de force}.

On the other hand, one can rely again on the relation between the anomaly and the index theorem. The  Dirac operator  now includes a coupling to  the spin connection, as well as possibly to (non-abelian) gauge fields \footnote{
Note that $\frac{1}{4} \o^{ab}_\m \g^{ab}=-\frac{i}{2} \o^{ab}_\m J^{ab}_{\rm spin}$, so that this terms has exactly the same structure as the gauge term $-i A_\m^\a t_\a$.
}:
\be
\Dsl = \g^c E^\m_c \Big( \del_\m -i A_\m^\a t_\a +\frac{1}{4} \o^{ab}_\m \g^{ab} \Big) \ .
\ee
The relevant index theorem is a generalization of \eqref{generalindex} to curved manifolds~:
\be\label{Diracgenusindex}
{\rm ind}\, \big(i\Dsl_{2n+2}(A)\big) = \int \big[\wh A(R) {\rm ch}(i F)\big]_{2n+2} \ ,
\ee
where the Chern character ${\rm ch}(i F)$ and the Dirac genus $\wh A(R)$ are the formal sums of forms of different degrees, defined as
\be
{\rm ch}(i F)=\tr \exp\big( - \frac{1}{2\pi} F\big)
=\tr {\bf 1} -\frac{1}{2\pi} \tr F + \frac{1}{8\pi^2} \tr F^2 - \frac{1}{48 \pi^3} \tr F^3 +\ldots 
= \tr {\bf 1} + \sum_{n\ge 0} \frac{c_{2n} }{2\pi}\tr F^{n+1} \ ,
\ee
where $\tr {\bf 1} ={\rm dim}(\cR)$ is just the dimension of the fermion representation,
and
\ba
\wh A(R)&=&1 + \frac{1}{(4\pi)^2} \frac{1}{12}\tr R^2
+ \frac{1}{(4\pi)^4} \Big(\frac{1}{360}\tr R^4 +\frac{1}{288} \big( \tr R^2\big)^2\Big)\nonumber\\
&&+  \frac{1}{(4\pi)^6} \Big(\frac{1}{5670}\tr R^6 +\frac{1}{4320} \tr R^4 \tr R^2   +\frac{1}{10368} \big( \tr R^2\big)^3\Big) +\ldots \ ,
\ea
and the subscript $[\ldots]_{2n+2}$ in \eqref{Diracgenusindex} indicates that one must extract the form of degree $2n+2$.
We can then proceed exactly as formalized in \eqref{genindexdescent}, where here
\be
I_{2n+2}=2\pi\, \big[\wh A(R) {\rm ch}(i F)\big]_{2n+2} \ .
\ee
For example, in $d=2n=4$ dimensions one has
\be
I_6=2\pi \Big(  - \frac{1}{48 \pi^3} \tr F^3 + \frac{1}{(4\pi)^2} \frac{1}{12}\tr R^2\,  \big(-\frac{1}{2\pi} \tr F\big) \Big) \ .
\ee
The first term corresponds to the pure gauge anomaly extensively discussed before, while the second term corresponds to a so-called mixed gauge-gravitational anomaly. For any simple group the trace of a single generator vanishes and, hence, $\tr F$ is non-zero only for an abelian $U(1)$ gauge group where it is proportional to the sum of all $U(1)$ charges. Hence, in a non-trivial gravitational background, the vanishing of this anomaly requires this sum to vanish. This is indeed the case for the hypercharges of any family in the standard model. We also see (again) that there is no pure gravitational anomaly in 4 dimensions.

\subsection{Anomaly cancellation in ten-dimensional IIB supergravity}

Supergravity theories contain the metric / viel-bein and its superpartner, the gravitino which is a spin-$\frac{3}{2}$ field which can  be either chiral or real (Majorana), or both. In so-called extended supergravity theories there are two (or more) such gravitini, and they can have the same or different chiralites. In addition,  anti-symmetric tensor fields are present, and if they satisfy a certain self-duality constraint they are also chiral fields. Indeed such fields can be constructed from a pair of chiral spinors. Finally, there can also be chiral spin-$\frac{1}{2}$ fields, as well as various (non-chiral) scalar and vector fields.

One of the most prominent supergravity theories is the ten-dimensional IIB supergravity which is the low-energy limit of IIB superstring theory.\footnote{
The ``II" indicates that there are two (ten-dimensional) supersymmetries, and ``B" indicates that the two supercharges are spinors of the same chirality, while in IIA they would have opposite chirality. Hence IIA supergravity is non-chiral and has no anomalies. The latter can be obtained by dimensional reduction from eleven-dimensional supergravity and, indeed, in eleven dimensions one cannot have anomalies.
} 
It contains two positive chirality Majorana-Weyl gravitini, two negative chirality Majorana-Weyl spin-$\frac{1}{2}$ fields as well as one self-dual antisymmetric tensor field, in addition to the non-chiral fields.  All these chiral fields contribute additively to the anomaly. It is enough to identify the corresponding characteristic classes in 12 dimensions that are given for each chiral field (up to the factor $2\pi$) by the corresponding index.  There is no gauge group here and, hence, ${\rm ch}(iF)= 1$. We already know the index for the complex, chiral spin-$\frac{1}{2}$ field. Since here this is a real  (Majorana) chiral field, one just needs to divide the previous result \eqref{Diracgenusindex} by 2.

A spin-$\frac{3}{2}$ field can be obtained as a  spin-$\frac{1}{2}$ field with an extra vector index, and subtracting the  spin-$\frac{1}{2}$ part. The extra vector index can be treated as an $SO(d)$ gauge symmetry in the vector representation, resulting in a contribution to the index of \footnote{
Recall the generators \eqref{vectorgen} of the vector representation, so that $(R^{ab}J^{ab}_{\rm vect})_{cd}= 2i R^{cd}$. Then e.g. $\tr_{\rm vect} (R^{ab}J^{ab}_{\rm vect})^2= (2i)^2 R^{cd}R^{dc}$, where $R^{ef}=\frac{1}{2}R^{ef}_{\ \ \m\n}\d x^\m \wedge \d x^\n$.
} 
$\tr_{\rm vect} \exp\Big( \frac{1}{2\pi} \frac{1}{2} R^{ab}J^{ab}_{\rm vect}\Big)=\tr \exp\Big(\frac{i}{2\pi} R\Big)$, so that the index is
\be\label{gravitinoindex}
{\rm ind}(i D^{3/2}_{2n+2})=\int \Big[\wh A(R) \Big( \tr \exp\Big(\frac{i}{2\pi} R\Big)-1\Big)\Big]_{2n+2} \ ,
\ee
where the $-1$ simply subtracts the spin-$\frac{1}{2}$ contribution.
If such a spin-$\frac{3}{2}$ field would also couple to a gauge field, one would  need to include a factor ${\rm ch}(iF)$, but  this is not the case here. Moreover, if it is Majorana-Weyl, one needs again to divide by 2. Note that $\tr 1=\dd^a_a= d=2n$ just gives the dimension of the vector representation which is, of course, the space (-time) dimension. It the  follows that 
\be\label{expiRm1}
 \tr \exp\Big(\frac{i}{2\pi} R\Big)-1
 =
 2n -1- \frac{1}{2(2\pi)^2}\tr R^2 +\frac{1}{24(2\pi)^4} \tr R^4 - \frac{1}{720 (2\pi)^6} \tr R^6 +\ldots .
 \ee
 
For the self-dual antisymmetric tensor the relevant index turns out to be
\be\label{selfdualtensor}
{\rm ind}(i D^A_{2n+2})=-\frac{1}{2} \int\Big[ \frac{1}{4} L(R)\Big]_{2n+1} \ ,
\ee
where $L(R)$ is the so-called Hirzebruch polynomial, given by
\ba\label{Hirze}
L(R)&=& 1-\frac{1}{(2\pi)^2} \frac{1}{6} \tr R^2 
+\frac{1}{(2\pi)^4} \Big(- \frac{7}{180} \tr R^4 +\frac{1}{72} (\tr R^2)^2 \Big) \nonumber\\
&& +\frac{1}{(2\pi)^6} \Big(- \frac{31}{2835} \tr R^6 +\frac{7}{1080} \tr R^4 \tr R^2
-\frac{1}{1296} (\tr R^2)^3 \Big) +\ldots
\ea
The factor $\frac{1}{4}$ in \eqref{selfdualtensor} originates as a $\frac{1}{2}$ from the reality constraint and another $\frac{1}{2}$ due to the second chirality projector.  The sign reflects the fact that the antisymmetric tensor field is bosonic rather than fermionic.
The extra factor $\frac{1}{2}$ takes into account the different sizes of the spinor representation between $2n+2$ and $2n$ dimensions. Let us then  explicitly give the relevant 12-dimensional characteristic classes for the different anomalies in 10 dimensions.

Recall that the relevant $I_{12}$ are given by the integrand on the rhs of the index theorems, times a factor $2\pi$.  Specifically, in ten dimensions, and in the absence of a gauge group, we have  for a positive chirality Majorana spinor, positive chirality Majorana gravitino, and for a self-dual antisymmetric tensor \footnote{
Note that for the Majorana spin-$\frac{1}{2}$ without gauge group one simply has  $I_{12}^{1/2, {\rm Maj}}=2\pi \frac{1}{2} \wh A(R)_{12}$, and that the coefficient $-\frac{11}{126}$ of the $\tr R^6$ term in the bracket of $I_{12}^{\rm gravitino}$ arises as
 $2\pi \frac{1}{2}\Big( 1 + \ldots + \frac{1}{(4\pi)^6}\frac{1}{5670}\tr R^6\Big) 
 \Big(10-1 + \ldots - \frac{1}{720 (2\pi)^6} \tr R^6 \Big) =  2\pi \frac{1}{2} \frac{1}{(4\pi)^6}\Big( \frac{9}{5670} - \frac{64}{ 720}\Big)\tr R^6+\ldots = 2\pi \frac{1}{2} \frac{1}{(4\pi)^6}\Big( -\frac{11}{126}\Big)\tr R^6+\ldots $.
}
\ba
I_{12}^{1/2, {\rm Maj}}&=&2\pi\times \frac{1}{2}\times
 \frac{1}{(4\pi)^6} \Big(\frac{1}{5670}\tr R^6 +\frac{1}{4320} \tr R^4 \tr R^2   +\frac{1}{10368} \big( \tr R^2\big)^3\Big)\nonumber\\
 I_{12}^{\rm gravitino}&=&2\pi \times \frac{1}{2} \times{1\over (4\pi)^6} \left(
-{11\over126}\tr R^6
+{5\over96}\tr R^4\tr R^2
-{7\over1152}(\tr R^2)^3
\right)
 \nonumber\\
 I_{12}^{\rm antisym}&=&  2\pi \times {1\over8 (2\pi)^6}\left({31\over2835}\tr R^6
-{7\over1080}\tr R^4\tr R^2+{1\over1296}(\tr R^2)^3\right)\ .
 \ea
  Then, summing the contributions from $n_{3/2}$ positive chirality Majorana gravitini, $n_{1/2}$ positive chirality Majorana spin-$\frac{1}{2}$ and $n_A$ self-dual antisymmetric tensor fields, one gets the total anomaly polynomial
 \ba
 I_{12}(n_{1/2}, n_{3/2},n_A)&=& \frac{1}{128 (2\pi)^5} 
 \Bigg(\frac{ n_{1/2} -495 n_{3/2} +992 n_A}{5670}\tr R^6\, \nonumber\\
&& \hskip-1.5cm+ \frac{ n_{1/2} + 225 n_{3/2} -448 n_A}{4320}
\tr R^4\, \tr R^2 +\frac{n_{1/2} -63 n_{3/2} + 128 n_A}{10368} \, (\tr R^2)^3  \Bigg) \ .
 \ea
 The theory has gravitational anomalies unless all 3 terms vanish. This condition constitutes a set of 3 homogeneous linear equations for the 3 unknowns $n_{1/2}, n_{3/2}$ and $n_A$ which generically has no non-zero solution. However it is easy to see that here the system not only admits a solution, but the solution also is very simple, namely
 \be\label{IIBsol}
 n_A=1\ ,\quad n_{3/2}= 2\ ,\quad n_{1/2}=-2 \ ,
 \ee
 or any integer multiple of it. These numbers \eqref{IIBsol} correspond exactly to the field content of the ten-dimensional IIB supergravity~: one self-dual antisymmetric tensor, two positive chirality Majorana-Weyl gravitini and two negative chirality Majorana-Weyl spin-$\frac{1}{2}$ fields!
 

\newpage

\section{Conclusions}
\setcounter{equation}{0}

In these notes we have studied anomalies in quantum field theory from several complementary perspectives. We have seen how classical symmetries can fail to survive quantization when regularization is required. This phenomenon manifests itself equivalently as the non-conservation of a quantum current, as anomalous Ward identities, or as the non-invariance of the functional integral measure. We have provided a clear dictionary how these different manifestations correspond to each other.

The abelian anomaly provides the simplest example. Through both Fujikawa's method to compute the Jacobian determinant in the functional integral, and through the explicit computation of the corresponding triangle diagrams, it can be seen that the anomaly is  finite, local, and unavoidable. Its relation to the Atiyah--Singer index theorem in four dimensions already highlights its topological character. This abelian anomaly concerns global chiral symmetries, and only implies that certain selection rules no longer hold.

When chiral fermions are coupled to gauge fields, similar anomalies are present, but now they threaten the consistency of the theory itself. In this case, violation of gauge invariance would spoil unitarity and renormalizability, making anomaly cancellation a fundamental requirement. The remarkable fact that the anomaly is one-loop exact and constrained by the Wess-Zumino consistency conditions allows this requirement to be formulated as a purely group theoretical condition on the fermion representations.

The Standard Model offers a remarkable realization of an anomaly free theory~: the assignments of quark and lepton quantum numbers are such that all gauge anomalies cancel within each fermion generation. This non-trivial cancellation, first identified by Bouchiat, Iliopoulos and Meyer, constitutes one of the most profound structural features of the theory.

On the more formal side, the Wess-Zumino consistency condition was reinterpreted as the BRST closedness of the anomaly, and then relevant anomalies correspond to BRST cohomology classes at ghost number one on the space of local functionals in $d$ dimensions. This cohomology was shown to be uniquely characterized  by formal characteristic classes in $d+2$ dimensions. This in turn hinted at a relation with index theorems in $d+2$ dimensions, and following \cite{AGG} we have sketched how this allows to obtain the anomalies without doing a single loop computation, including also the generalization to gravitational anomalies. Even more amazing than the anomaly cancellation in the Standard Model is the highly non-trivial anomaly cancellation in ten-dimensional IIB supergravity.

Anomalies occupy a distinctive place in quantum field theory: they emerge from quantum effects but are governed by symmetry and topology, and they both violate classical invariance and enforce fundamental consistency conditions.




\end{document}